\documentclass[3p]{elsarticle}
\usepackage{lineno,hyperref}
\modulolinenumbers[5]
\usepackage[utf8]{inputenc}
\usepackage{graphicx}
\usepackage{amsmath}
\usepackage{siunitx}
\usepackage{longtable,tabularx}
\usepackage{float}
\usepackage{setspace}

\usepackage{caption}

\usepackage{subcaption}
\usepackage{multirow}
\usepackage{lscape}
\usepackage{multirow}
\usepackage{booktabs}
\usepackage[normalem]{ulem}
\usepackage[table,xcdraw]{xcolor}
\renewcommand{\figurename}{Fig.}
\begin{document}
\begin{frontmatter}
\title{Physics-Aware Neural Network Flame Closure for Combustion Instability Modeling in a Single-Injector Engine}
\author[]{Zeinab Shadram\corref{cor1}\footnote{Graduate Research Assistant, Department of Mechanical and Aerospace Engineering}}
\cortext[cor1]{Corresponding author} \ead{zshadram@uci.edu}
\author[]{Tuan M. Nguyen\footnote{Assistant Specialist, Department of Mechanical and Aerospace Engineering, current position: Postdoctoral researcher at Sandia National Laboratories}}
\author[]{ Athanasios Sideris \footnote{Professor, Department of Mechanical and Aerospace Engineering}}
\author[]{ William A. Sirignano \footnote{Professor, Department of Mechanical and Aerospace Engineering} }
\address{University of California Irvine, Irvine, CA, USA, 92697}
		\begin{abstract}
		Neural networks (NN) are implemented as sub-grid flame models in a large-eddy simulation of a single-injector liquid-propellant rocket engine with the aim to replace a look-up table approach. The NN training process presents an extraordinary challenge. The multi-dimensional combustion instability problem involves multi-scale lengths and characteristic times in an unsteady  flow problem with nonlinear acoustics, addressing both transient and dynamic-equilibrium behaviors,  superimposed on a turbulent reacting flow with very narrow, moving flame regions. Accurate interpolation between the points of the training data becomes vital. A major novel aspect of the proposed NNs is that they are trained to reproduce relevant portions of the information stored in a flamelet table by using only limited data from a few CFD simulations of a single-injector liquid-propellant rocket engine under different dynamical configurations.  This is made possible by enriching the training set with contrived data resulting from the physical characteristics of the combustion model and also by including the flame temperature as an extra input to the NNs that are trained to model other flame variables of interest.  These physics-aware NN-based closure models are first tested offline by comparing them directly with the flamelet table and then are successfully implemented into CFD simulations in place of the flamelet table and verified on various dynamical configurations.
		The results from those tests compare favorably with counterpart table-based CFD simulations.  Computational advantages of the approach are discussed.
		\end{abstract}
	\begin{keyword}
		{Machine Learning}\sep {Neural Network} \sep {Combustion Instability} \sep{Turbulent Combustion}
	\end{keyword}
	\end{frontmatter}

		 \section*{Nomenclature} 
	{\renewcommand\arraystretch{1.0}
		\noindent\begin{longtable*}{@{}l @{\quad=\quad} l@{}}
			$a_\gamma$ & ratio of specific heat capacity coefficient, \si{\per\kelvin}\\
			$C$ & progress variable\\
			$e_f$& flame internal energy, \si{\meter \squared \per \second\squared}\\
			$P$ & pressure, \si{\kilo \pascal}\\
			$R$ & gas constant, \si{\joule \per \kilogram \per \kelvin}\\
			$T$ & flow temperature, \si{\kelvin} \\
		   $T_f$ & flame temperature, \si{\kelvin} \\
			$U$ & axial velocity \si{\meter\per \second}\\
			$Z$ & mixture fraction\\
			$Z''^2$ & mixture fraction variance from the mean\\
			$\gamma$ & ratio of specific heat capacity\\
			$\lambda$ & thermal conductivity, \si{\watt \per \meter \kelvin}\\
						$\rho$ & density \si{\kilo\gram\per\meter\cubed}\\
			$\dot{\omega}_C$& progress variable reaction rate \textcolor{black}{(PVRR)}, \si{\kilogram \per
				\meter\cubed\per \second}\\
					$\dot{\omega}_i$& i-th species reaction rate, \si{\kilogram \per
				\meter\cubed\per \second}\\
					$\chi$& Scalar dissipation rate\\
			$\Phi$ & flamelet input set\\
			$\Psi$ & flamelet output set
	\end{longtable*}}	 
	\subsection*{Superscripts}
	{\renewcommand\arraystretch{1.0}
		\noindent\begin{longtable*}{@{}l @{\quad=\quad} l@{}}
			$\widetilde{\textcolor{white}{-}}$ &density-weighted Favre average\\
			$\overline{\textcolor{white}{-}}$& Reynolds average\\
			$\widehat{\textcolor{white}{-}}$& empirical mean of a set
	\end{longtable*}}	 
	\section{Introduction} \label{Intro}
In today's research world, huge availability of data has one eminent benefit, which is the use of machine learning to generate data-driven models. By means of machine learning and exploitation of this data, it is possible to explore potentially more descriptive models that help us achieve a better understanding of any phenomena.
In recent years, there are numerous works exploring the application of different machine learning techniques in fluid mechanics for reactive as well as non-reactive flow modeling. Different techniques such as classification, clustering, and regression analysis have been applied to various problems in fluid mechanics and combustion studies \cite{annurev-fluid-010719-060214}.
Clustering is used mainly in the tabulation of probability density function (pdf)-based turbulent flame models through self-organizing maps algorithms \cite{Ranade, FRANKE2017245, BLASCO2}. Principal component analysis (PCA) algorithms have been used for chemistry reduction in turbulent flame models \cite{ WAN2020119,  ALQAHTANI2021142, BABAEI20144622}.
The Neural Network (NN) approach, which is a regression based  machine learning method, has been applied to model turbulence flow with chemical reactions in \cite{okstateSAN2018681, CMexample2, CMexample_2016}. Convolutional NNs have been developed to estimate unfiltered variables from filtered ones from LES simulation with turbulent combustion \cite{Nikolau1}. The same technique has also been used to estimate the sub-grid reaction scale rates in \cite{CNNRR}.
The modeling of sub-grid quantities such as scalar dissipation rate is also discussed in \cite{spray1}.
More specifically, in combustion modeling, \cite{SEN2010566, SEN201062} used NN for the coupling of the linear eddy mixing sub-grid model with the LES. Owoyele et al. \cite{Oppel2019} developed a set of NNs to model diesel spray flame. 

The ultimate goal of utilizing machine learning is to use the data from experiments and computationally expensive CFD simulations to develop low-cost NN-based models. 
Among initial steps toward this goal, articles like \cite{IHME_SCHMITT_PITSCH_PCI32, Shadram_Journal_1,  sandiaFlamezhang} developed NN-based models utilizing data from CFD simulations with flamelet closure. Flamelet, first introduced analytically by Williams \cite{Williams75}, is defined as the thin reactive diffusive layer embedded in an otherwise non-reacting flow field. Peters \cite{Peters84} later proposed the idea of generating flamelet libraries based on presumed sub-grid distribution to treat turbulent combustion closure. From a data science perspective, these flamelet libraries are essentially sets of input/output blocks that can be isolated and replaced by NN-based models.
 Thus, the flamelet model provides a proper stepping stone for exploring the use of machine learning in the combustion studies. The NN-based models developed in \cite{IHME_SCHMITT_PITSCH_PCI32, Shadram_Journal_1, sandiaFlamezhang} successfully replace the flamelet table in their CFD testbeds. In particular, the NN-based flamelet model in \cite{sandiaFlamezhang} has been developed to be used with the Sandia Flame D. NNs have been used in \cite{IHME_SCHMITT_PITSCH_PCI32} for representing a flamelet model in a stably burning flame in Sydney bluff-body swirl-stabilized flame.

Recently, Shadram et al. \cite{Shadram_Journal_1} explored NN capability of replacing flamelet libraries as a sub-grid flame closure model to simulate longitudinal combustion instability in a single injector rocket configuration called the Continuously Variable Resonance Combustor (CVRC) \cite{Yu2,Yu3}. Highly turbulent flow field, with a Reynolds number of 400000 based on the mean oxidizer mass flow rate, occurs in the CVRC test-bed.  The flame is also subjected to intense acoustical fluctuation that significantly alters its behavior depending on the cycle \cite{Tuan1, Tuan4, Menon1}.
Pressure and thus the flame variables change drastically when the instability occurs. The NN training process presents an extraordinary challenge due to the  multi-scale lengths and characteristic times in an unsteady flow problem with nonlinear acoustics, addressing both transient and dynamic-equilibrium behaviors, superimposed on a turbulent reacting flow with very narrow, moving flame regions. Accurate interpolation between the points of the training data becomes vital. Shadram et al.  \cite{Shadram_Journal_1} addressed these challenges by using relatively large NNs, and proposing a training set selection method that requires full access to the pressure-dependent flamelet table. In that work, the NNs  were validated by testing on the table, and verified in multiple stability configurations of CVRC. 
Similar to previous work, the results in \cite{Shadram_Journal_1} suggests that although NN requires a higher computational cost than a look-up table representation of a flamelet table, it requires much less memory. This feature provides the flexibility of using graphics processing unit (GPU) to reduce the associated computational cost \cite{GPUERA}. 
The accuracy of the NNs in \cite{Shadram_Journal_1} are limited by the training data derived for the flamelet combustion model and thus inherent all the limitation associated with the model assumptions. In a novel step closer toward this goal, the NNs in this work are trained directly using the CFD data, which only observes parts of the flamelet space. Our model are then verified by testing the NNs on a number of different simulations. The designed NNs are also tested on the whole flamelet table library, which includes the unobserved parts as well as the training data. The objective is to develop new NNs flame closures that can be more universally applicable in any reacting flow simulations while removing as much possible limitations found in our previous work.

Section \ref{prelim} provides background on the flamelet model and the numerical scheme for our CVRC CFD analysis.
Section \ref{NNtext} discusses the NN design and testing process in detail. In Section \ref{Results}, first, the proposed NN-based flamelet models are validated by comparing them to table-based flamelet models, see Subsection \ref{NNtestSEC}. Then, the NN-based models are verified by comparing the NN-based CFD simulations with the table-based ones at different stability configurations in Subsection \ref{onlinetest}. The paper is concluded in Section \ref{conclu}. 

\section{Computational Setup} \label{prelim}
Combustion instability is a potentially destructive acoustical phenomenon inside a combustion chamber, where an unstable high-amplitude pressure oscillation  is excited and sustained by a high rate of combustion energy release. 
Study and theorization of liquid-propellant rocket engine (LPRE) combustion instabilities have been a major research subject for decades \cite{CrocCheng2, culick1995overview, Poinsot2, SigPop2, SigPop3}. One key component of these studies is the numerical analysis done through CFD simulations, which requires the development of descriptive models that can represent a complicated phenomenon such as combustion in a turbulent reacting flow with unsteady pressure. Research has been conducted to generate such descriptive models \cite{SAGHAFIAN2015652, Menon1}.

\begin{figure}[hbt!]
  	\centering
  	\begin{subfigure}{.49\textwidth}
  		\centering
  		\includegraphics[width=1\textwidth]{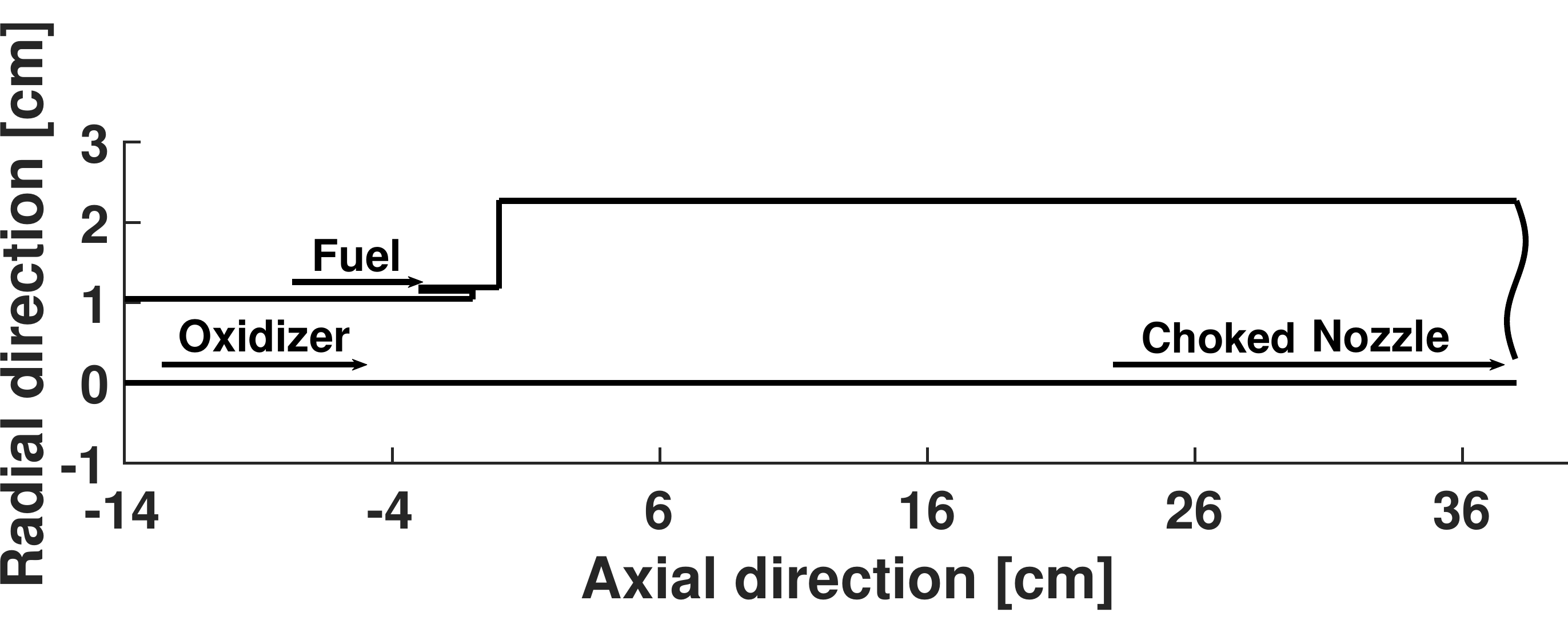}
  		\caption{ Schematic of CVRC configuration \cite{Tuan1}}	
  		\label{CVRC}		
  	\end{subfigure}
  		\begin{subfigure}{.49\textwidth}
  		\centering
  		\includegraphics[width=1\textwidth]{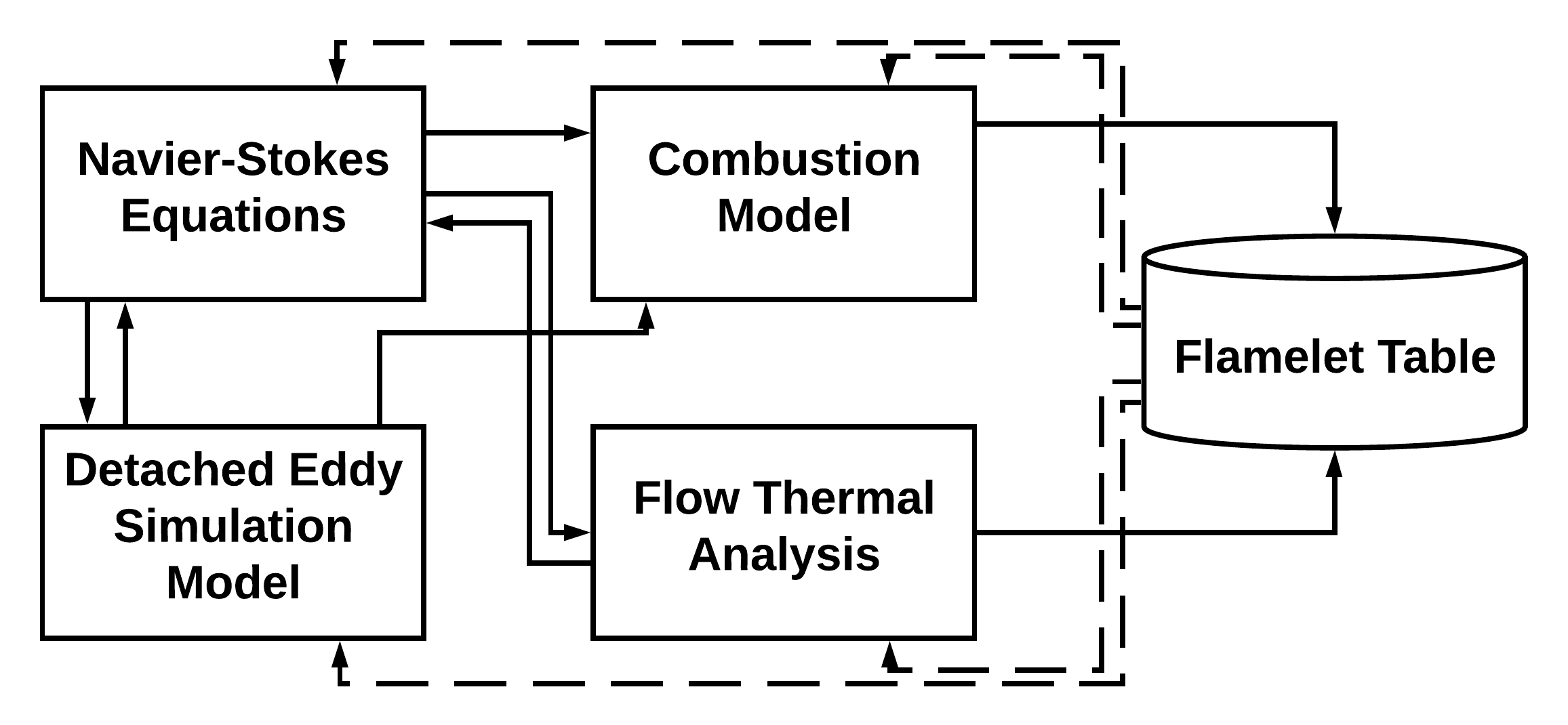}
  		\caption{Top level architecture of CFD code}
  		\label{toplevel}
  	\end{subfigure}
  	\caption{Overview of the computational domain for the CVRC experiment and the CFD code architecture}
  \end{figure}

In this work, we focus our modeling efforts on the combustion instability phenomenon, which occurs in the single element model rocket combustor called CVRC. 
A schematic of the CVRC experiment is shown in \figurename{~\ref{CVRC}} together with the CFD solver's outline. Different stability behaviors are observed for different oxidizer post lengths. At its most unstable operation, CVRC consists of a 14-\si{\centi\meter} oxidizer post with a 38-\si{\centi\meter} combustion chamber. In this configuration, methane is injected through the outer concentric tube at 300~\si{\kelvin}, and the oxidizer, composed of 58\% $H_2O$ and 42\% $O_2$,  is injected through the inner tube at 1030~\si{\kelvin}. The equivalence ratio in this globally fuel-lean flow is 0.8, which results from a constant mass flow rate of fuel and oxidizer  at $0.027$~\si{\kilogram\per\second} and $0.32$~\si{\kilogram\per\second}, respectively. Although the CRVC experiment uses gaseous injection, it is an accepted benchmark for computational methods that address combustion instability in Liquid Propellant Rocket Engines (LPREs).
  
Both spontaneous and triggered instabilities in this single-injector coaxial dumped combustor have been analyzed using CFD simulations in the past \cite{Tuan1, Tuan2, Tuan3}.
The analysis resolves both types of unsteadiness, turbulent flow fluctuations and large-amplitude acoustic oscillations. In our work, the flamelet theory is used as a sub-grid model because it introduces the reasonable set of multi-physical phenomena that exist at sub-grid dimensions, such as chemical kinetics and molecular diffusion of heat and mass. Turbulence is treated using delayed detached-eddy simulation. Due to the intense local extinction and reignition observed in the CFD calculations \cite{Tuan4}, the Compressible Flamelet Progress Variable (CFPV) \cite{Pecnik} is implemented.
For all simulations performed in this work, we use an in-house structured finite-difference solver, as shown in \figurename{~\ref{toplevel}}. The mesh structure is based on the mesh used in the 3D calculations in \cite{Menon1} and consists of  137,494 grid points. The smallest radial and axial grid sizes are 0.05 \si{\milli\meter} and  0.2 \si{\milli\meter}, respectively, which are located around the mixing shear layer or near any walls.
The overall accuracies in time and space are fourth-order and second-order, respectively.
At both reactant inlets, a constant mass flow rate boundary condition is implemented using the Navier-Stokes characteristic boundary conditions \cite{Poinsot4}.
A computationally efficient short-choked-nozzle \cite{CrocSig} outlet boundary condition is used.
The walls are assumed impermeable with the no-slip boundary condition. For the thermal aspect, two different boundary conditions are analyzed; in one case leading to spontaneous instability, the walls are assumed to be adiabatic. The second case assumes an iso-thermal wall temperature, which leads to triggering instability behavior \cite{Tuan3}. \cite{Tuan3, TuanThesis} give a complete description of the current solver and numerical setup validation .

\subsection{Flamelet and NN modeling} 
In the flamelet concept, the chemical time scales are shorter than the turbulent time scales so that the flame can be viewed as a collection of laminar flamelets \cite{peters_2000}. This definition allows the chemistry computation to be performed independently of the main flow simulation and pre-process as flamelet libraries/tables. Since Peters \cite{Peters84} first proposed the turbulent flamelet closure method, there has been intense focus on developing flamelet models for non-premixed combustion with the most popular examples being the Steady Laminar Flamelet (SLF) \cite{Peters84}, Representative Interactive Flamelet (RIF) \cite{peters_2000}, and Flamelet Progress Variable (FPV) \cite{pierce_moin_2004} models. Regardless of the implementation strategy, one must first obtain the flamelet solutions. In this work, we solve the second order quasi-steady laminar flamelet equations in the mixture fraction space: 
	
\begin{equation}
-\frac{\rho \chi}{2} \frac{\partial^2  \psi_i}{\partial Z^2}=\dot{\omega}_i 
\end{equation}  
where $\psi_i$ can be any reactive scalar quantities such as species mass fractions and temperature.

In the current CFPV model \cite{Saghafian1}, presumed PDFs are used to relate the laminar flamelet solutions in the mixture fraction space to their Favre-averaged/mean counterparts. The $\beta$ PDF is assumed for the mixture fraction while the Dirac $\delta$ PDF is assumed for both the progress variable and pressure. The Favre-averaged thermo-chemical quantities ($\tilde{\psi}_i$) at each pressure value are pre-processed as lookup libraries using the convolution: 

\begin{equation} \label{equa:flame1}
\tilde{\psi}_i(\tilde{Z}, \widetilde{Z''^2}, \tilde{C},\bar{p}) = \int_{0}^{1} \int_{0}^{C} \int_{p_o}^{p} \psi_i (Z,C) \beta ({Z}, {Z''^2}) \delta ({C}) \delta ({p}) dZ dC dp
\end{equation} 

\noindent
where $Z$ is the mixture fraction, $C$ is the progress variable. During the LES calculations, the solver only requires transporting a reduced set of governing equations for $\tilde{Z}, \widetilde{Z''^2}, \tilde{C}$, with $\tilde{C}$ being the only reactive scalar. Therefore, flamelet model avoids the well-known stiffness problem associated with finite rate combustion models, especially for skeletal mechanisms such as the 27-species and 72 reactions one used in this work. 
This presents no physical error in the expected situations where mixing is generally the rate-controlling process, but detailed chemistry is needed to determine accurately the flame temperature and product composition.
Complete description of the flamelet governing equations as well as in-depth examination of the CFPV model impact in predicting combustion instability in the CVRC are given in \cite{Tuan4}. 

There are 7 flamelet tables, which describe the flame thermo-chemical properties, needed for the CFD calculations:
	$\tilde{\Psi}=[\widetilde{\dot{\omega}}_C,\widetilde{T}_f,\widetilde{e}_f, \widetilde{R},\widetilde{\lambda},\widetilde{\gamma}
	,\widetilde{a}_\gamma]$. From a data science perspective, the seven flamelet libraries as formulated above are a set of well characterized outputs needed for CFD calculations with a reasonably well defined set of inputs: $\Phi=[\widetilde{Z},B,\widetilde{C},\overline{P}]$, where parameter $B=\frac{\widetilde{Z"^2}}{\widetilde{Z}(1-\widetilde{Z})}$ is a surrogate variable that resembles the variance of the mixture fraction to describe the combustion state in the turbulent flow. Therefore, reduced manifolds such as flamelet libraries are well suited for Neural Network closure model development.
  
Note that despite their advantages, the flamelet models still possess many limitations. The first weakness of the flamelet model is its turbulent closure approach , where sub-grid Probability Density Functions (PDF) are assumed for transported variables. Furthermore, flamelet theory usually assumes either an axisymmetric or planar strain field, while three-dimensional structure exists in realistic flow field. In addition, flamelet theory assumes that fuel or oxidizer would only exist on one side of the flame, yet, combustible mixtures can exist on both sides in a diffusion flame. Detailed discussions on the challenges toward improving flamelet models are provided in \cite{Sirignano_coufl2019, SirignanoWSSCI, ClaudiaWSSCI,sirignano_2021,sirignano_2}.

Fully regression-based data-driven models such as NNs, are designed to predict the output data provided to them as training examples, with the least error. Therefore, their predictive capability is limited by the properties of this data. Selecting the richest and most informative training data set is a crucial step in developing NN-based models. 
At current stages of exploring the potential of using NNs in combustion, where the NNs are trained based on currently available models, one should note that the predictive accuracy of the resulting models is limited by the accuracy of the target model.

One such way to improve the NN model proposed in this work is to provide it with higher fidelity training data, such as the table data coming from the Transported PDF method \cite{Pope_PDF} together with In-Situ Adatptive Tabulation (ISAT) method \cite{Pope_ISAT}. The Transported Probability Density Function (TPDF) \cite{Haworth,PopovPope} is arguably the best closure model for chemistry-turbulence interaction, as it does not require any additional model for the chemical source terms. However, because of the high dimensionality of its argument, the model requires Monte-Carlo simulations of at least 30-50 notional particles in a cell. The PDF simulations are, thus, usually very computationally expensive even with a simple chemistry model \cite{Haworth}. Fortunately, ISAT can be used to accelerate chemistry calculation in a TPDF calculation by generating on-the-fly tables that store thermo-chemical states from the previous timestep. Therefore, manifolds produced from such calculations do not suffer from simplifying assumption of the flamelet model. Furthermore, because of the adaptive tabulation strategy, we can eliminate the need to design sample selection strategy, as presented in section \ref{nn_selection} below. Furthermore, successful NN models generated from such dataset can significant reduce the computational cost while maintaining reasonable accuracy compared to the TPDF method.

\section{Neural Network Design} \label{NNtext}
\begin{figure}[hbt!]
	\centering
	\includegraphics[width=.8\textwidth]{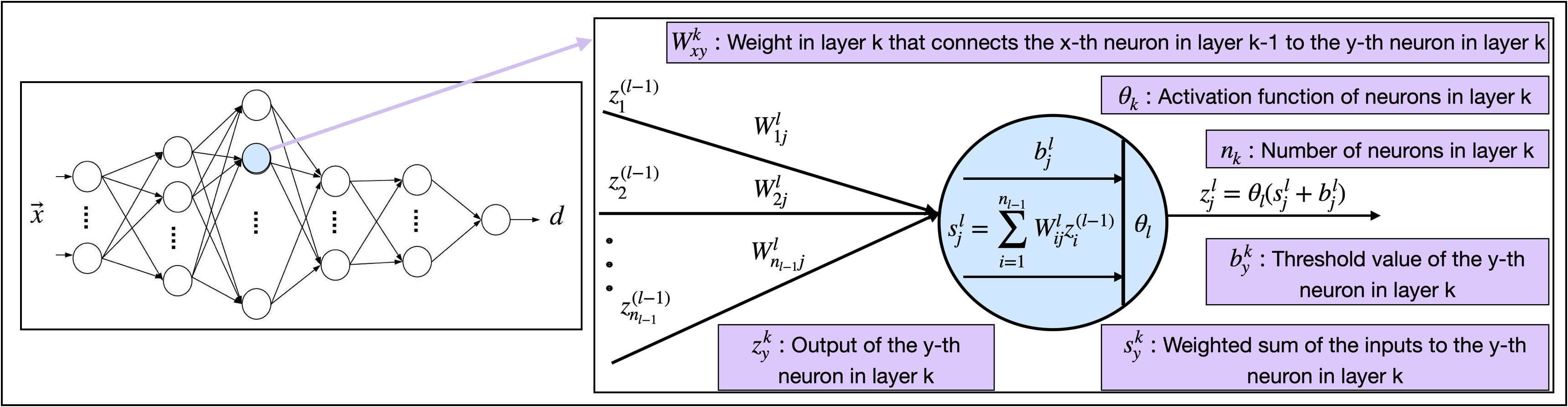}
	\caption{ Single output NN structure and schematic of a neuron as a computing unit in the NN}	
	\label{NN2}		
\end{figure}

Neural networks are trained to emulate an input-output mapping based on a provided set of samples.  Feed forward neural networks consist of connected layers of computational units called neurons. Each neuron performs a nonlinear function (activation) on a linear combination of the outputs of neurons in the previous layer via weights that collectively constitute the parameters of the NN. Figure \ref{NN2} provides a schematic of a neuron as a computational unit. 
The neurons in the final layer generate the output of the NN. Regression-type NNs, which are of interest in this work, minimize a cost function, which is the difference between the outputs of the NN-based model and the targets (training samples), in a least mean squared error sense over the parameters of the NN. 
This minimization is accomplished via the iterative Backpropagation method, which is an efficient implementation of gradient descent \cite{BP}. The cost function, $R$, of the regression NN for $N$ training samples is defined as: $R=\frac1N\sum_{i=1}^{N}(d_i-y_i)^2$, where $d_i$ is the sample output associated with an input $x_i$, and $y_i$ the predicted output calculated by the NN. To find appropriate weights $W_l$ for layer $l$, one minimizes $R$ with respect to $W_l$, requiring that:
\begin{eqnarray}
	\frac{\partial R}{\partial W_l}=0 \Rightarrow	\frac{1}{2N}\sum_{i=1}^{N}(d_i-y_i)\frac{\partial y_i}{\partial W_l}=0 \label{opt1}
\end{eqnarray}
 The Backpropagation algorithm recursively computes $\frac{\partial y_i}{\partial W_l}=0 , l=L,L-1,\ddots ,1$ in Eq.~\eqref{opt1}. In its simplest form,
  this iterative algorithm starts from assigning a random initialization of the weights and proceeds with calculating the cost function and updating the weights in the direction of the negative gradients, i.e. $-\frac{\partial R}{\partial W_l}$, with a  step size $\alpha>0$ that is called the learning rate. More specifically,
\begin{eqnarray}
	W_l(k+1)=W_l(k)-\alpha \frac{\partial R(W_1, W_2,...,W_L)}{\partial W_l} \label{updrule}
\end{eqnarray}
Using a random subset of the training dataset at each iteration results in the method of stochastic gradient descent (SGD). More advanced methods for selecting the step size at each iteration  are comprehensively discussed in \cite{SDGII}.

Here, the designed NNs (one for each of the flame variables) are meant to replace the flamelet table in the CFD structure shown in \figurename{~\ref{toplevel}}; thus, the input set to the NN is $\Phi$  and the output set is from $\Psi$. 
As discussed in Sec.~\ref{Intro}, the ultimate goal of using NN is to develop a fully data-driven model from experimental and numerical data. 
This goal can be achieved if a rich enough data set of training samples is provided that covers different regions of flame dynamics. Since each NN is essentially designed to learn the pattern among these samples, the richer the training data set is the more comprehensive model can be developed. 
In this work, we use only the data that are generated from a single injector configuration. However,  in a multi-dimensional system like combustion, one configuration observes only a  manifold or a subset and not the whole dynamic. The availability of comprehensive data is essential in developing data-driven models. However, comparing the NNs  trained on limited data helps to test the extent of the modeling power of NN beyond the training region and paves the way for further and more complicated methods.
\subsection{NN Development: Sample Selection} \label{nn_selection}
A challenge that is particularly associated with learning from large data sets (numerical and or experimental data) is the selection of a sufficiently representative training set that does not lead to a high computational cost for the training process. 
As an example, a CFD simulation of spontaneous instability, in which the pressure signal grows from a relatively low pressure amplitude to a large amplitude limit cycle, requires simulation of 10 ms of the system to capture the whole dynamic. This simulation, the so-called transient case \cite{Shadram_Journal_1}, is used mainly as the training set source.
Storing data from this simulation at 200 \si{\kilo\hertz} leads to a pool of data with around  275 million points. 
 
In order to select the best training set at this stage, the framework of the flamelet table is utilized. In the table-based CFD set-up, the flame variables for a given input vector are calculated by linearly interpolating the values stored in the flamelet table. We then form the training set from the table values and corresponding inputs that have been called through the CFD simulation of each configuration.
As mentioned above, by adding more simulations and experiments to the main pool of data, a richer set can be developed at the cost of more complex data management at the training stage. Here, we mainly use  the transient  simulation of CVRC with 14-\si{\centi\meter} oxidizer post. This data set leads to developing a relatively good model for every variable except for PVRR, due to lack of data in very high pressure regions. To remedy this situation, data from another simulation that is started after the occurrence of  the instability, i.e. after high limit cycles have set-in, the so-called dynamic equilibrium case is also included in the training set for PVRR. 

 The observed points by the transient case amount to less than 3.3\% of the flamelet table.
Using only the transient case, significant regions of the table are not observed in the training set. To have a more reliable model, it is important that our model meets the physics-related constraints of the system.
The training set generated from sampling the CFD simulation data does not necessarily include the boundary points of the flamelet table,
 and these points define the flame dynamics of the sub-grid counterflow model. For example, for every pressure, turbulence level, and progress variable, the species mass fractions and temperature are known when the flow is purely oxidizer or purely fuel, which are indicated by $\widetilde{Z}=0$ and $\widetilde{Z}=1$. Accordingly, the value of all other flamelet outputs is known at these two points, regardless of the availability of training data from these regions. Moreover, the training set can be enriched by our knowledge of the flow when the progress variable is zero. At this point, there is no combustion, and the flamelet model is governed by the flow dynamics. Therefore, the flow data can be added to the training set.

 \subsection{NN Development: Algorithm and Structure}
 The training set based on the transient CFD simulation is used to train the NNs for each output table variable. To this end, three main NN structures are proposed based on the trade-off between increasing accuracy and decreasing the  computational cost in data retrieval. A NN with a lower number of neurons and layers but a sufficiently small error is desired for each variable.

While the proposed structures are different for different variables, the training process is common.  In all the NNs, the activation function in hidden layers is selected to be leaky rectified linear $LReLU(x)=max(x,0.001x)$ \cite{ReLu}. In the output layer, that function is linear.  The NN weights are initialized randomly by the Xavier initialization algorithm \cite{pmlr-v9-glorot10a}  and updated via the Adaptive moment estimation (Adam) update rule \cite{murphy2013machine}  used in the back-propagation procedure.  Adam is a well-known adaptive update rule, in which the learning rate for a weight at each iteration is determined by the running average of the magnitudes of recent gradients for that weight and the second moment of these gradients. The learning rate is initialized with 0.001.

A mini-batch, with a 30\% of the data, training mode is also used, as it is helpful to avoid local-minima in the training process, and it more efficiently handles large data sets.   The issue of overfitting can be observed even in simple regression problems, in which fitting a very high order polynomial leads to a smaller fitting error but a huge test error. To avoid overfitting and to improve the performance of the NN, we employed  Tikhonov regularization during the training process.
 
 The flamelet model assumes second-order ODEs for the species mass fractions and for the flame temperature. 
Other outputs of the flamelet model are essentially functions of species mass fractions and flame temperature. Thus, in developing NN-based models for those variables, flame temperature is also considered as an input. Accordingly, we have managed to significantly improve the accuracy of our NN-based models for $\Psi_1=[\widetilde{e}_f, \widetilde{R},\widetilde{\lambda},\widetilde{\gamma}
,\widetilde{a}_\gamma]$. Including the flame temperature as an input along with the other flamelet inputs ($\Phi=[\widetilde{Z},B,\widetilde{C},\overline{P}]$) not only helps to improve the accuracy but also ensures that the flamelet outputs are consistent when utilized in a CFD simulation by maintaining correct physical dependencies between different flame variables. A disadvantage of this approach is that the modeling error in  $\widetilde{T_f}$ can propagate in other variables as well.

\begin{figure}[hbt!]
	\centering
	\begin{subfigure}{.49\textwidth}
		\centering		
		\includegraphics[width=.95\textwidth]{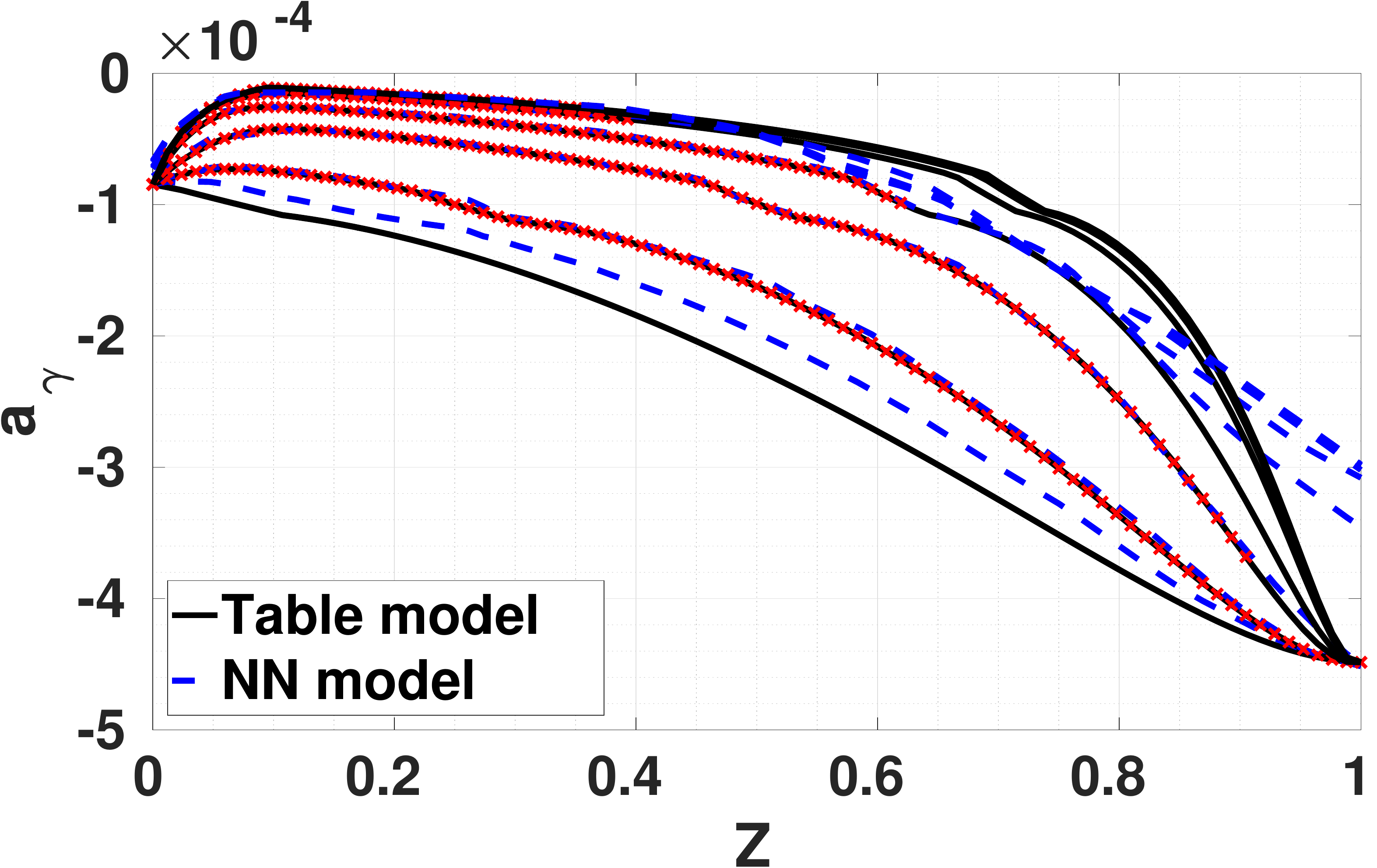}
		\caption{no training enhancement, standard input }\label{Offline_Res/bad/A_gamma_B1_table}	
	\end{subfigure}	
	\begin{subfigure}{.49\textwidth}
	\centering		
	\includegraphics[width=.95\textwidth]{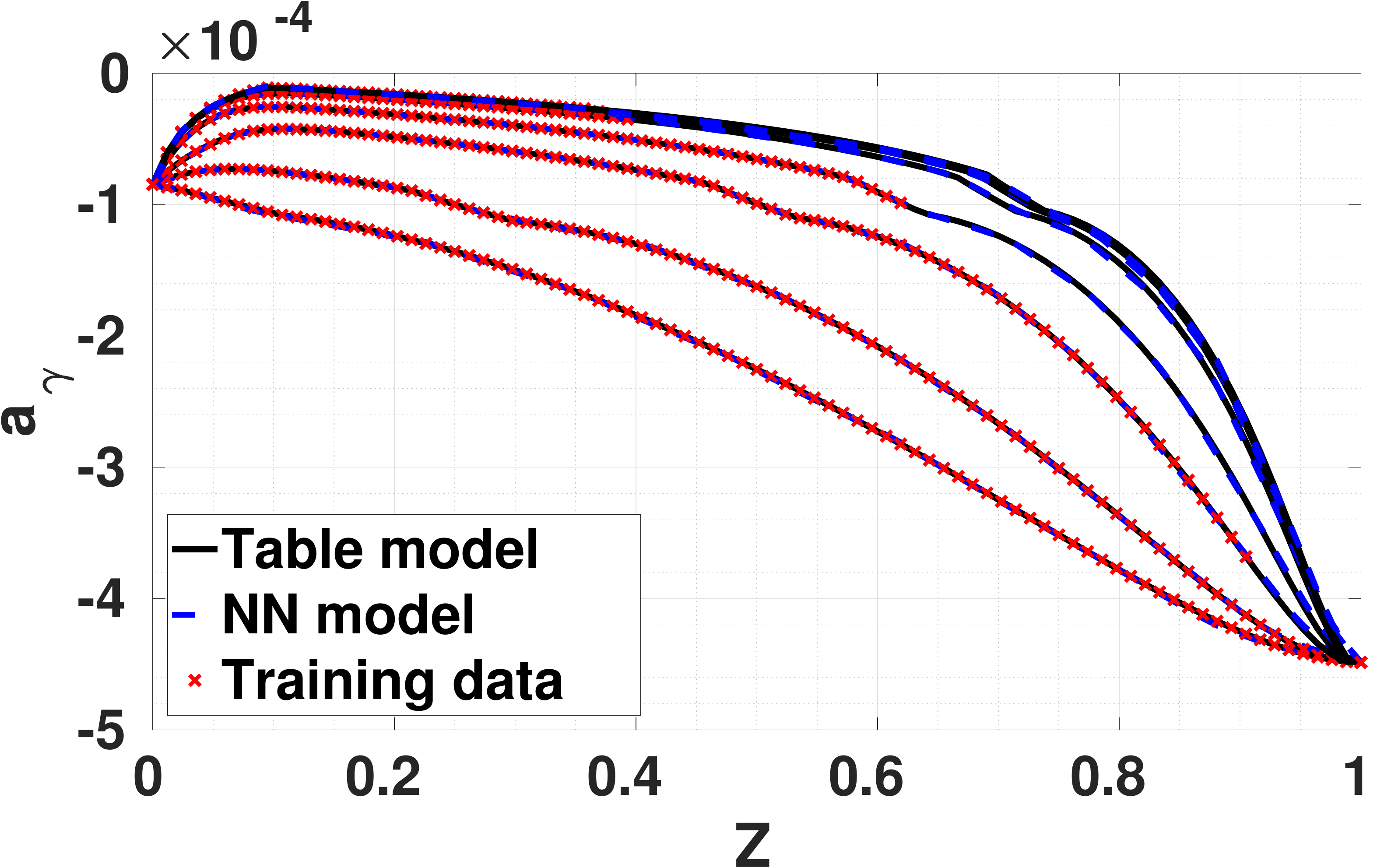}
	\caption{training enhancement, temperature added input}\label{Offline_Res/combo/A_gamma_B1_table}	
\end{subfigure}	
	\caption{Analysis on the effect of training data set enhancement and adding temperature to the input set on the performance of NN models for $a_\gamma$ at B=0 when $C$ varies.}\label{T_effect}
\end{figure}

Including flame temperature as an additional input improves accuracy in NN prediction. The highest accuracy improvement is for the linear expansion coefficient of heat capacity ratio ($a_\gamma$), as shown by the comparison in \figurename \ref{T_effect}. The difference between CFD and flamelet temperatures, and consequently the difference between flamelet and flow internal energies, are very significant in the compressible regime. In our case, they are caused by the strong pressure gradient imposed on the flow field due to combustion instability. Therefore, the NN model ability to capture accurately $a_\gamma$ behavior is crucial in this work. A NN developed on the original data set is shown in \figurename \ref{Offline_Res/bad/A_gamma_B1_table}.  \figurename \ref{Offline_Res/combo/A_gamma_B1_table} shows the results in which the training set is enhanced with values of $a_\gamma$ at $Z=0$, $Z=1$ and $C=0$, and flame temperature is added as an extra input to the model. Adding temperature as an extra input to the model helps the NN to learn the dynamic that was not observed in the original training data set. Significant improvement in the model prediction is observed for the interval of mixture fraction between 0.6 to 0.8; where the NN model faithfully captures the steep gradient dynamics when flame temperature is included as an extra input.  
 
 The NN designed for PVRR is  the most expensive structure among all variables. Due to the very nonlinear dynamic of PVRR, finding an accurate NN model is a challenging task; this is discussed in more detail in Sec.~\ref{PVRRsec}. Our study did not indicate a considerable improvement in coupling PVRR inputs with $\widetilde{T_f}$. The large structure for PVRR is selected after testing many other smaller structures which failed to capture its dynamic.
The training error and testing error for the flame variables and PVRR are provided in \tablename~\ref{Selected_NNstruc} and further discussed in Sec.~\ref{NNtestSEC}. 

\begin{table}[hbt!]
	\centering
	\begin{tabular}{c c c c c}
	\hline
		\textbf{Quantity}&\textbf{ NN structure} &\textbf{ FLOP} & \textbf{Training error (\%)} & \textbf{Testing error(\%)} \\ \hline
		$\widetilde{T_f}$      & 4-15-10-15-1                & 375  & 0.59               & 4.83    \\ 
		$\widetilde{e_f}$     & 5-15-10-15-1                & 390  & 0.0566             & 0.57     \\ 
		$\widetilde{R}$       & 5-15-10-15-1                & 390  & 0.0526             & 0.65      \\ 
		$\widetilde{\lambda}$   & 5-15-10-15-1                & 390  & 0.21               & 3.57        \\ 
		$\widetilde{\gamma}$   & 5-15-10-15-1                & 390  & 0.0371             & 0.48   \\ 
		$\widetilde{a}_\gamma$ & 5-15-10-15-1                & 390  & 0.22               & 1.41       \\ 
		$\widetilde{\dot{\omega}}_C$     & 4-15-20-25-35-25-20-15-5-1  & 3490 & 1.49               & 32.86    \\ \hline
	\end{tabular}
	\caption{Selected structure for each NN for the flame variables} \label{Selected_NNstruc}
\end{table}
 The structure of the NN proposed for each variable is provided in the second column of \tablename~\ref{Selected_NNstruc}.
In the third column, the computational cost of generating one output from each of the proposed NNs is given in terms the number of floating points operations (FLOP). Implementing these NNs inside CFD costs around 5815 FLOPs for calculating the outputs of the NN-based flamelet model at each point in time and space, which is the least NN-related computational cost that we could achieve among all of our several experiments, including our previous work \cite{Shadram_Journal_1}. As discussed in the literature, the computational cost issue related to the NN can be resolved by the utilization of GPU. However, in this work, CPU is used as the computational resource. 

\section{Results and Analysis}\label{Results}

The designed NNs are validated through different levels of testings using both offline and online data.  Section ~\ref{NNtestSEC} presents comparison between NN results trained from CFD data and the flamelet table itself to assess their performance. The objective of this activity is to identify the optimized NN structure. Note that such optimized NN should not be taken in the most absolute sense, but rather as the most optimized for the current configuration. The NNs developed from the first test are then implemented into the main CFD solver. CFD results between NN based and table based, which is where the training data are obtained, are compared. Finally, NNs simulation of triggered instability is performed.
In this configuration, a 1030-K isothermal wall boundary condition is applied to dampen the instability of the dynamic equilibrium. Different perturbation waves are applied to the reactant flow rate, thus re-triggering the instability. The readers are referred to Nguyen et al.  \cite{Tuan3} for an extensive analysis the triggering instability phenomenon. Here, a setup with excitation through two periods of sinusoidal variation
in mass flow rate is applied for the NN-based simulation. The main objective is to understand the potential of NNs for capturing correct behavior on data that they have not been trained on.

\subsection{ NN Validation by being Tested Flamelet Table}\label{NNtestSEC}
To analyze the performance of each NN, their outputs are compared with the corresponding outputs of the flamelet table, on the training set and on the whole table.
 To compare the set of the outputs, the overall relative error ($e_o$) defined in Eq.~\eqref{eodef} is used:
\begin{eqnarray}
	e_o=\frac{{||\vec{d}-\vec{y}||_2}}{{||\vec{d}||_2}} \label{eodef}
\end{eqnarray}
Here, $\vec{d}$ represents the outputs of the truth model (the flamelet table here), and $\vec{y}$ represents the outputs of the NN-based model. The training 
and testing errors are provided in \tablename~\ref{Selected_NNstruc}, fourth and fifth columns. $e_o$ is essentially the root mean squared (rms) of the absolute error normalized by the rms of the target set. PVRR has the highest error in both cases. The variables in the $\Psi_1$ vector, which have flame temperature in their input set, are modeled with less than 0.25\% in training and 4\% error  in test error. The flame temperature is modeled with an error of less than 0.6\% in training and less than 5\% in testing.

As discussed before, only parts of the flamelet library are included in the training set obtained from the CFD simulation. The distribution of the input values is not uniform, and the performance of the NN in the training stage and consequently testing is skewed by these distributions. For example, the frequency of occurrence of inputs in the laminar region or regions with lower turbulence is higher than that of points with higher turbulence levels. Therefore, it is to be expected that the performance of NN in the laminar zone will be better than in the rest of the table. 

\begin{figure}[hbt!]
	\centering		
	\begin{subfigure}{.49\textwidth}
 		\centering
\includegraphics[width=1\textwidth]{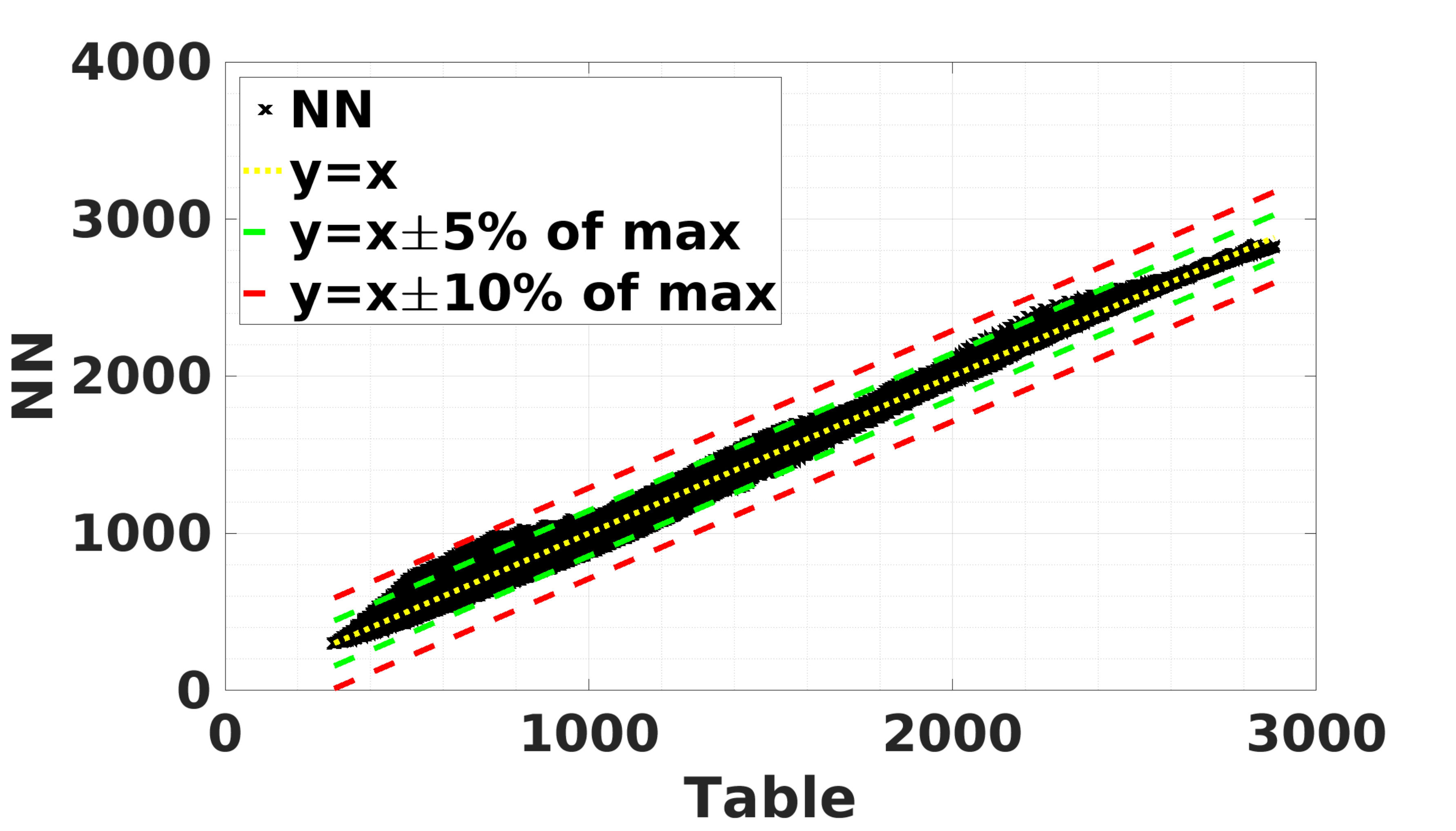}
\caption{ $\widetilde{T_f}$ (\si{\kelvin}) }\label{TfNN_corr}
	\end{subfigure}				
\begin{subfigure}{.49\textwidth}
\centering
\includegraphics[width=0.98\textwidth]{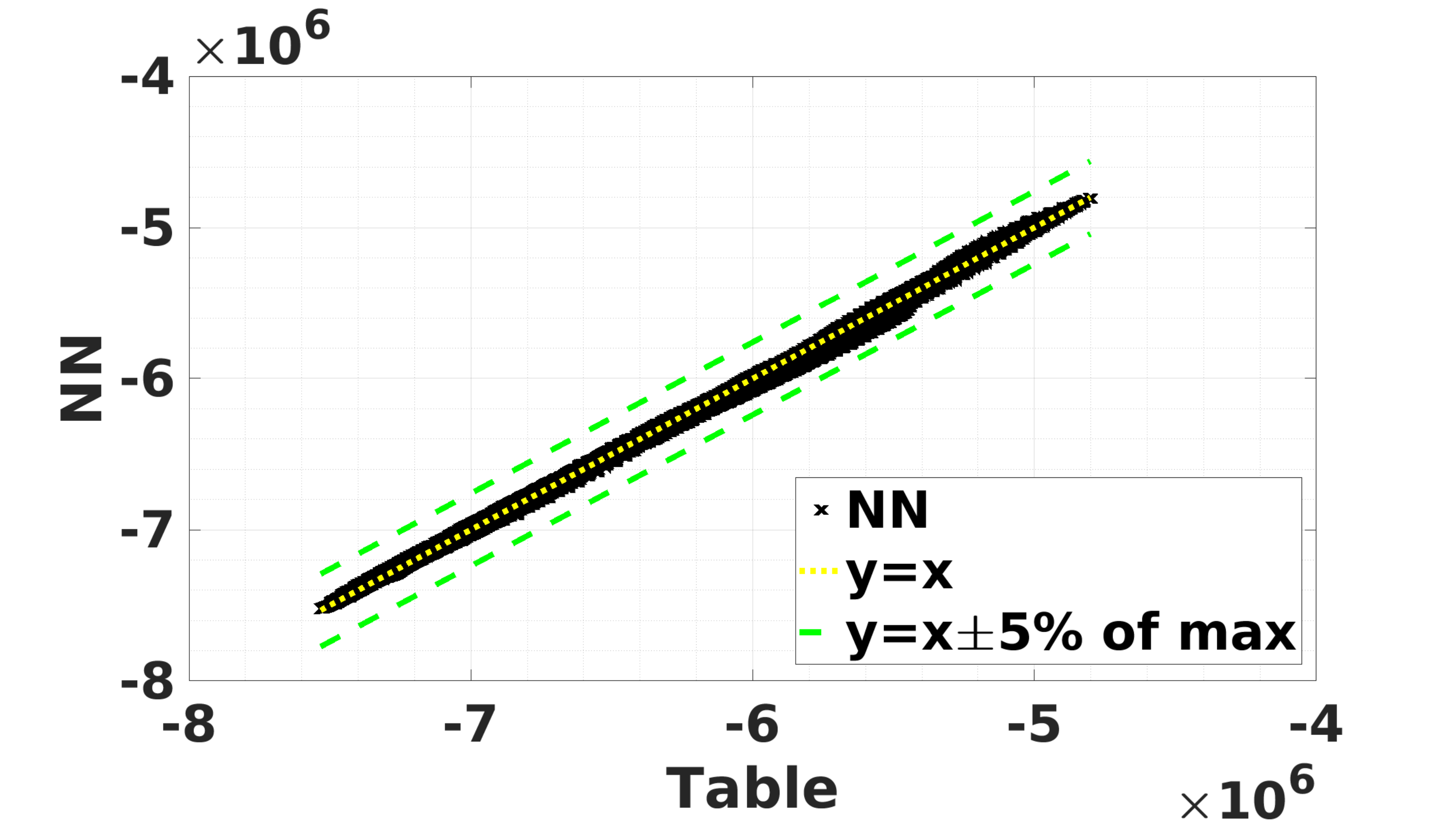}
\caption{ $\widetilde{e}_f$ (\si{\meter\squared\per\second\squared})}\label{ALLIEfNN}			\end{subfigure}				
\begin{subfigure}{.49\textwidth}
\centering
\includegraphics[width=0.98\textwidth]{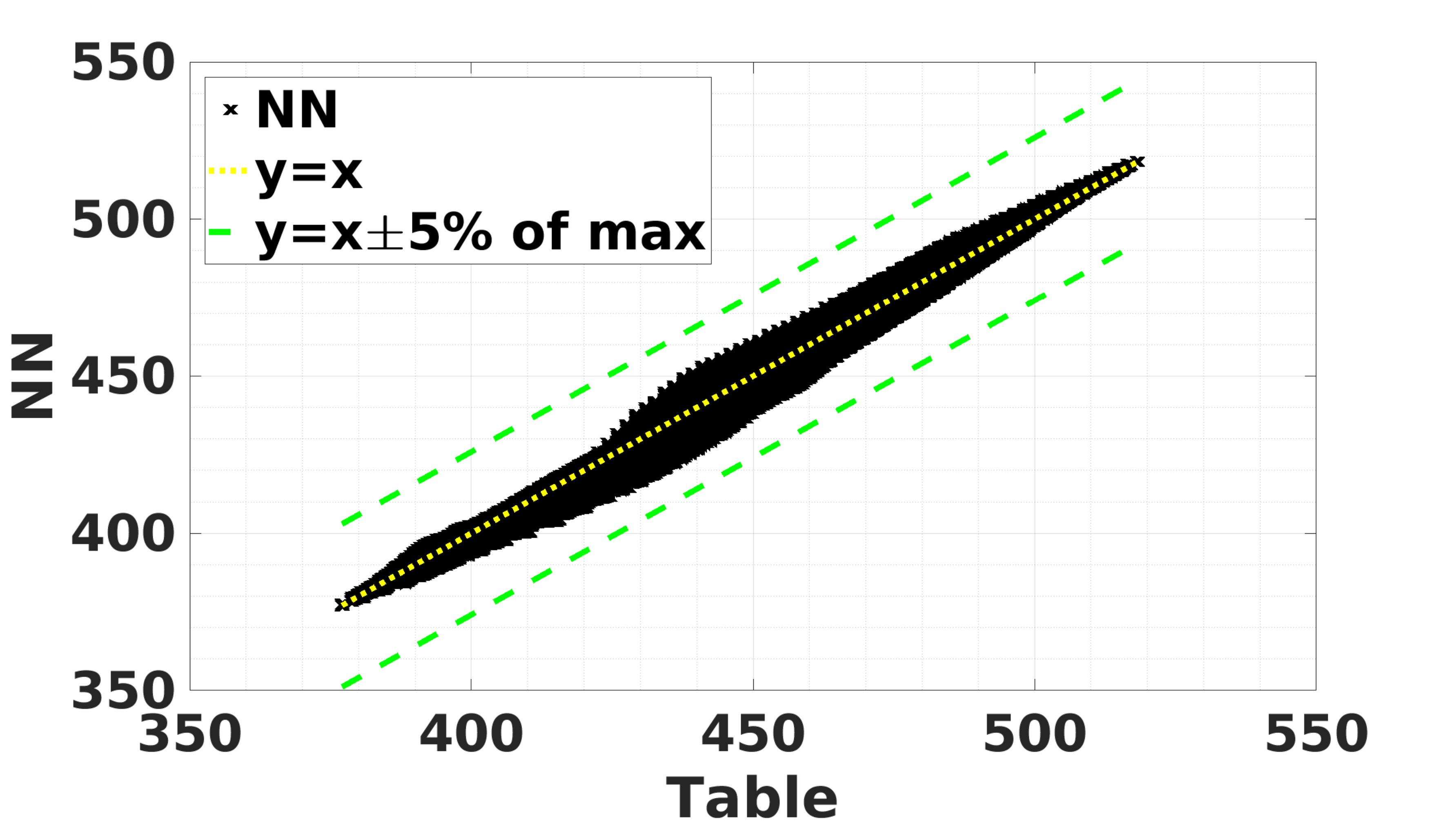}
\caption{$\widetilde{R}$ (\si{\joule\per\kilo\gram\kelvin})}\label{ALLRNN}
	\end{subfigure}				
	\begin{subfigure}{.49\textwidth}
	\centering		
\includegraphics[width=0.98\textwidth]{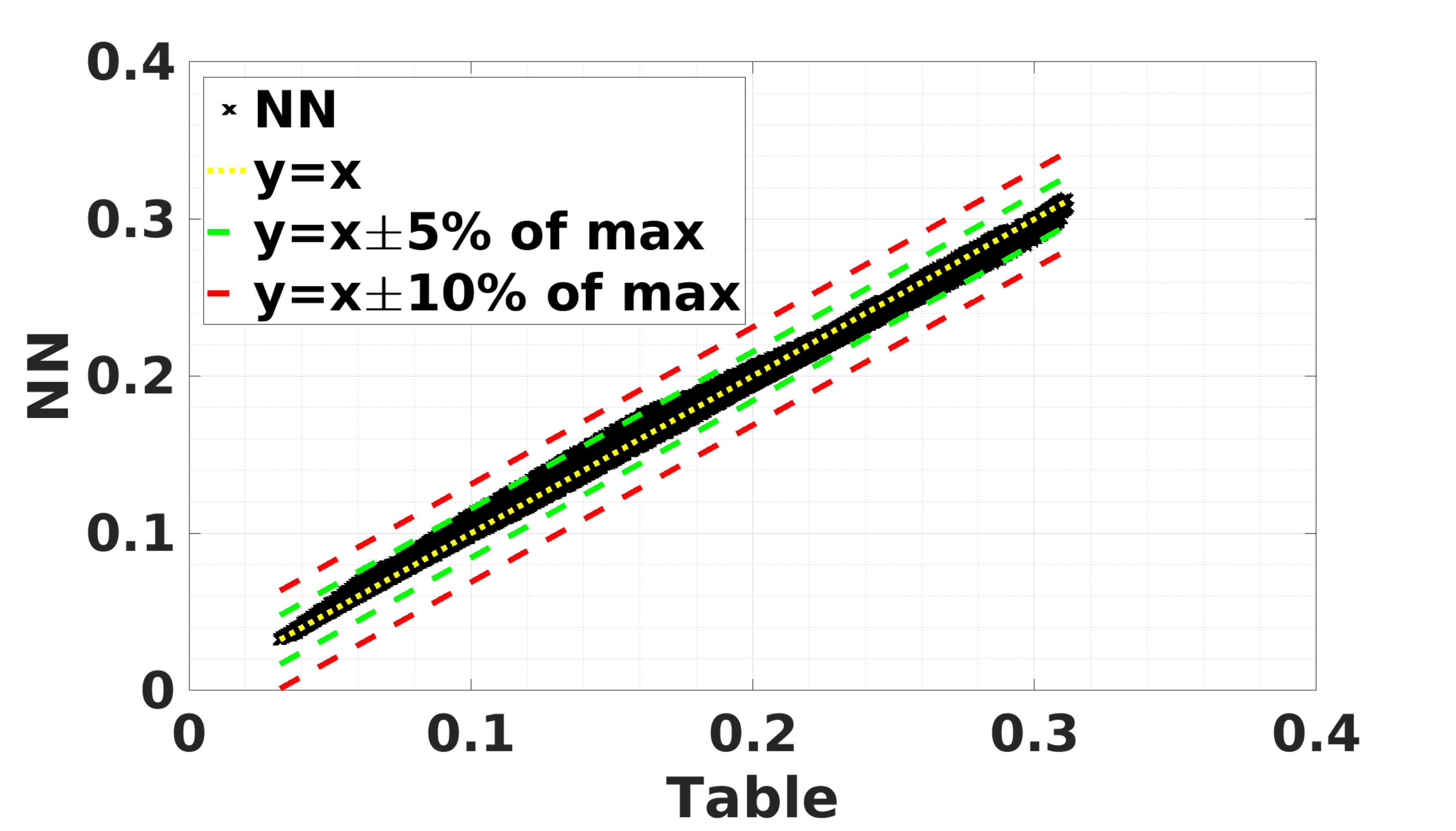}
\caption{$\widetilde{\lambda}$ (\si{\watt\per\meter\per\kelvin})}\label{ALLlambaNN}
	\end{subfigure}				
\begin{subfigure}{.49\textwidth}
\centering		
\includegraphics[width=0.98\textwidth]{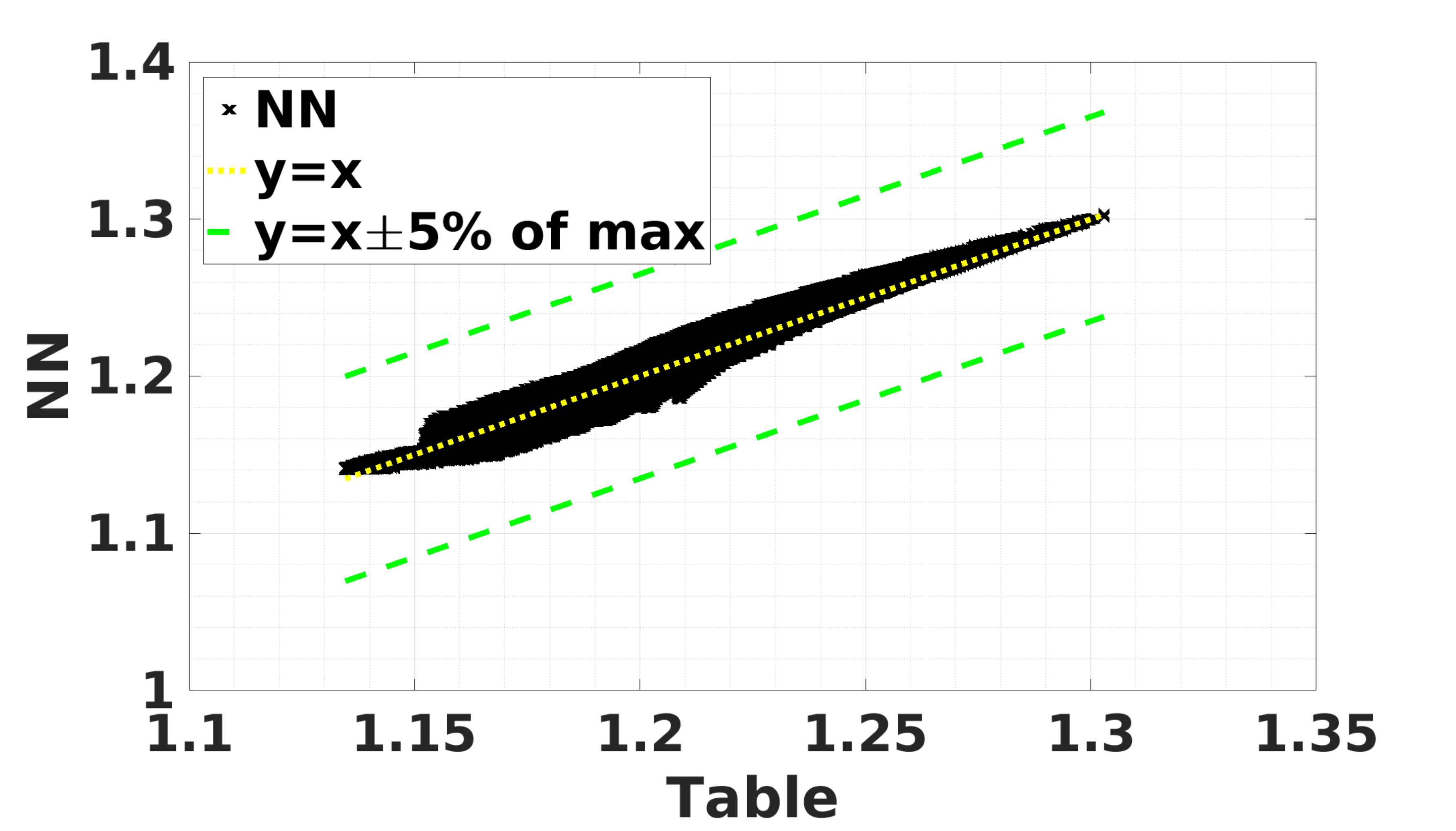}
\caption{$\widetilde{\gamma}$ }\label{ALLgammaNN}
	\end{subfigure}				
\begin{subfigure}{.49\textwidth}
\centering
\includegraphics[width=0.98\textwidth]{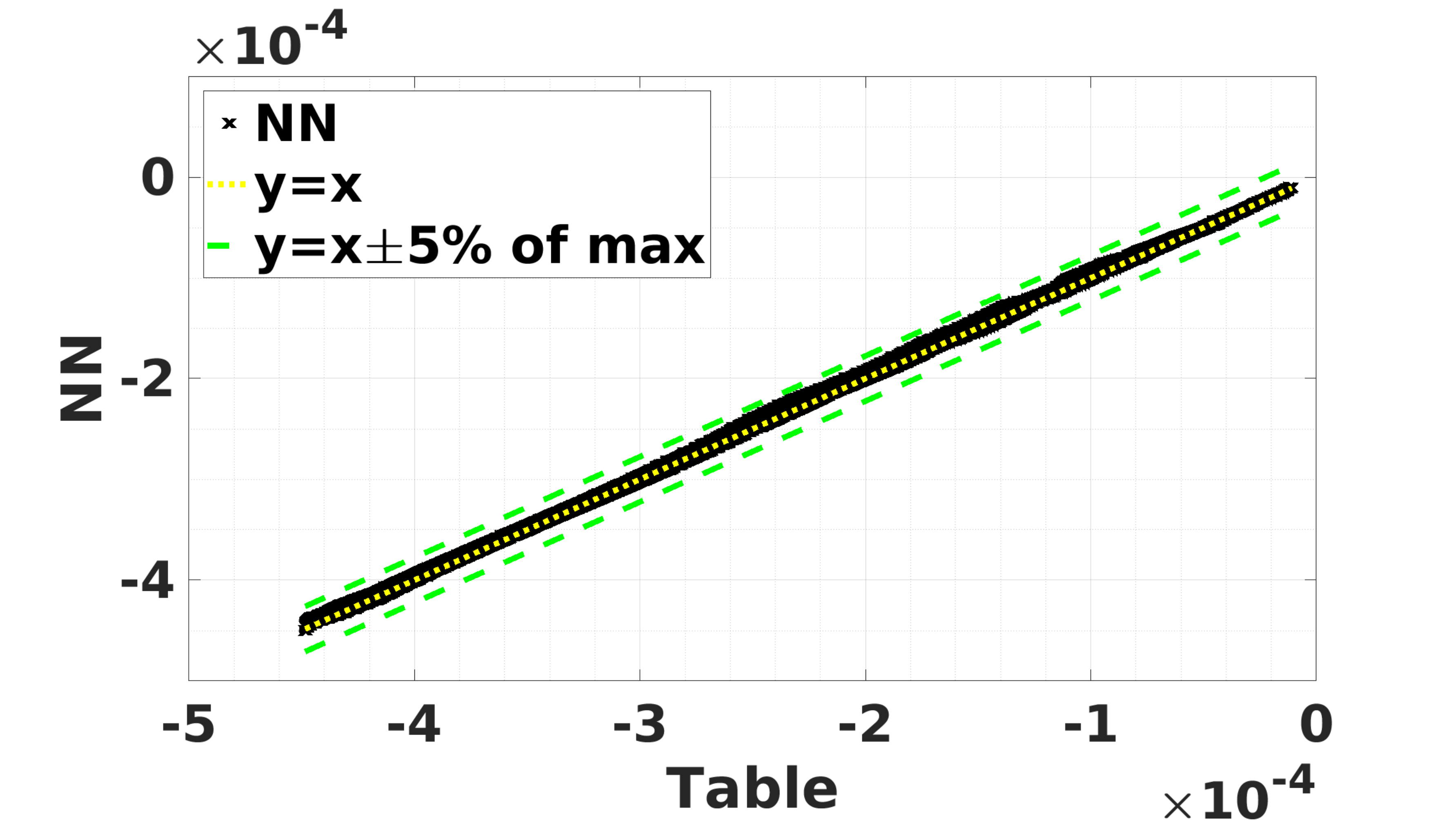}
\caption{ $\widetilde{a}_\gamma$ (\si{\per\kelvin}) }\label{ALLA_gammaNN}		
\end{subfigure}										
			\caption{ Comparison of NN-based models and flamelet table data}\label{AllofflineNN1}				
	\end{figure}

In \figurename~\ref{AllofflineNN1},  we compare the outputs from the NN for flame temperature, internal energy, gas constant, heat capacity ratio, its coefficient, and heat conductivity with the corresponding outputs from the flamelet table. Flame temperature is used as an extra input in the NN model for these variables (except itself). 
 In each graph, a set of guidelines (dotted yellow line and dashed lines) are provided to measure the performance of the neural network in the training and testing stages. The dotted yellow line shows the 45-degree line, which is the loci of the desired data in a perfectly fitted model. The dashed guidelines reflect error bounds  defined as a percentage of the maximum of the absolute value of each target variable, as shown in \figurename~\ref{AllofflineNN1}. The black crosses in \figurename~\ref{AllofflineNN1} show the points from the testing stage.

\subsubsection{Flame Temperature}\label{Temperaturesec}
\begin{figure}[hbt!]
		\centering
	\begin{subfigure}{.49\textwidth}
		\centering		
		\includegraphics[width=.95\textwidth]{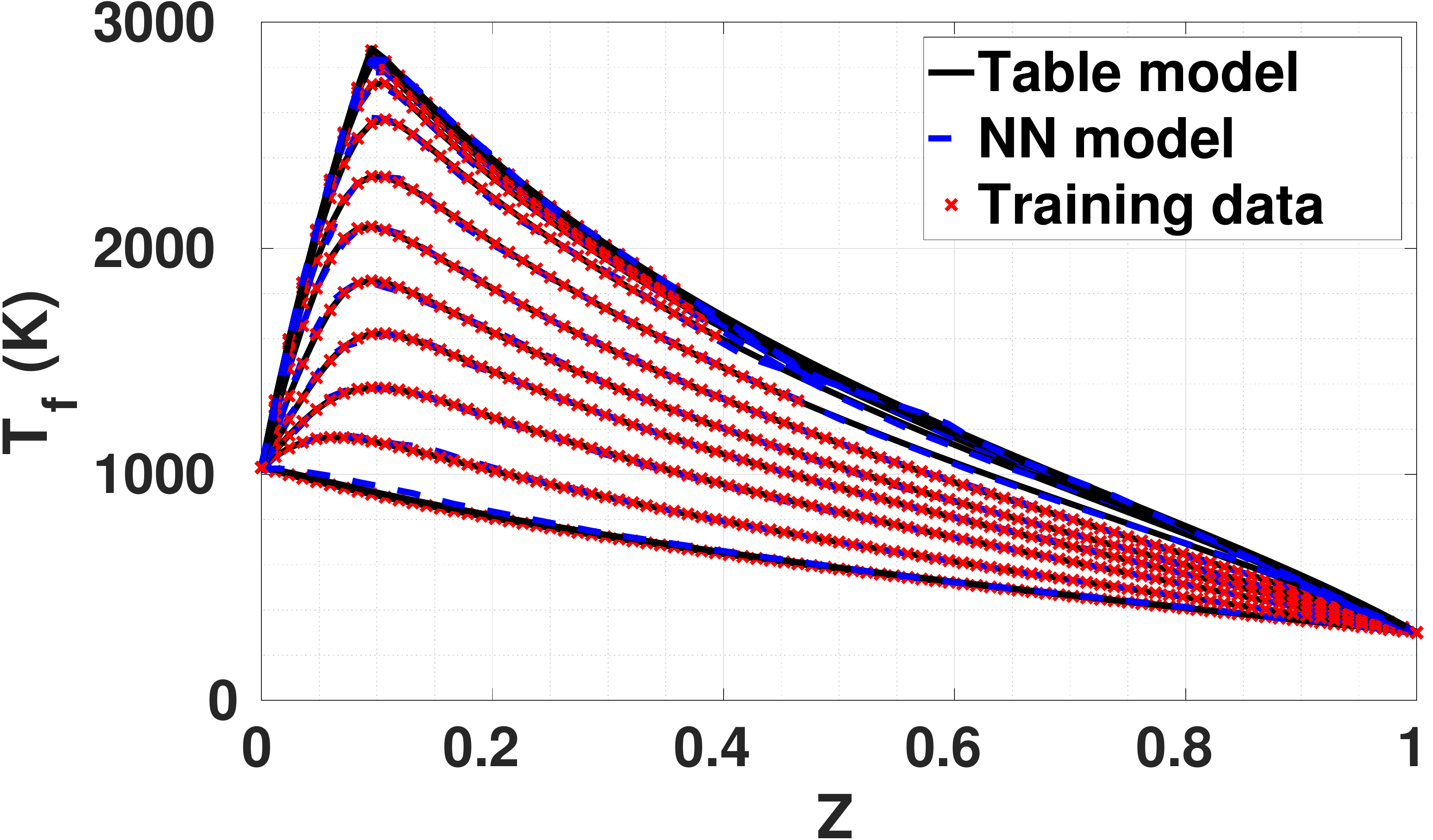}
		\caption{$B=0$, various $\widetilde{C}$ values}\label{Offline_Res/combo/Tf_B1_table}	
			\end{subfigure}	
	\begin{subfigure}{.49\textwidth}
				\centering		
		\includegraphics[width=.95\textwidth]{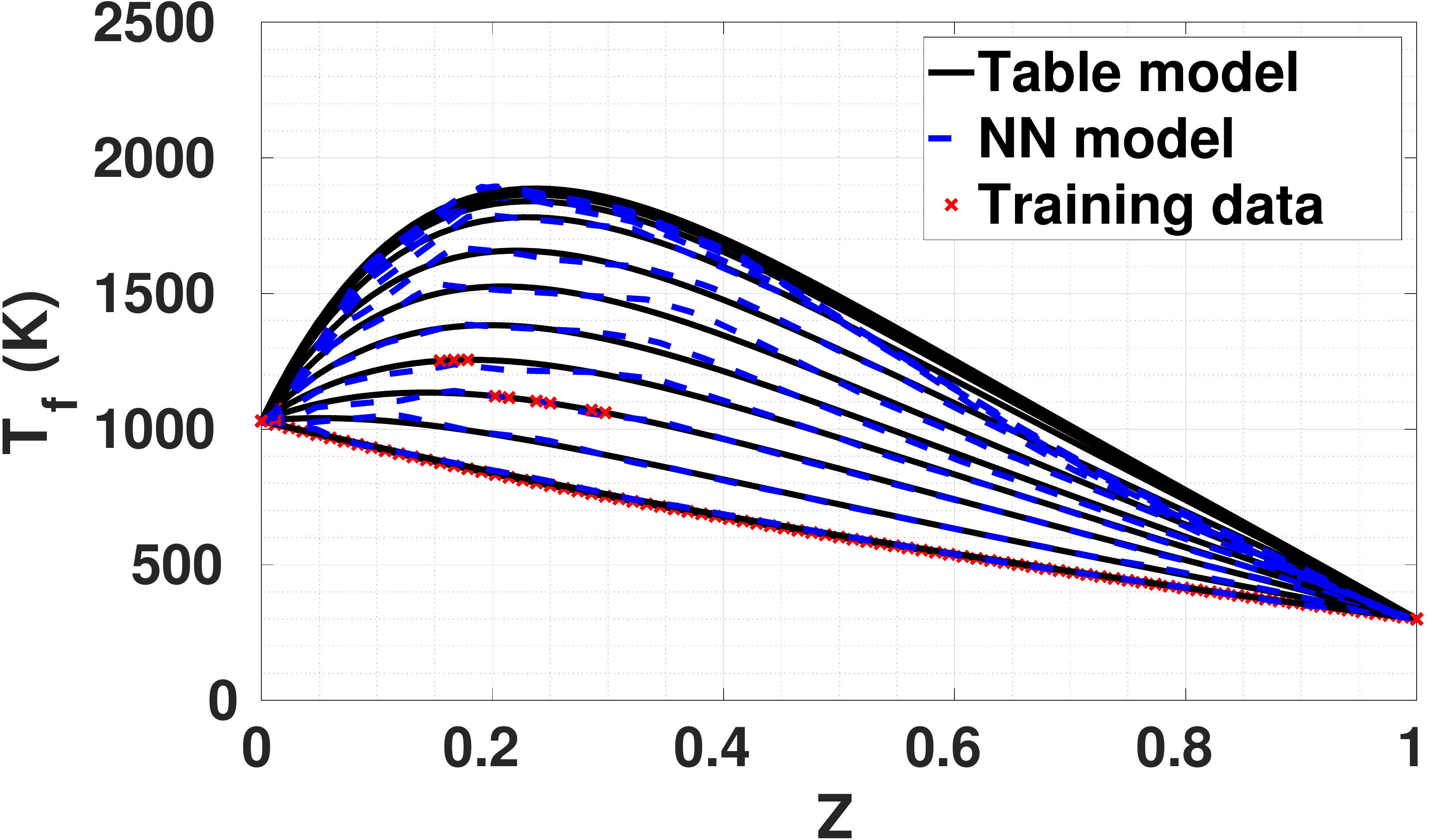}
		\caption{$B=0.27$, various $\widetilde{C}$ values}\label{Offline_Res/combo/Tf_B8_table}	
			\end{subfigure}	
	\caption{Comparison of $\widetilde{T_f}$ between the NN-based model and flamelet tables: dependence on $B$ and $\widetilde{C}$}\label{T_f_err_ZvsC}
\end{figure}

As mentioned above, flame temperature has a significant effect on the analysis of combustion. 
 Therefore, a deeper analysis of the model's strengths and weaknesses could help in further improving the models.  In the following, we compare the flamelet table data with the temperature calculated from both NN-based and the table-based models at different values of the parameter $B$. The effect of pressure on the errors in modeling flame temperature was insignificant; therefore, we only show the flamelets at a relatively high pressure point, 20 \si{atm} in \figurename~\ref{T_f_err_ZvsC}. In each sub-figure, flame temperature is plotted as a function of mixture fraction for different values of progress variable (solid black curves). The maximum value of temperature has a monotonically increasing relationship with values of progress variable, i.e.
 when $\widetilde{C}$ increases, maximum values of temperature increase, and curves converge together for higher $\widetilde{C}$ values. The dark blue curves  show the counterpart of each flamelet solution from NN-based models. On each curve, points used in the training set are shown with  red cross markers.

The NN models for flame temperature are generated after reinforcing the training set with data from regions with high $\widetilde{Z}$ values. In those regions, because there is no combustion,
empirical evidence shows that $\widetilde{T_f}$ depends linearly on $\widetilde{Z}$.
Essentially, there is a threshold in $\widetilde{Z}$ value, which 
is a  function of $\widetilde{C}$, $B$, and $\overline{P}$, above which this linear dependency sets-in. To determine the threshold values,
 we linearly interpolate $\widetilde{T_f}$ values at specified $\widetilde{Z}$ points where linear behavior is observed in the original data set. This way of reinforcing  the training set helps to account for the lack of data in regions with high $\widetilde{Z}$ values. The CFD simulations correspond to a fuel lean configuration, and therefore most of the collected  data has lower $\widetilde{Z}$ values.  The lack of training points with high $\widetilde{Z}$ values is correlated with deviations of NN-based models from the table-based models when $\widetilde{Z}$ is near to one.

 In \figurename~\ref{Offline_Res/combo/Tf_B1_table}, the laminar flamelet model is shown. The availability of training data in the laminar zone is highest among different $B$ values. It can be observed from \figurename~\ref{T_f_err_ZvsC} that where there is no training data, the NN-based curves deviate from the table curve. 
 By increasing the turbulence level (parameter $B$), the availability of training points on the curve reduces, and the deviation of the NN-based curves increases \figurename~\ref{Offline_Res/combo/Tf_B8_table}. 
 Better performance can be observed in the lower mixture ratio side, where there is more data available. 
 
\subsubsection{Progress Variable Reaction Rate}\label{PVRRsec}
\begin{figure}[hbt!]
	\centering										
	\begin{subfigure}{.49\textwidth}								
		\centering
		\includegraphics[width=.98\textwidth]{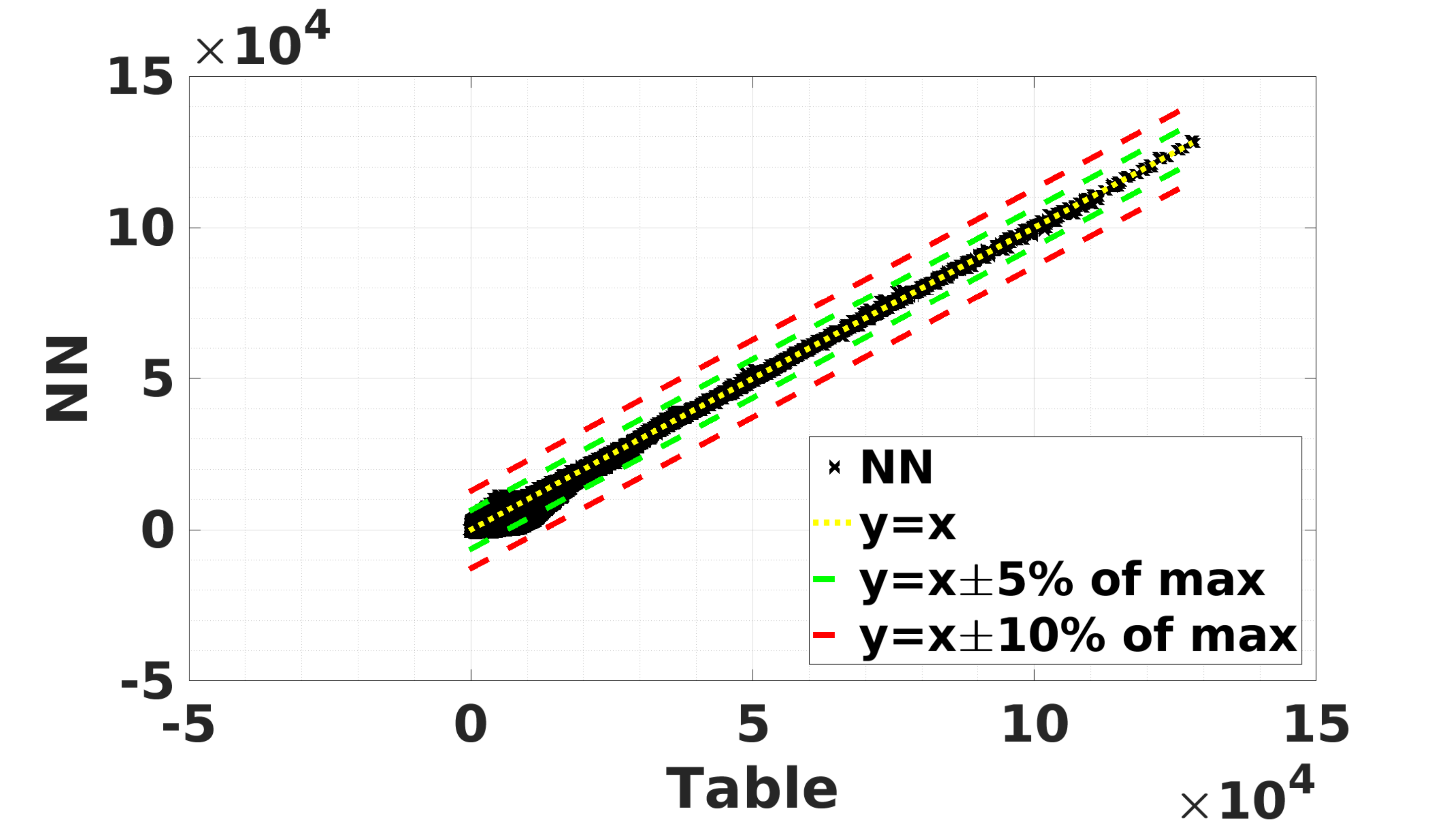}
	\caption{ $\widetilde{\dot{\omega}}_C$ (\si{\kilogram\per\meter\cubed\per\second}) }
	\label{PVRRNN_corr}				
	\end{subfigure}		
	\begin{subfigure}{.49\textwidth}	
	\centering
	\includegraphics[width=1\textwidth]{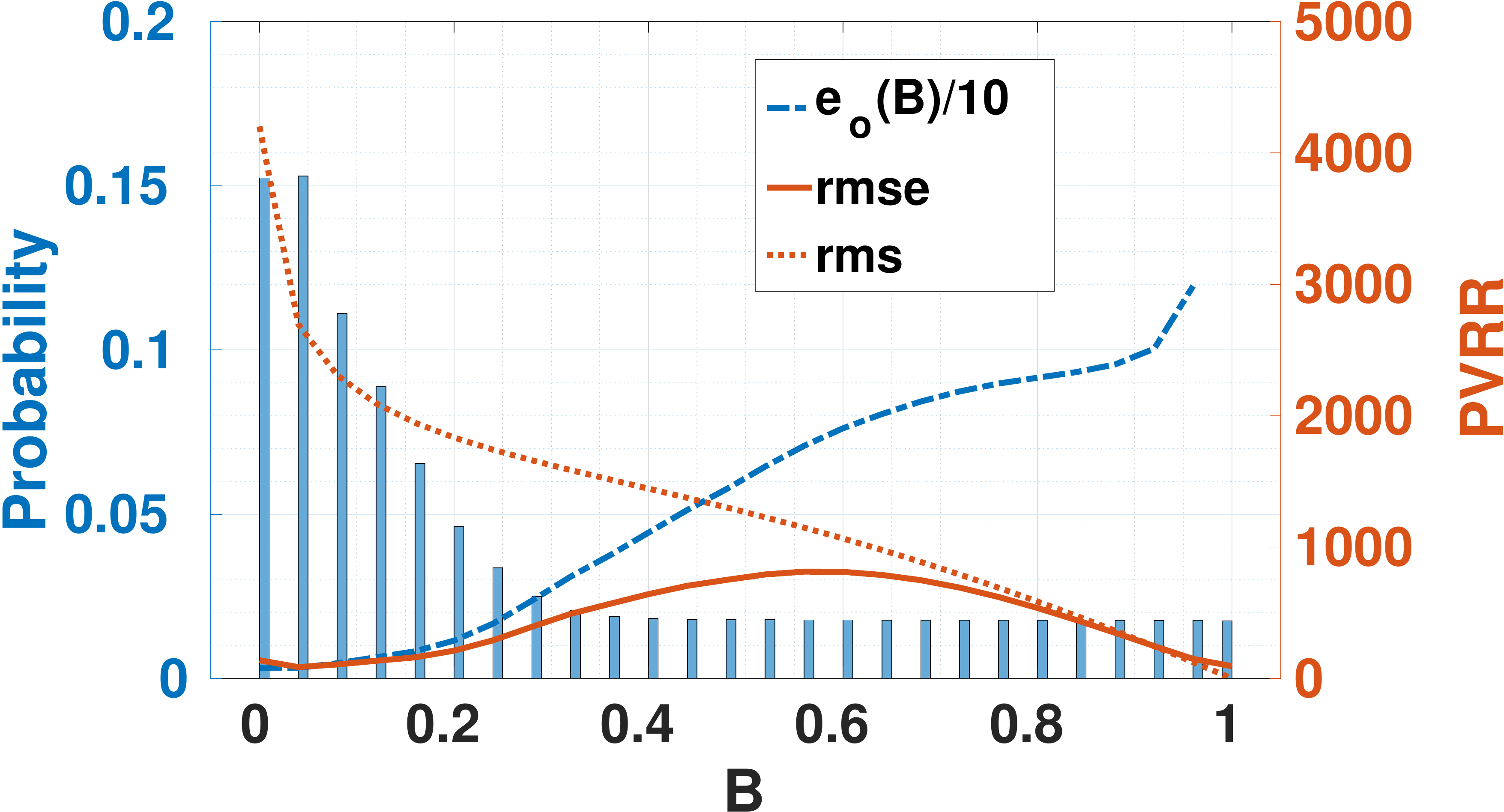}		
		\caption{ rmse, rms, and $e_o$  as functions of $B$ and $B$'s histogram}\label{PVRRBNN_corr}
	\end{subfigure}		
\caption{PVRR NN performance analysis}\label{NNperPVRR}
\end{figure}

PVRR is a significant output of the flamelet model as it represents the combustion energy driving the transport equation for the progress variable in the flamelet model. In this section, we investigate the performance of the NN-based model for PVRR. The magnitude of PVRR is drastically changed with pressure and, thus, it has multiple scales. This great range of magnitude affects the performance of the NN-based model. In the training process, we minimize the least squared-error in absolute manner; therefore, data with a larger magnitude dominates the learning process, and the design becomes more affected by those points. The NN-based estimation of PVRR is compared with its table-based values in \figurename~\ref{PVRRNN_corr}. The black crosses show the test values. The testing points fall in the 5\% guidelines, only for PVRR with higher magnitudes. The points with smaller values, however, have a higher error range and fall in the 10\% guidelines.
In the analysis of combustion instability, the points associated with higher energy dominate the physics; so, better performance for the  NN-based model for points with higher magnitude works in our favor. 
This result was achieved by reinforcing the training set with data from the dynamic equilibrium simulation, in which more points with higher pressure are available. 

In the following, we investigate the flamelet solutions for PVRR in more detail. PVRR is highly affected by pressure. There is no data available with pressure values lower than 10 \si{atm}, and as a result, the averaged relative error is higher for the lower pressure range. Parameter $B$ is also affecting the NN-based model performance significantly. The rmse and rms of PVRR as a function of parameter $B$ is provided in \figurename~\ref{PVRRBNN_corr}. The probability distribution of parameter $B$ in the training set is also provided on the left axis of \figurename~\ref{PVRRBNN_corr}. Around 60\% of the training set has a $B$ value lower or equal to 0.2. The one-tenth scaled relative error  as a function of $B$ is also provided on the left axis in \figurename~\ref{PVRRBNN_corr}. For the laminar flame, where $B=0$, the relative error is around 3\%. Then it grows, as the probability distribution decreases in higher ranges of $B$ values. The rms of PVRR also decreases with increasing the parameter $B$. The absolute error increases when the availability of data decreases and the distance of a point from $B$ values with higher availability increases. When the rms becomes small, the rmse also decreases, while the relative error still goes higher.

\begin{figure}[hbt!]
	\centering
	\begin{subfigure}{.49\textwidth}
		\centering		
		\includegraphics[width=.95\textwidth]{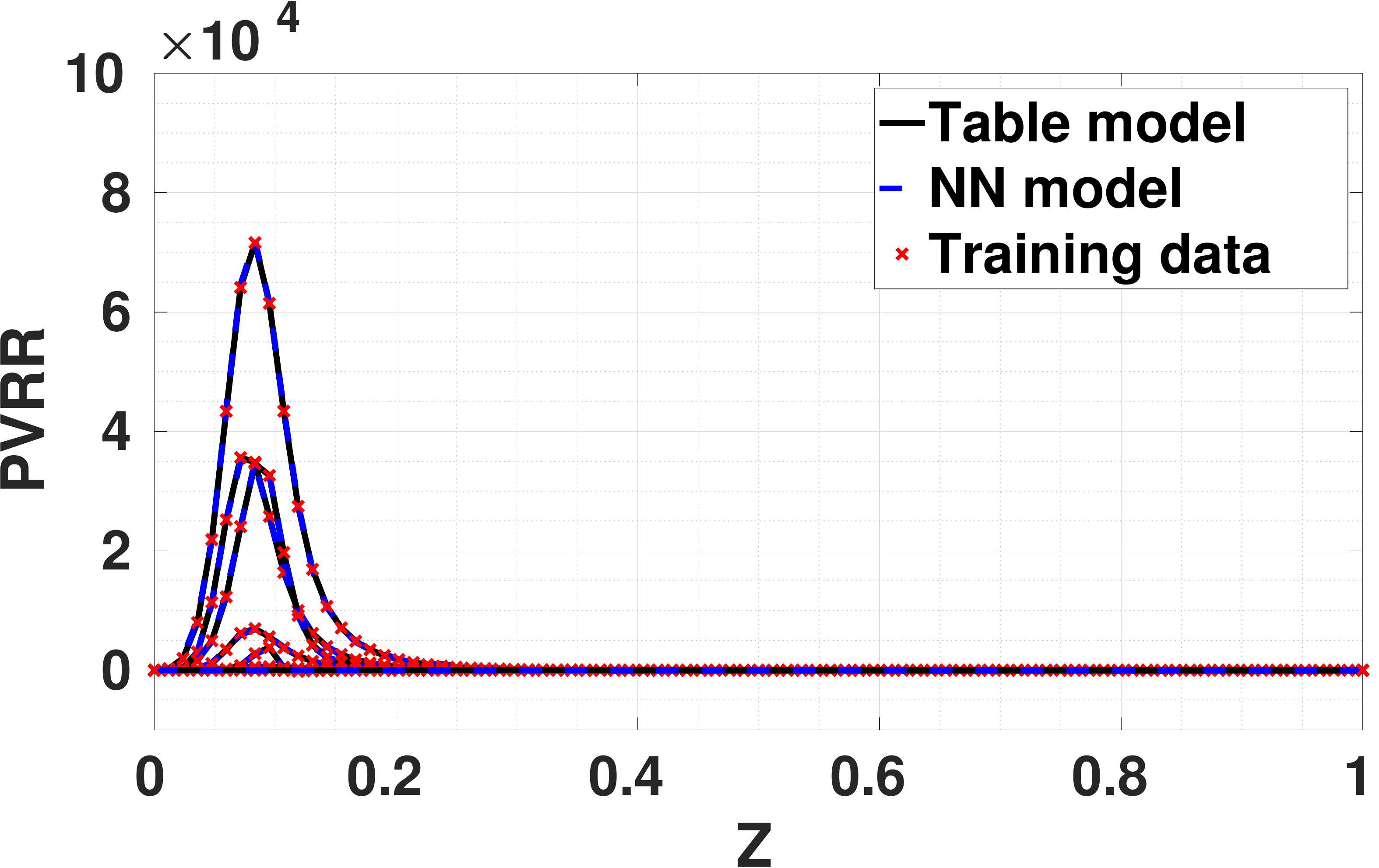}
		\caption{ $B=0$, various $\widetilde{C}$ values}\label{Offline_Res/combo/ProdC_B1_table}	
			\end{subfigure}	
		\begin{subfigure}{.49\textwidth}
					\centering		
			\includegraphics[width=0.95\textwidth]{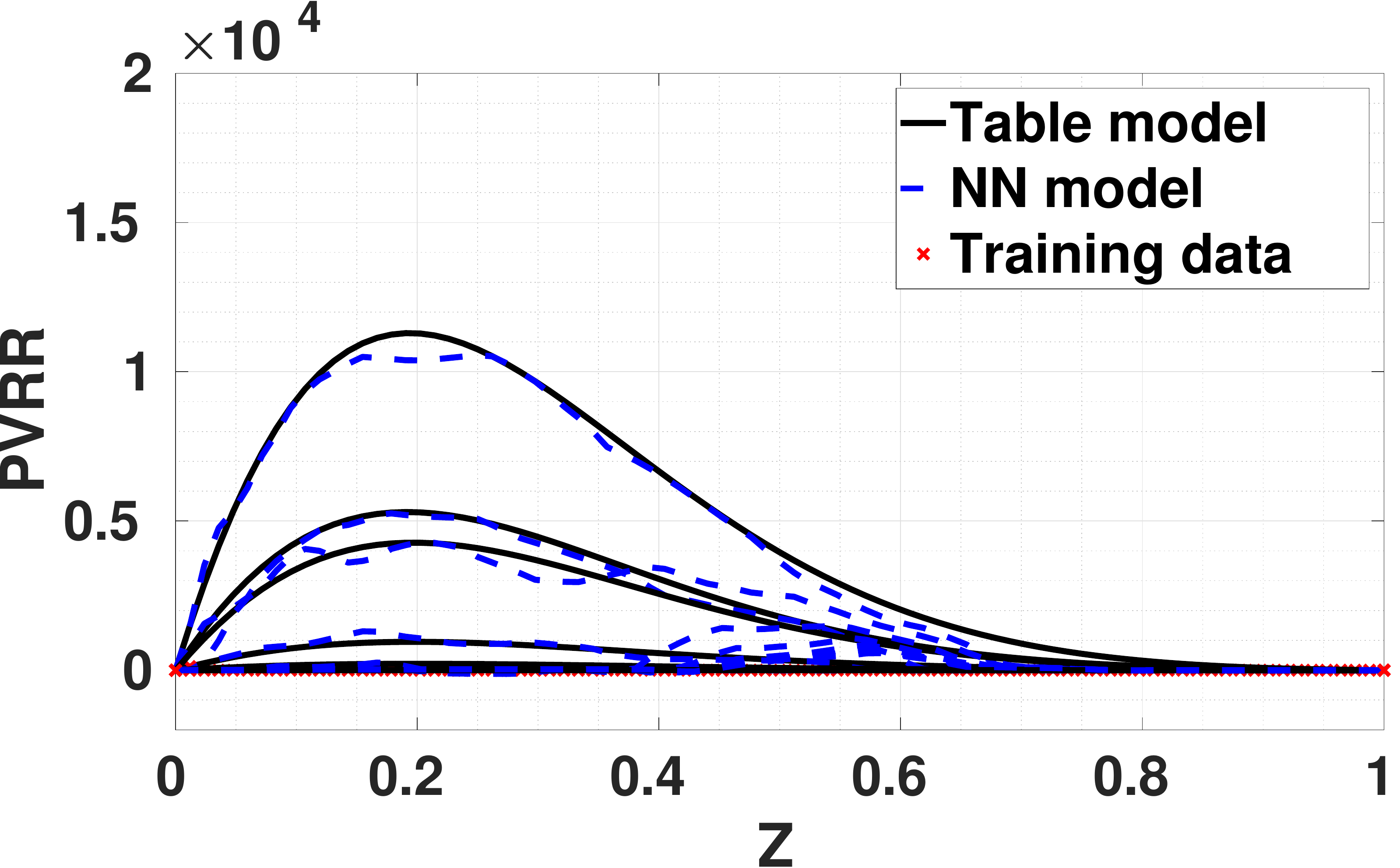}
			\caption{$B=0.27$, various $\widetilde{C}$ values}\label{Offline_Res/combo/ProdC_B8_table}	
				\end{subfigure}	
	\caption{Comparison of PVRR (\si{\kilogram\per\meter\cubed\per\second}) between the NN-based model and flamelet tables: dependence on $B$ and $\widetilde{C}$}\label{ProdC_err_ZvsC}
\end{figure}

In \figurename~\ref{ProdC_err_ZvsC}, we  look into PVRR values of the flamelet table at 20 \si{atm} for different values of parameter $B$ and progress variable. Different flamelet solutions of PVRR are provided with the black curves at different $\widetilde{C}$ values in each plot. 
The behavior of PVRR changes drastically with the progress variable. Maximum values of PVRR occur at $\widetilde{C}$ values around 0.15.
The solid dark blue  line shows the NN-based results. The training data are shown with a red cross when read from the flamelet table.
For the laminar flame ($B=0$), there is more data available, and the NN-based model follows the flamelet solutions particularly well for lower $\widetilde{Z}$ values, see \figurename~\ref{Offline_Res/combo/ProdC_B8_table}. The NN-based model starts  to deviate when $\widetilde{Z}$ goes to one, and there is not much data available due to fuel lean composition of the CFD-based data pool. 

In \figurename~\ref{ProdC_err_ZvsC}, by increasing the $B$ values, the availability of training data decreases, and performance degrades. The estimation maintained a decent level of accuracy up to $B=0.27$ (see  \figurename~\ref{Offline_Res/combo/ProdC_B8_table}) where the relative error is around 10\%; yet, the results diverge more for higher $B$ values. PVRR is the most challenging variable to model in the study of turbulent flame with unsteady pressure.
\subsection{In-Situ Test of NN-based Models in CFD}\label{onlinetest}
To test the NN-based models further, we implement them into our CFD simulations by replacing the flamelet models. The configuration under study  has a 14-\si{\centi\meter} oxidizer post length, which is characterized by a very high level of instability, according to \cite{Tuan1}. Two simulations describing the dynamic equilibrium and triggering cases are analyzed.  The results of the NN-based simulations are compared with the table-based ones  to validate our NN-based models. 

\begin{figure}[hbt!]
	\begin{subfigure}{.49\textwidth}
		\centering
		\includegraphics[width=0.95\textwidth]{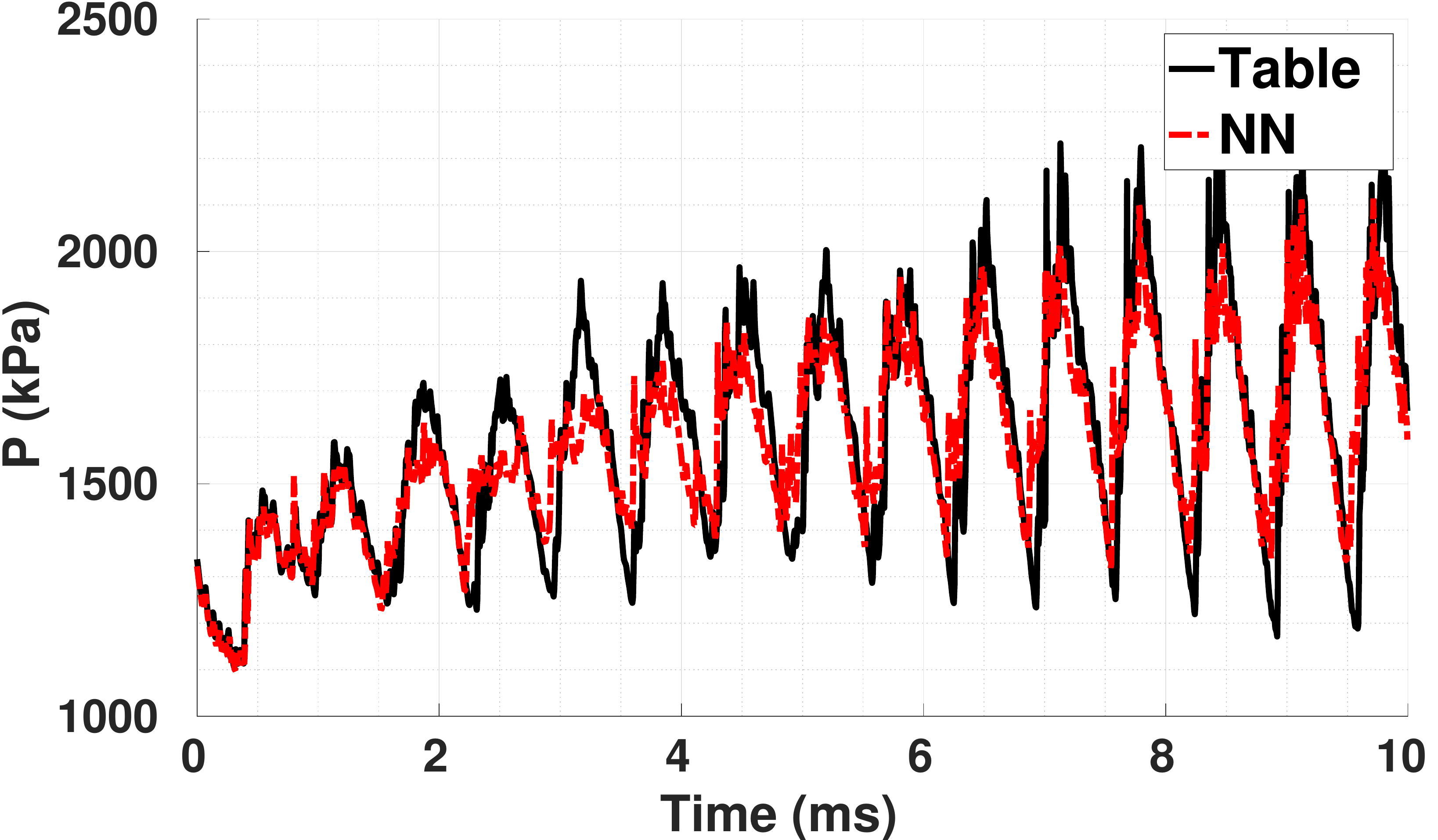}
		\caption{Oxidizer post, $r=0.5$ \si{\centi\meter}, $x=-10$ \si{\centi\meter}}\label{trans_Pcompop_28_34}		
	\end{subfigure}
	\begin{subfigure}{.49\textwidth}
		\centering
		\includegraphics[width=0.95\textwidth]{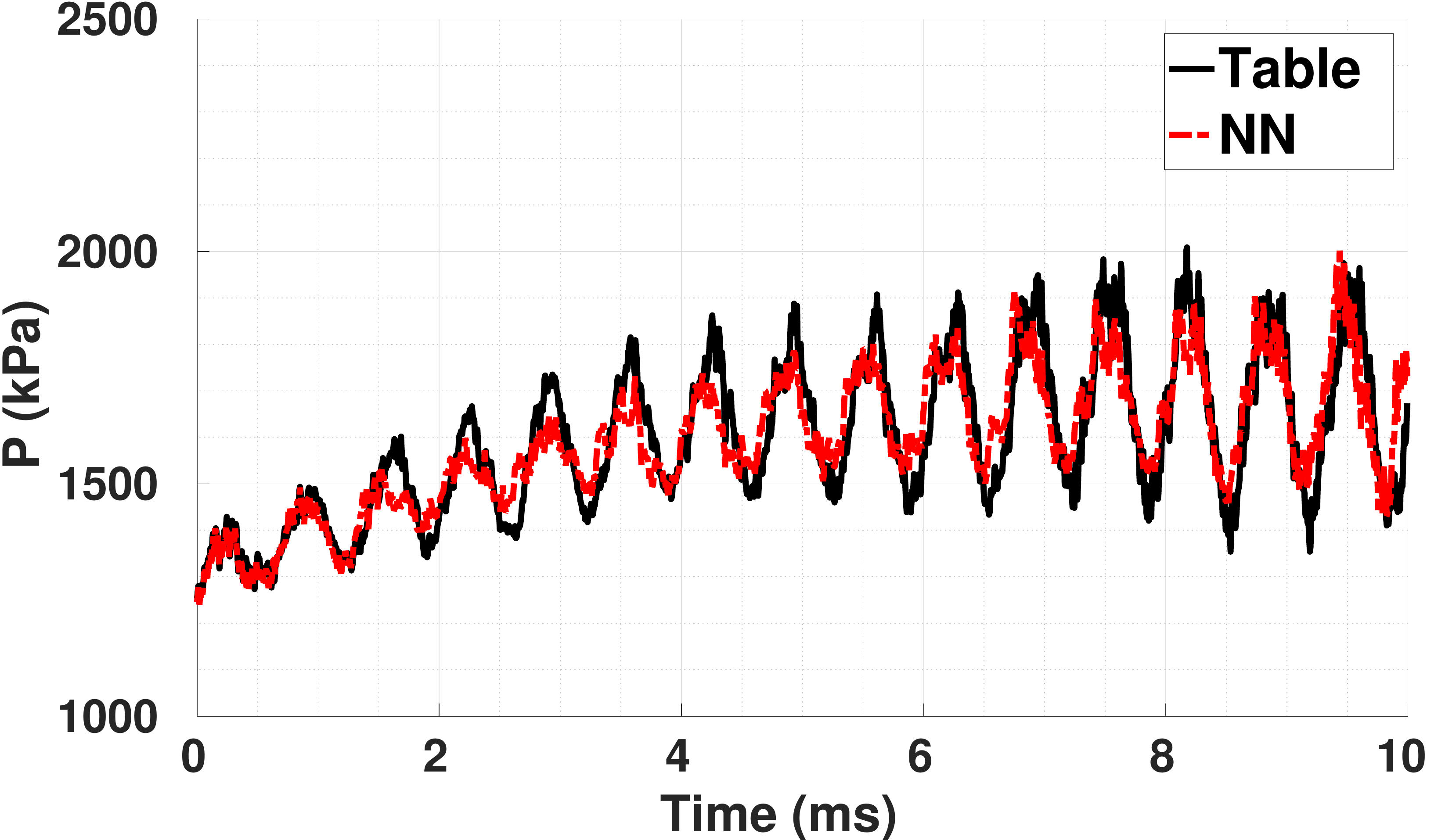}
		\caption{Shear layer, $r=1.13$ \si{\centi\meter}, $x=8$ \si{\centi\meter}}\label{trans_Pcompfp1_18_408}			
	\end{subfigure}			
	\caption{ TR: comparison of pressure time signals 
		at representative points} \label{trans_Psigcomp2}		
\end{figure}

Since the NN models are trained on a set of data generated from a transient (TR) CFD simulation, as a first step, we implemented the model into the transient simulation. The pressure signal at two points in the oxidizer post and in the shear layer are compared for the NN-based and table-based simulations in \figurename~\ref{trans_Psigcomp2}. The NN-based simulation is more accurate where the pressure signal reaches the high amplitude limit cycle. The success of our NN-based model in this simulation is manifested by its ability to follow the pressure 
 growth from a low pressure point to a high amplitude limit cycle.
 
Another measure of performance of the simulation is how well the instability criterion is estimated. 
Coupling between the combustion heat release and the acoustic pressure wave is the most important factor for the combustion instability. Here, the correlation between PVRR fluctuations and pressure fluctuation is used as a measure of coupling between combustion energy and the acoustic wave. The analysis in \cite{Shadram_Journal_1} showed that the modified Rayleigh index (\textit{mRI}), which is used here, resembles the same behavior as Rayleigh index.
The time-averaged local \textit{mRI} is defined in Eq.~\eqref{RIdef} over a time period ($\tau$) which is typically few cycles; where $p'$, and $\dot{\omega}_C'$ are the local fluctuations in $\overline{P}$ and PVRR, respectively. Also, $\widehat{P}$ and $\widehat{\dot{\omega}}_C$ are the global time averages of $\overline{P}$ and PVRR.
\begin{equation}
	\label{RIdef}
	mRI=\frac{1}{\tau}\int_{t_o}^{t_o+\tau}\frac{\widetilde{\gamma}+1}{\widetilde{\gamma}}\times\frac{p'}{\widehat{P}}\times\frac{\dot{\omega}_C'}{\widehat{\dot{\omega}}_C}dt 
\end{equation}
Next, the performance of NN-based models are analyzed in the dynamic equilibrium and triggering cases.
\subsubsection{Test on CFD: Dynamic Equilibrium}
\begin{figure}[hbt!]
	\begin{subfigure}{.49\textwidth}
		\centering
		\includegraphics[width=0.95\textwidth]{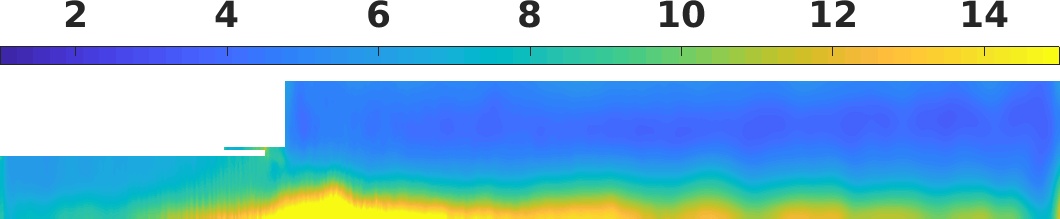}
		\caption{Overall relative error (\%)}\label{efullcont6}		
	\end{subfigure}
	\begin{subfigure}{.49\textwidth}
			\centering
	\includegraphics[width=0.95\textwidth]{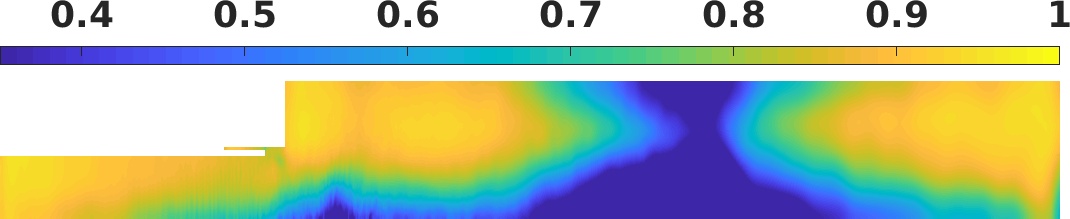}
	\caption{Fluctuation correlation}\label{corcont6}
\end{subfigure}
	\caption{DE: distribution of $e_o$(\%) and correlation of pressure signals fluctuation between the NN- and table-based simulations}\label{corcomp2}
	\end{figure}

First, we analyze the results from the dynamic equilibrium (DE) scenario, in which the simulation starts from an initial condition where instability has already set-in.
In our analysis, we look at the overall relative error calculated at each spatial point by computing the norm of the difference of  the pressure time waveforms from the NN-based and table-based CFD simulations. This error is shown in \figurename~\ref{efullcont6}. In the combustor, most of the points have an $e_o$ less than 5\%. Higher errors occur on the centerline near the dump plane, where the numerical scheme is very complicated. 
Moreover, subtracting the mean pressure from each signal, we can also look into the correlation of the fluctuation of the pressure signals between the NN-based simulation and table-based one. This correlation is calculated using Pearson's formula \cite{refPearson} and is shown in \figurename~\ref{corcont6}. The signals are highly correlated through the combustor except for regions in the vicinity of the pressure node, where the pressure fluctuation is very noisy.

\begin{figure}[hbt!]
	\begin{subfigure}{.49\textwidth}
				\centering
		\includegraphics[width=0.95\textwidth]{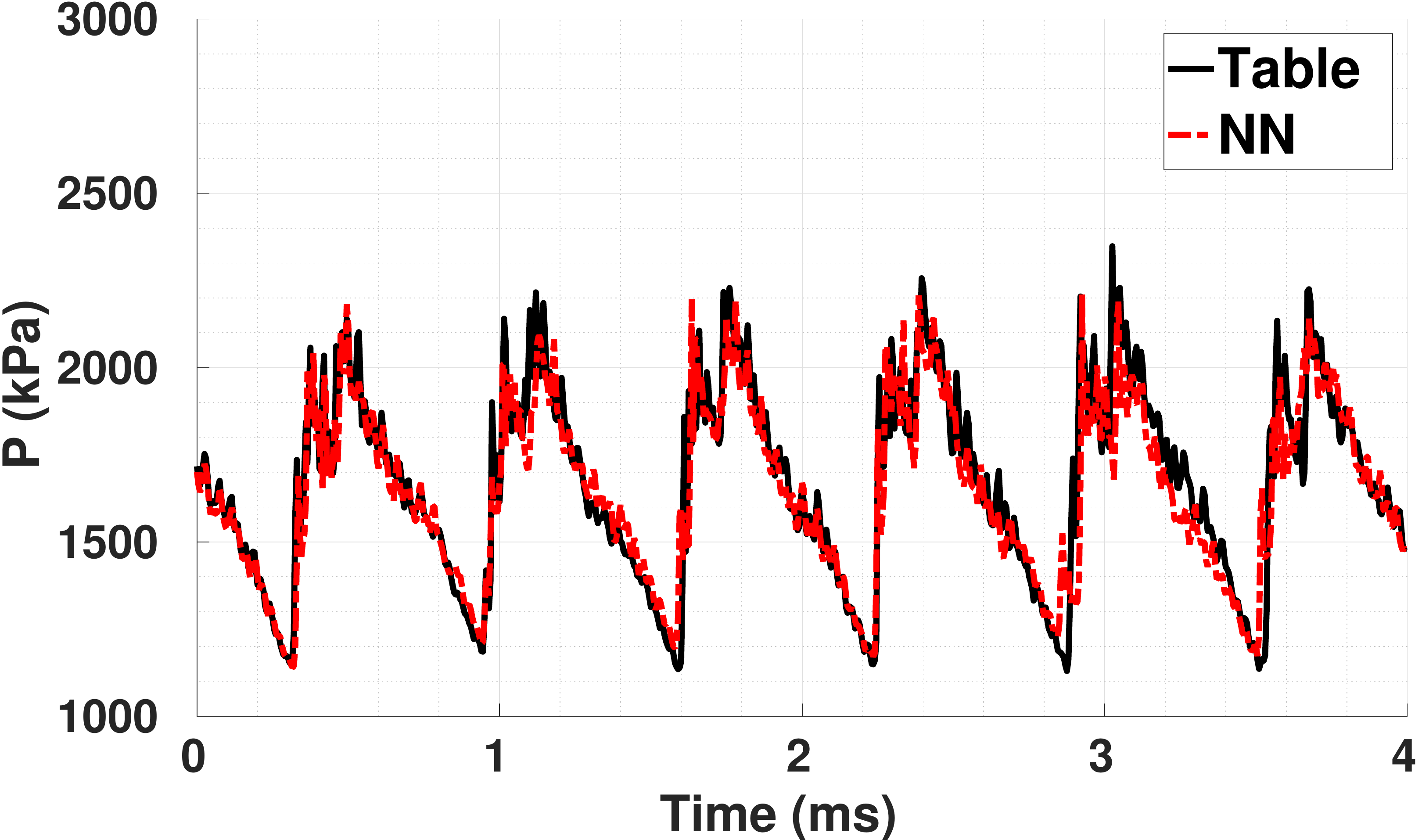}
		\caption{Oxidizer post, $r=0.5$ \si{\centi\meter}, $x=-10$ \si{\centi\meter}}\label{Pcompop_28_34}		
	\end{subfigure}
	\begin{subfigure}{.49\textwidth}
				\centering
\includegraphics[width=0.95\textwidth]{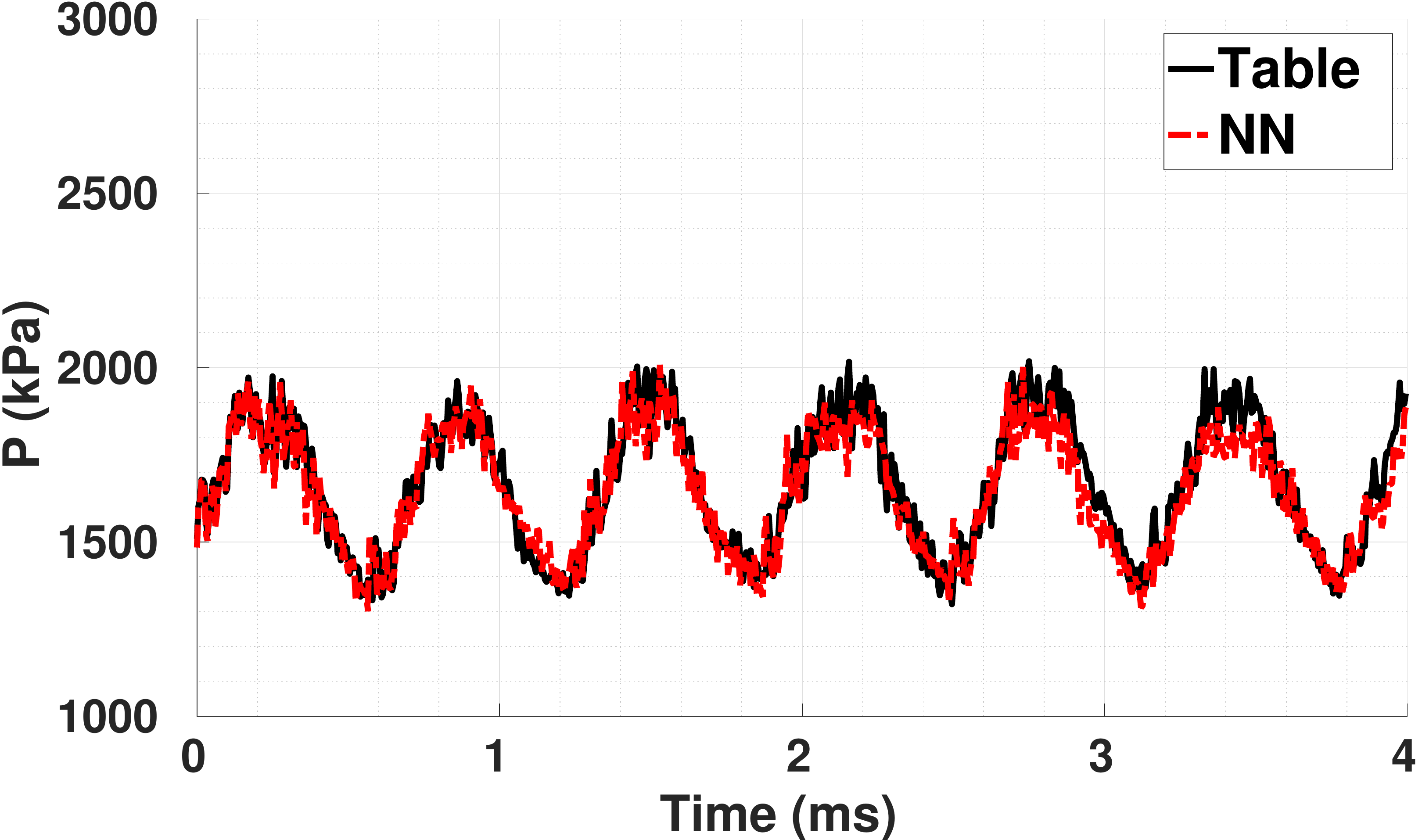}
\caption{Shear layer, $r=1.13$ \si{\centi\meter}, $x=8$ \si{\centi\meter}}\label{Pcompfp1_18_408}			
	\end{subfigure}			
	\caption{ DE: comparison of pressure time signals 
		at representative points} \label{Psigcomp2}		
\end{figure}

\begin{figure}[hbt!]
	\begin{subfigure}{.49\textwidth}
		\centering
		\includegraphics[width=.95\textwidth]{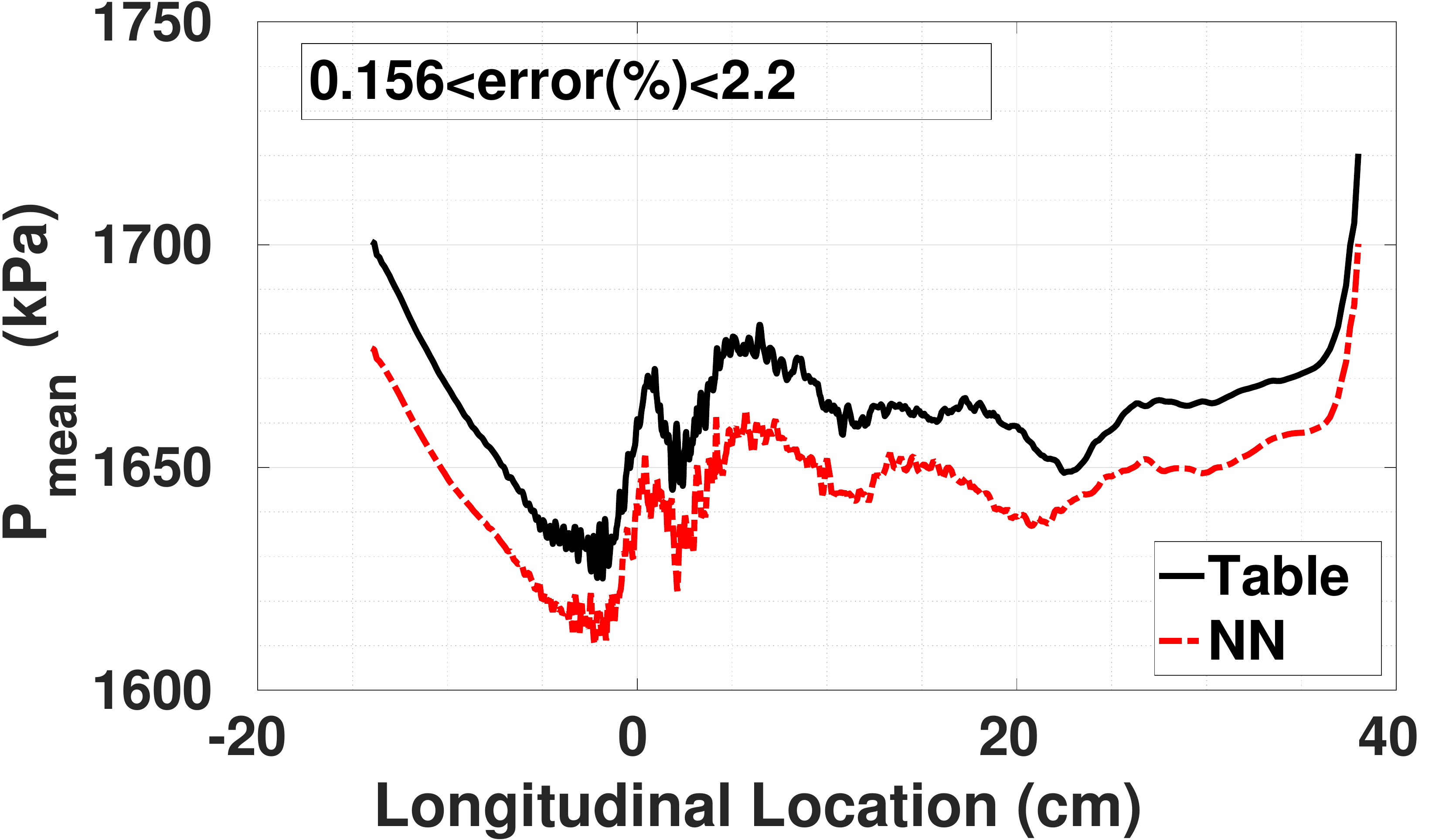}
		\caption{Mean pressure (\si{\kilo\pascal})}\label{PMSmean6}
			\end{subfigure}				
			\begin{subfigure}{.49\textwidth}			
			\centering
			\includegraphics[width=.95\textwidth]{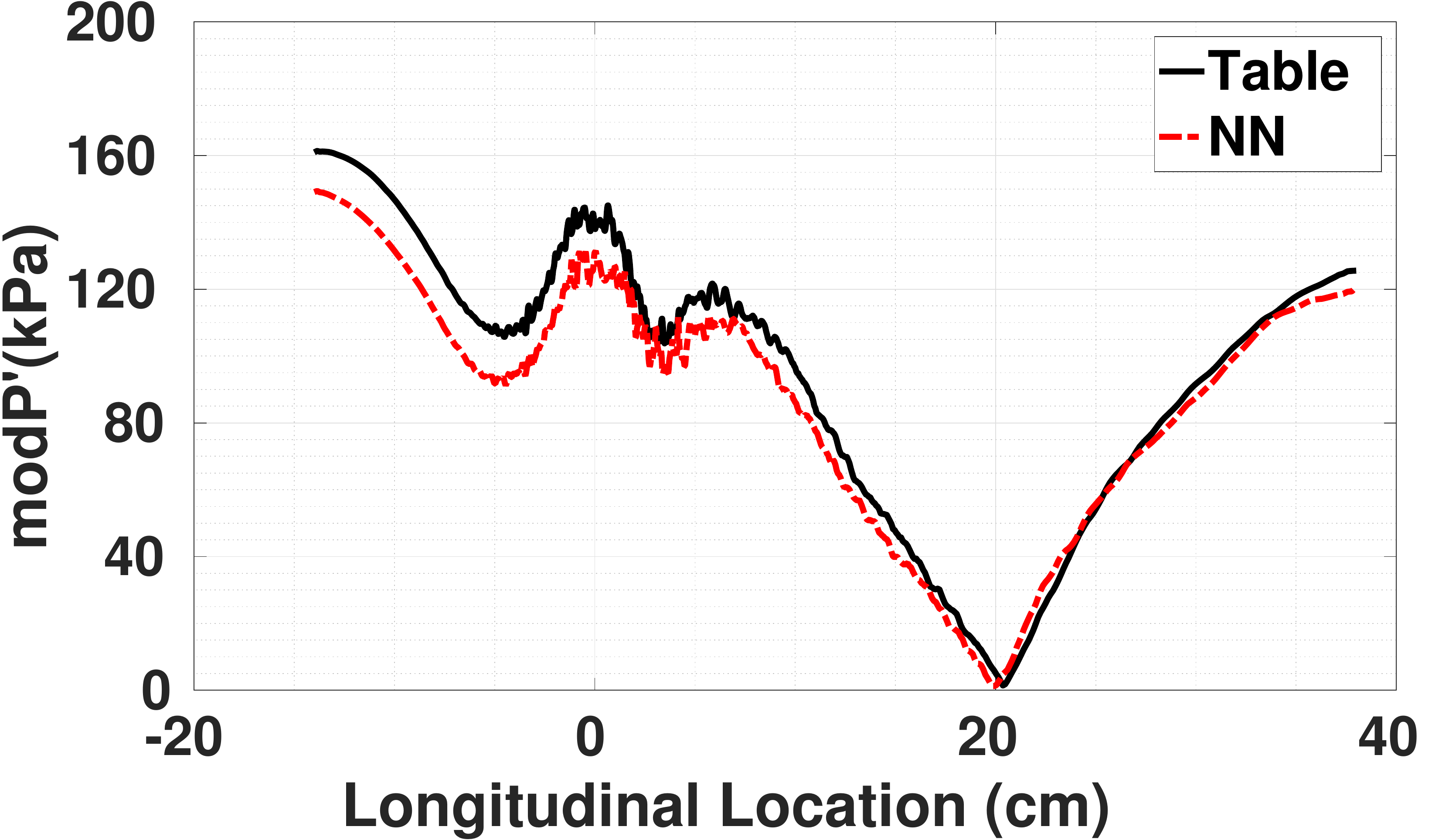}
			\caption{First mode shape (\si{\kilo\pascal}) }\label{PMSfirst6}	
		\end{subfigure}				
	\begin{subfigure}{.49\textwidth}
		\centering
		\includegraphics[width=.95\textwidth]{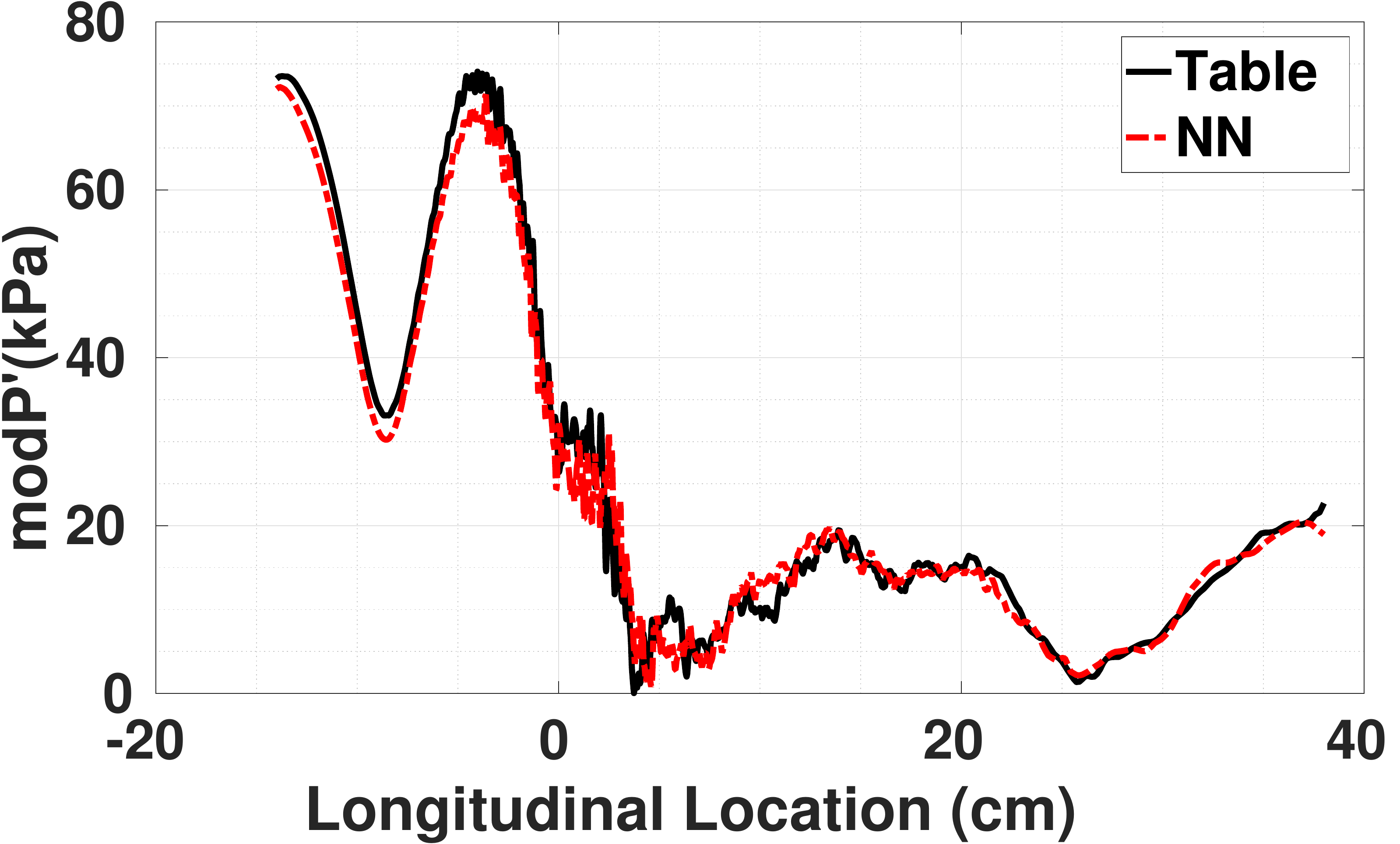}
		\caption{Second mode shape (\si{\kilo\pascal})}\label{PMSsec6}		
	\end{subfigure}		
	\begin{subfigure}{.49\textwidth}
		\centering
		\includegraphics[width=.95\textwidth]{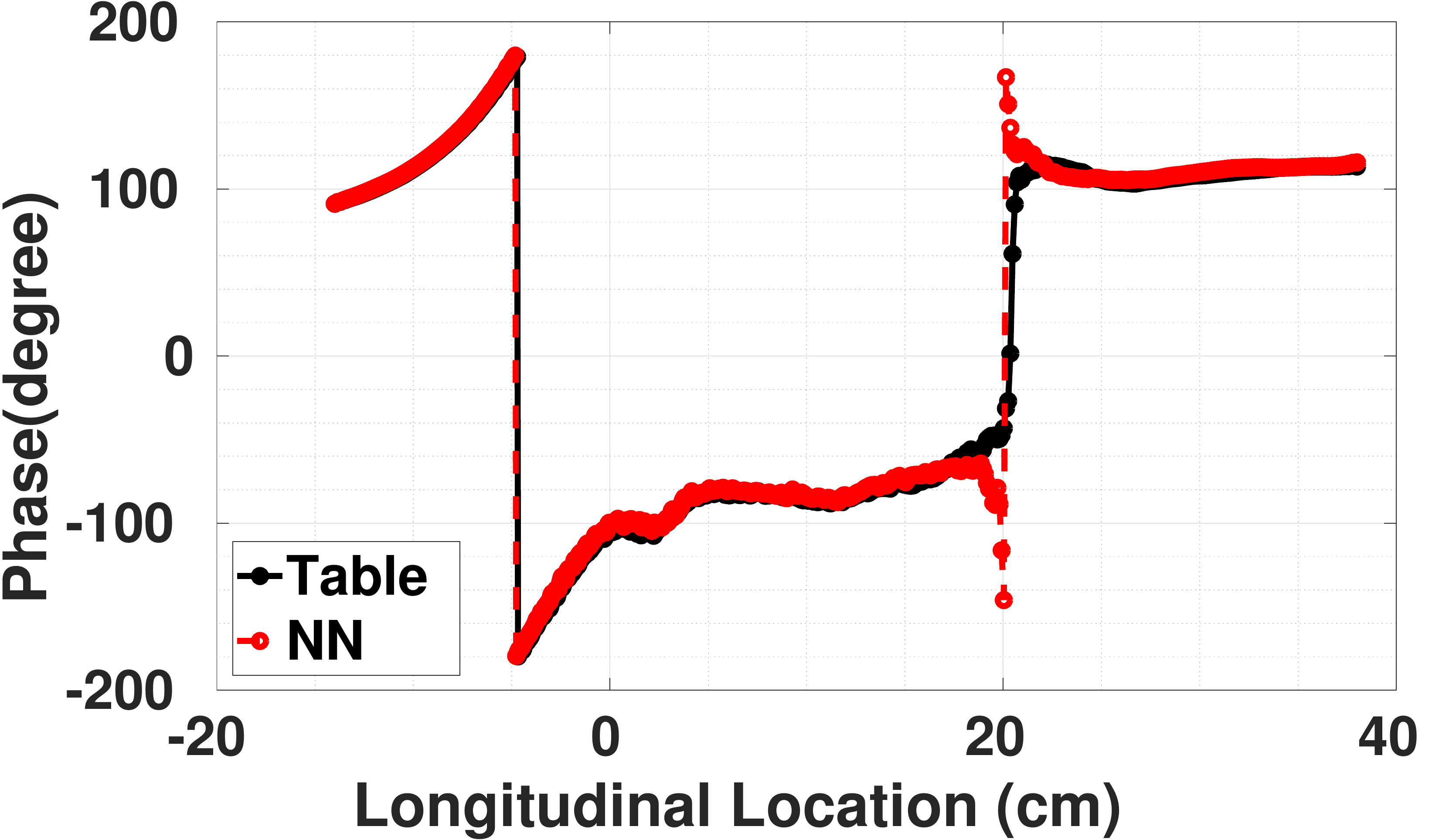}
		\caption{Phase of the first mode shape (\si{\kilo\pascal}) }\label{PMSphasefirst6}			
	\end{subfigure}		
	\caption{DE: comparison of pressure mean, the first longitudinal mode and its phase, and the second longitudinal mode shapes between NN-based and table-based simulations along  the  centerline}\label{modshcompnew}
\end{figure}

Next, we look at the pressure time waveforms at two representative locations, one in the oxidizer post and another in the shear layer, see \figurename~\ref{Psigcomp2}. In both locations, there is a great consistency between the table-based and NN-based simulations. The longitudinal mode shapes of pressure at the centerline are plotted in \figurename~\ref{modshcompnew}. They represent the modulus of the Fourier spectrum first and second peaks of pressure signals at each grid point along the centerline. The phase of the first longitudinal mode is also compared at each grid point for the NN-based and table-based simulations in \figurename~\ref{PMSphasefirst6}. There is a great agreement in all of these graphs between NN-based and table-based simulations, with the exception of some inconsistencies in the vicinity of 20 \si{\centi\meter} location, which is a pressure node. The pressure waveform at the node can be considered as a noise signal, as the amplitude of the first harmonic is close to zero. The frequencies of the first and the second longitudinal modes across the combustor are around 1500 \si{\hertz} and 3000 \si{\hertz}, respectively.

\begin{figure}[hbt!]
	\begin{subfigure}{.49\textwidth}
		\centering
		\includegraphics[width=0.95\textwidth]{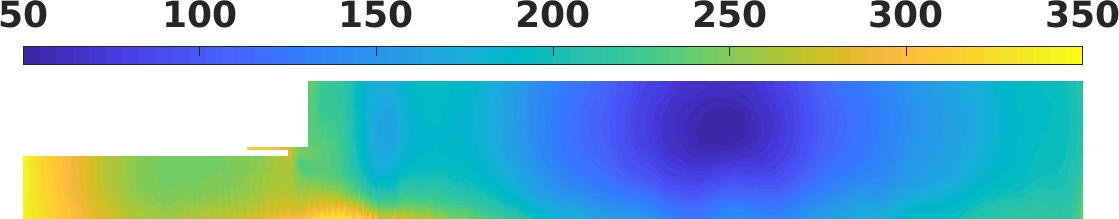}
		\caption{Table: rms}\label{dynrmspowTAB}		
		\centering
		\includegraphics[width=0.95\textwidth]{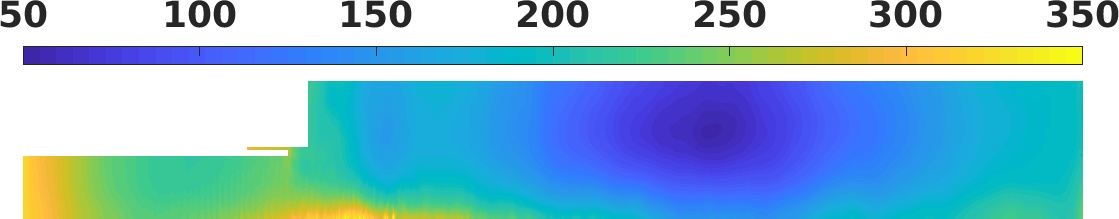}
		\caption{NN: rms}\label{dynrmspowNN6}
	\end{subfigure}
	\begin{subfigure}{.49\textwidth}
	\centering
	\includegraphics[width=0.95\textwidth]{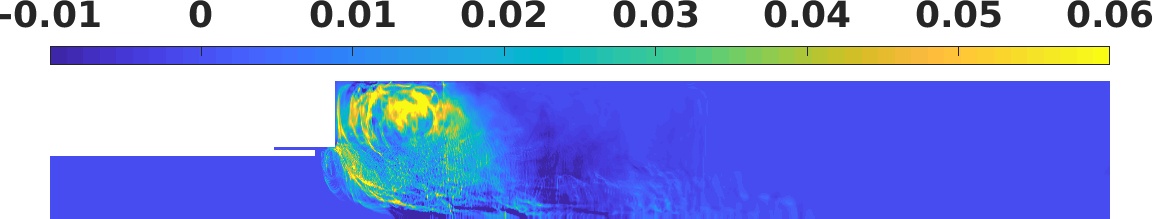}
	\caption{ Table: \textit{mRI}}\label{WCRIrr}		
			\centering
	\includegraphics[width=0.95\textwidth]{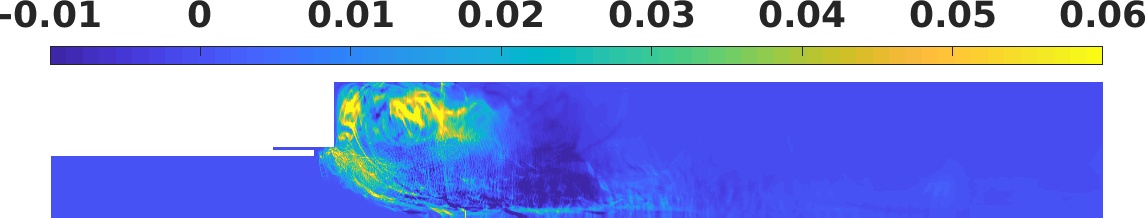}
	\caption{ NN: \textit{mRI} }\label{WCRInna}									
\end{subfigure}		
	\caption{ DE:  The rms of pressure signal fluctuations are compared between the simulations based  on NN  (\ref{dynrmspowNN6}) and table(\ref{dynrmspowTAB}). Also the \textit{mRI} from the NN-based (\ref{WCRInna}) and the table-based (\ref{WCRIrr}) simulations are compared}\label{rmsratcomp3}
\end{figure}

The signal power of pressure fluctuations across the grid points of the combustor are compared between the NN-based and table-based simulations in \figurename~\ref{dynrmspowNN6} and \figurename~\ref{dynrmspowTAB}, respectively. The dark blue region in the combustor shows the vicinity of the pressure node where the oscillations have noisy behaviors. The main difference between the two simulations occurs in the middle of the oxidizer post, in which the NN-based simulation underestimates the rms. This is attributed to the inaccuracies in the PVRR model. In the oxidizer post, the PVRR should be zero or  very close to zero, as there is no combustion occurring there. However, the training data was skewed toward higher values of PVRR, as they are important in the modeling of combustion instabilities. Generating an optimized training set for PVRR is a next step of this work.

Furthermore,  the modified Rayleigh index is compared between the NN-based and table-based simulations in \figurename~\ref{WCRIrr} and \figurename~\ref{WCRInna}, respectively. The high positive values in the left upper corner of the combustor show the driving effects of combustion on pressure instabilities. The two simulations are more consistent in the upper zone of the combustor. As it was discussed above and in \cite{Shadram_Journal_1}, the behavior near the centerline is more complicated due to numerical considerations. 

\begin{figure}[hbt!]		
	\begin{subfigure}{.49\textwidth}
		\centering		
		\includegraphics[width=0.95\textwidth]{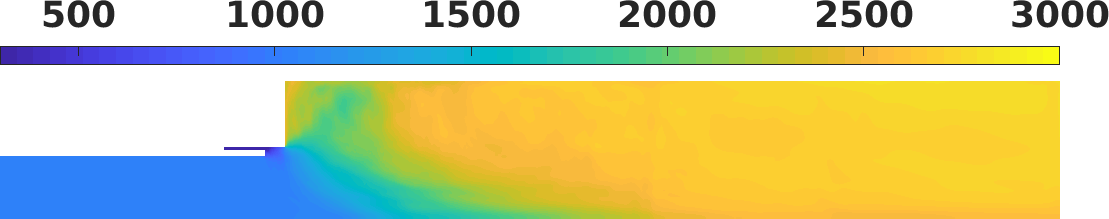}
		\caption{$\widehat{T}$ (\si{\kelvin}), table }\label{WTsigcotimeavgdyn8}	
			\centering		
		\includegraphics[width=0.95\textwidth]{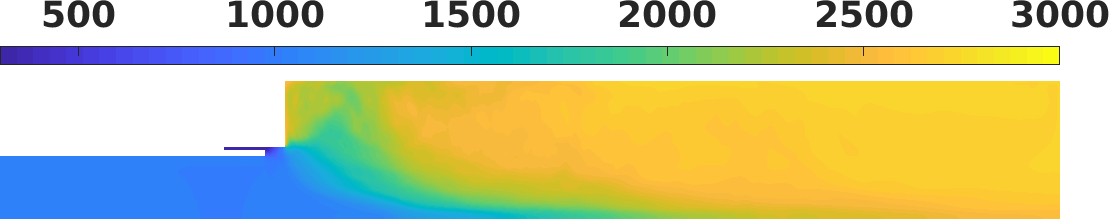}
		\caption{$\widehat{T}$ (\si{\kelvin}), NN}\label{NNTsigcotimeavgdyn8}				
		\centering		
		\includegraphics[width=0.95\textwidth]{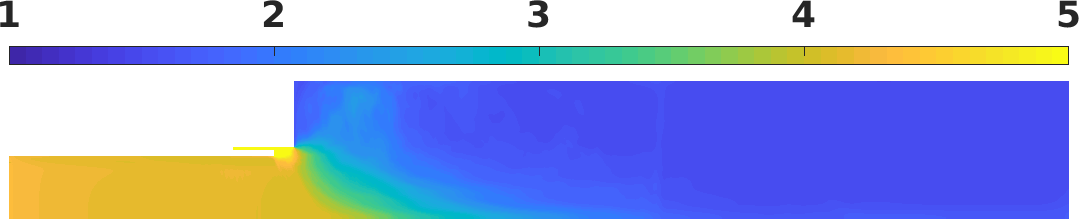}
		\caption{$\bar{\rho}$ (\si{\kilogram\per\meter\cubed}), table}\label{Wrhosigcotimeavgdyn8}			
		\centering		
\includegraphics[width=0.95\textwidth]{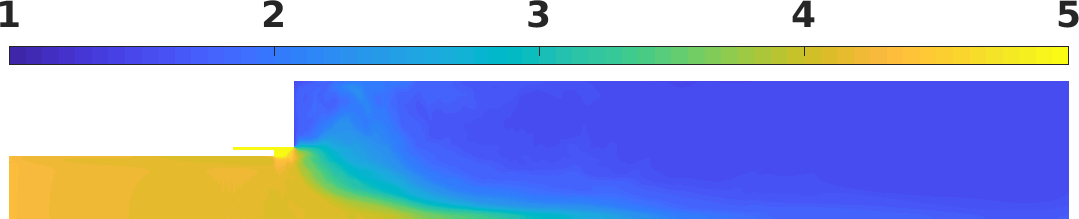}
\caption{$\bar{\rho}$ (\si{\kilogram\per\meter\cubed}), NN}\label{NNrhosigcotimeavgdyn8}		
		\centering		
	\end{subfigure}				
	\begin{subfigure}{.49\textwidth}		
				\includegraphics[width=0.95\textwidth]{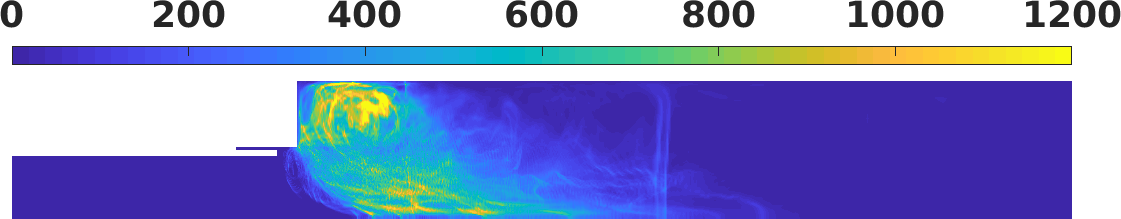}
		\caption{$\widehat{\dot{\omega}}_C$ (\si{\kilogram\per\meter\cubed\per\second}), table}\label{WProdCsigcotimeavgdyn8}					
		\centering
		\includegraphics[width=0.95\textwidth]{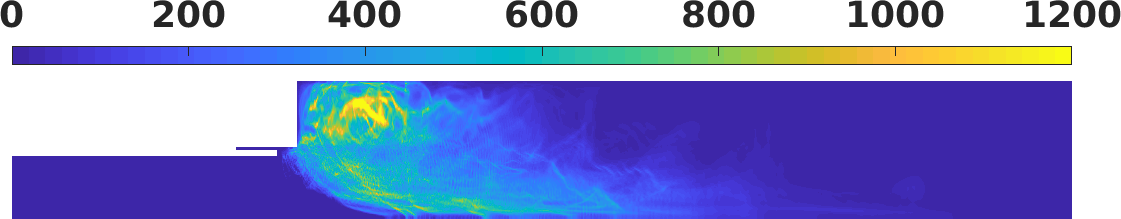}
		\caption{$\widehat{\dot{\omega}}_C$ (\si{\kilogram\per\meter\cubed\per\second}), NN}\label{NNProdCsigcotimeavgdyn8}
		\centering
			\includegraphics[width=0.95\textwidth]{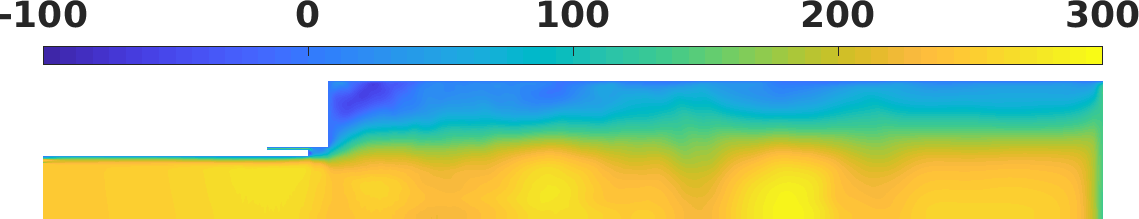}
			\caption{$\widehat{U}$ (\si{\meter\per\second\squared}), table }\label{WUsigcotimeavgdyn8}	
			\centering		
			\includegraphics[width=0.95\textwidth]{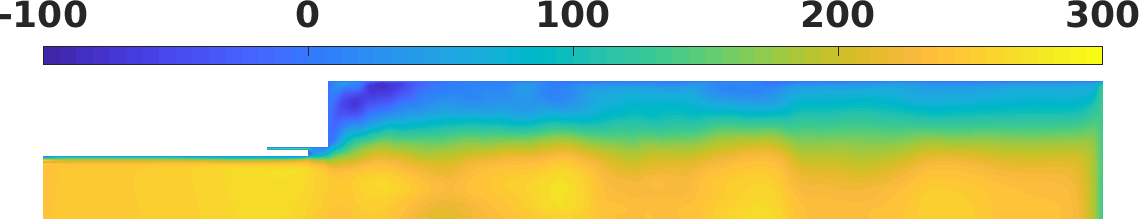}
			\caption{$\widehat{U} (\si{\meter\per\second\squared})$, NN}\label{NNUsigcotimeavgdyn8}										
	\end{subfigure}			
	\caption{ DE: time-averaged temperature, PVRR, density, and axial velocity from the NN- and table-based simulations}	\label{timeavgcontcomp}					
\end{figure}

In \figurename~\ref{timeavgcontcomp}, time-averaged values of  flow temperature, PVRR, density, and axial velocity from the NN-based and the table-based simulations are compared. In all these variables, great consistency is observed between the two simulations. Mainly, PVRR shows differences on the centerline and axial velocity in the vicinity of the pressure node. In both of these locations, noisy behavior is observed due to their numerical and physical characteristics.
\subsubsection{Test on CFD: Triggered Oscillation}
\begin{figure}[hbt!]
	\begin{subfigure}{.49\textwidth}
		\centering
		\includegraphics[width=0.95\textwidth]{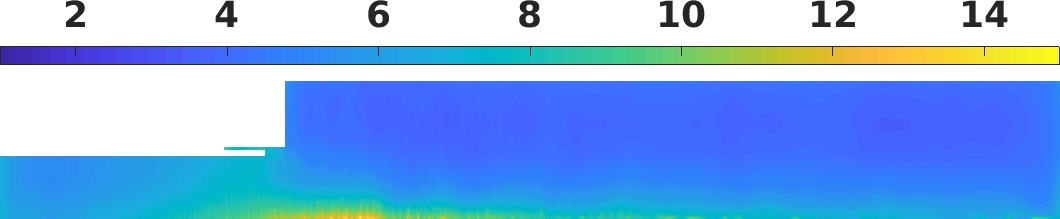}
		\caption{Overall relative error (\%)}\label{trig_efullcont6}		
	\end{subfigure}
	\begin{subfigure}{.49\textwidth}
		\centering
\includegraphics[width=0.95\textwidth]{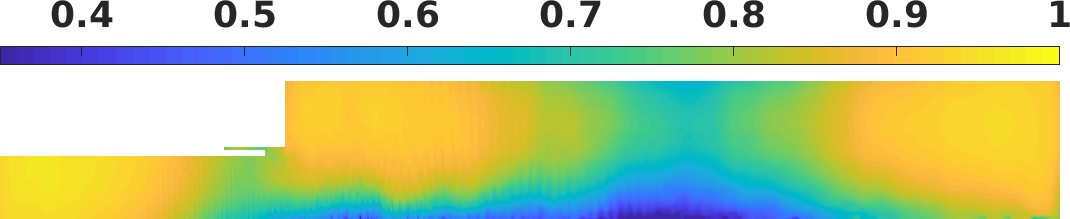}
\caption{Fluctuation correlation}\label{trig_corcont6}
	\end{subfigure}
	\caption{TG: distribution of $e_o$ (\%) and pressure fluctuation signal correlation between the NN- and table-based simulations}\label{trig_corcomp2}
\end{figure}

In the triggering (TG) scenario, heat loss is introduced by imposing an isothermal boundary condition on the chamber wall, and a sine wave perturbation on the propellant mass flow rate is applied for two periods. The top wall is modeled to be isothermal at 1030 \si{\kelvin}. The overall relative error between the pressure waveforms of the NN-based and table-based simulations
 is shown in \figurename~\ref{trig_efullcont6} for most of the different grid points in the combustor to be less than 5\%. We also look into the correlation of the pressure fluctuation signals between the NN-based  and table-based simulations using the Pearson formula. The correlations are very high, except on the centerline in the vicinity of the pressure node, as shown in \figurename~\ref{trig_corcont6}.

\begin{figure}[hbt!]
	\begin{subfigure}{.49\textwidth}
	\centering
	\includegraphics[width=0.95\textwidth]{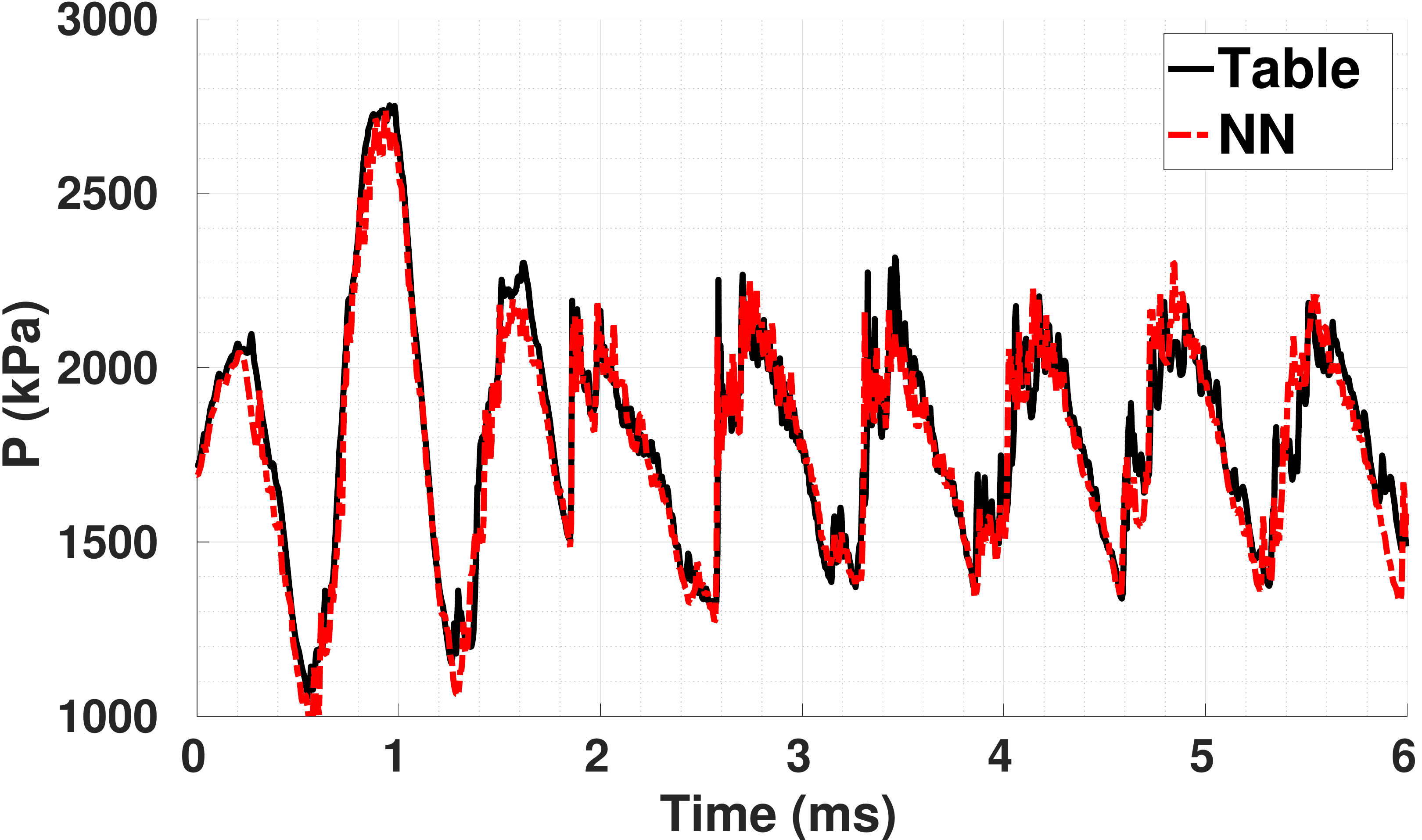}
	\caption{Oxidizer post, $r=0.5$ \si{\centi\meter}, $x=-10$ \si{\centi\meter}}\label{trig_Pcompop_28_34}		
\end{subfigure}
\begin{subfigure}{.49\textwidth}
	\centering
	\includegraphics[width=0.95\textwidth]{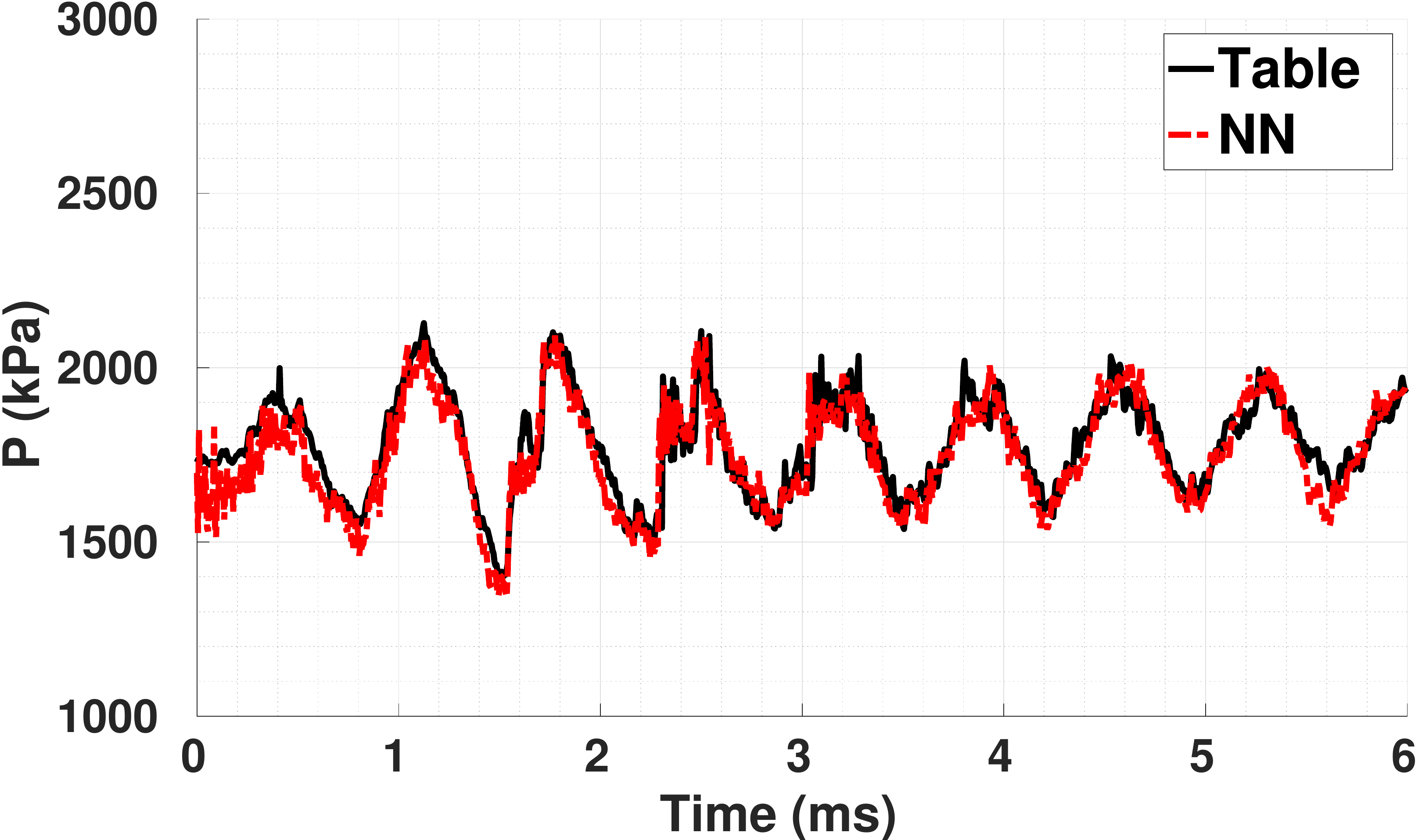}
	\caption{Shear layer, $r=1.13$ \si{\centi\meter}, $x=8$ \si{\centi\meter}}\label{trig_Pcompfp1_18_408}			
\end{subfigure}			
\caption{ TG: comparison of pressure time signals 
	at representative points} \label{trig_Psigcomp2}		
\end{figure}

Great consistency is observed in the pressure signal at the two representative locations  (in the oxidizer post and in the shear layer), \figurename~\ref{trig_Psigcomp2}. The pressure response to the perturbation in the mass flow rate can be observed in both simulations, with its effect on the response becoming damped after 2.5 \si{\milli\second}.

\begin{figure}[hbt!]
	\begin{subfigure}{.49\textwidth}
		\centering
		\includegraphics[width=.95\textwidth]{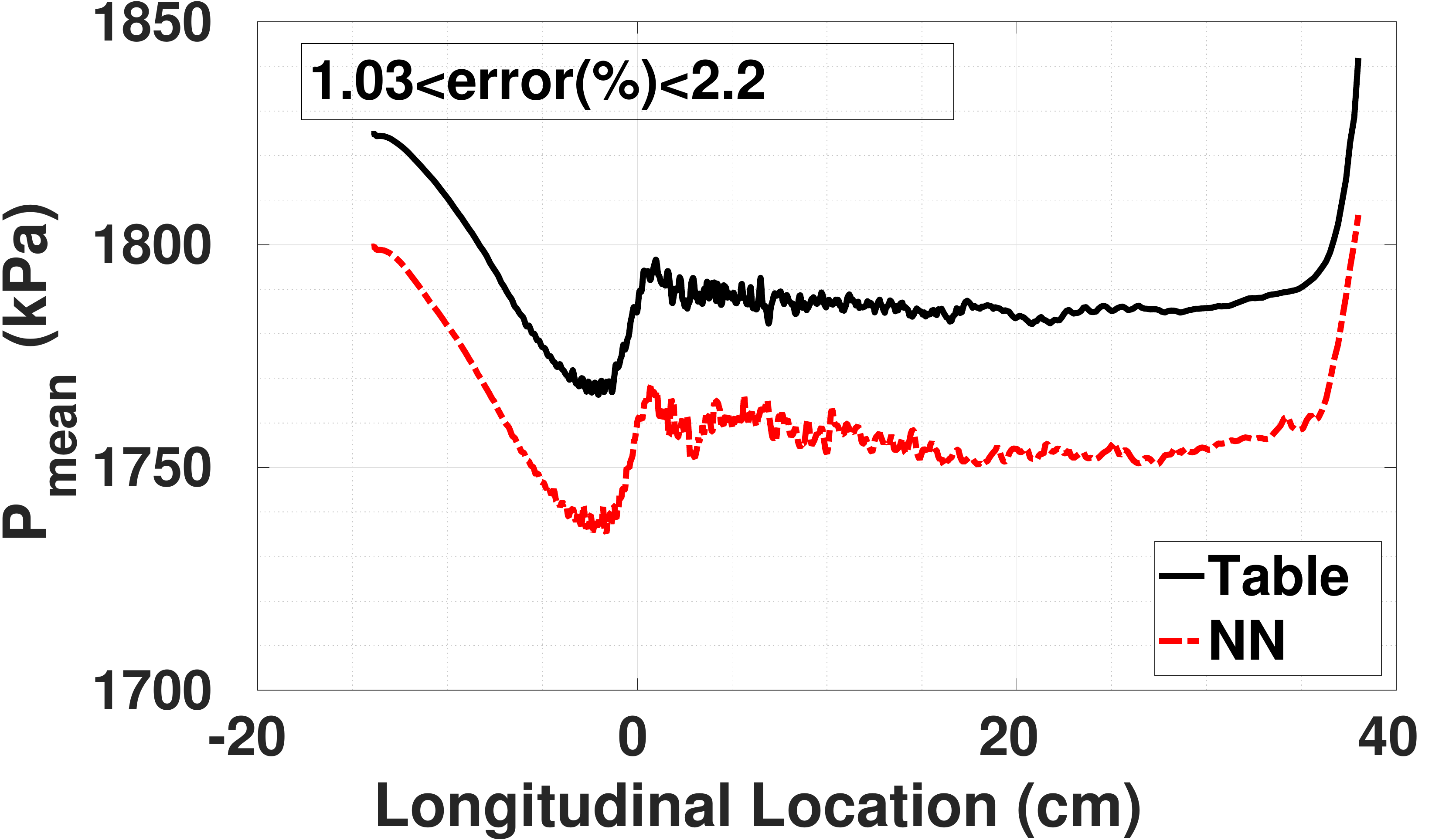}
		\caption{Mean pressure (\si{\kilo\pascal})}\label{trig_PMSmean6}
			\end{subfigure}		
			\begin{subfigure}{.49\textwidth}			
			\centering
			\includegraphics[width=.95\textwidth]{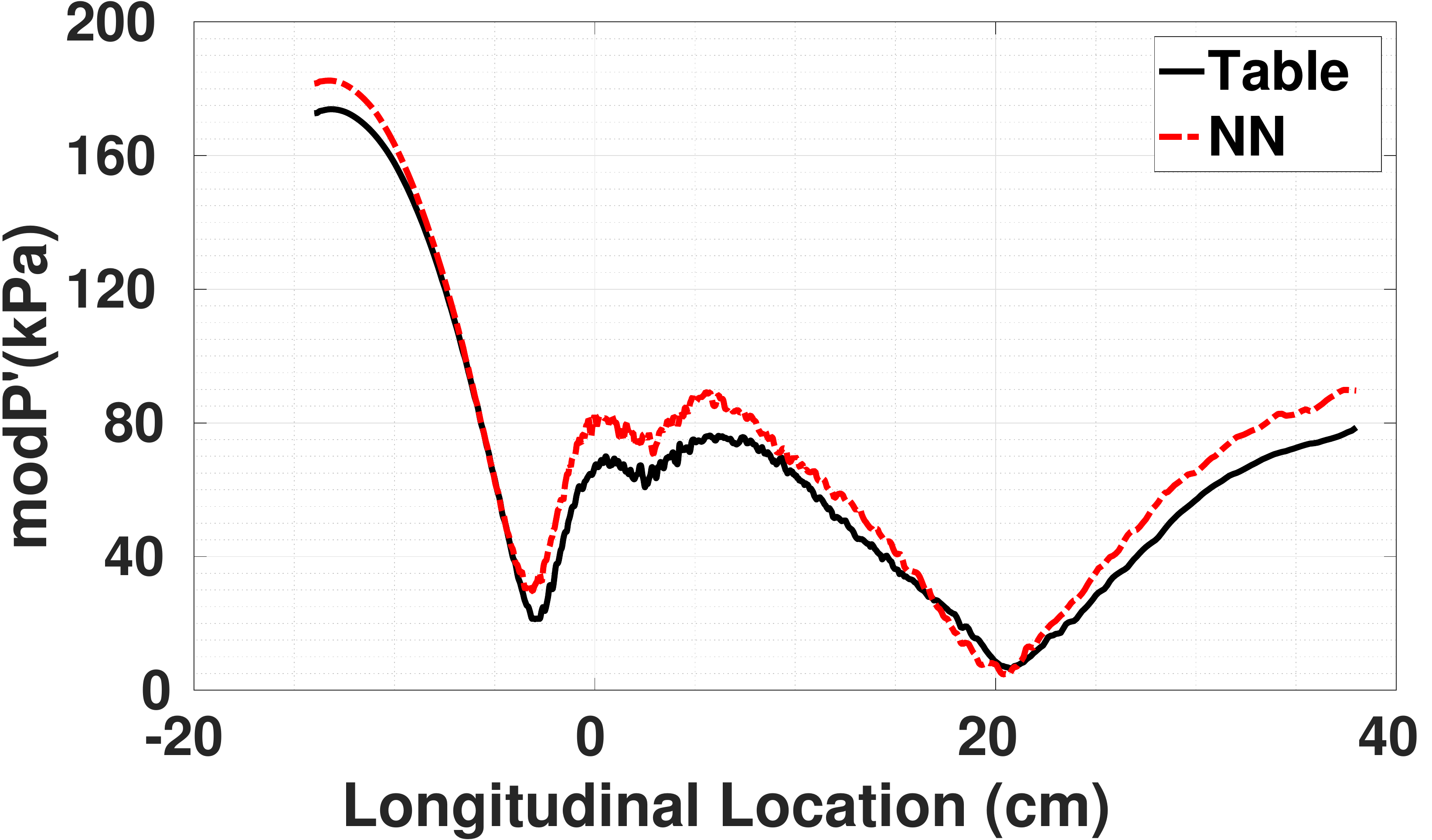}
			\caption{First mode shape (\si{\kilo\pascal}) }\label{trig_PMSfirst6}	
		\end{subfigure}		
	\begin{subfigure}{.49\textwidth}
		\centering
		\includegraphics[width=.95\textwidth]{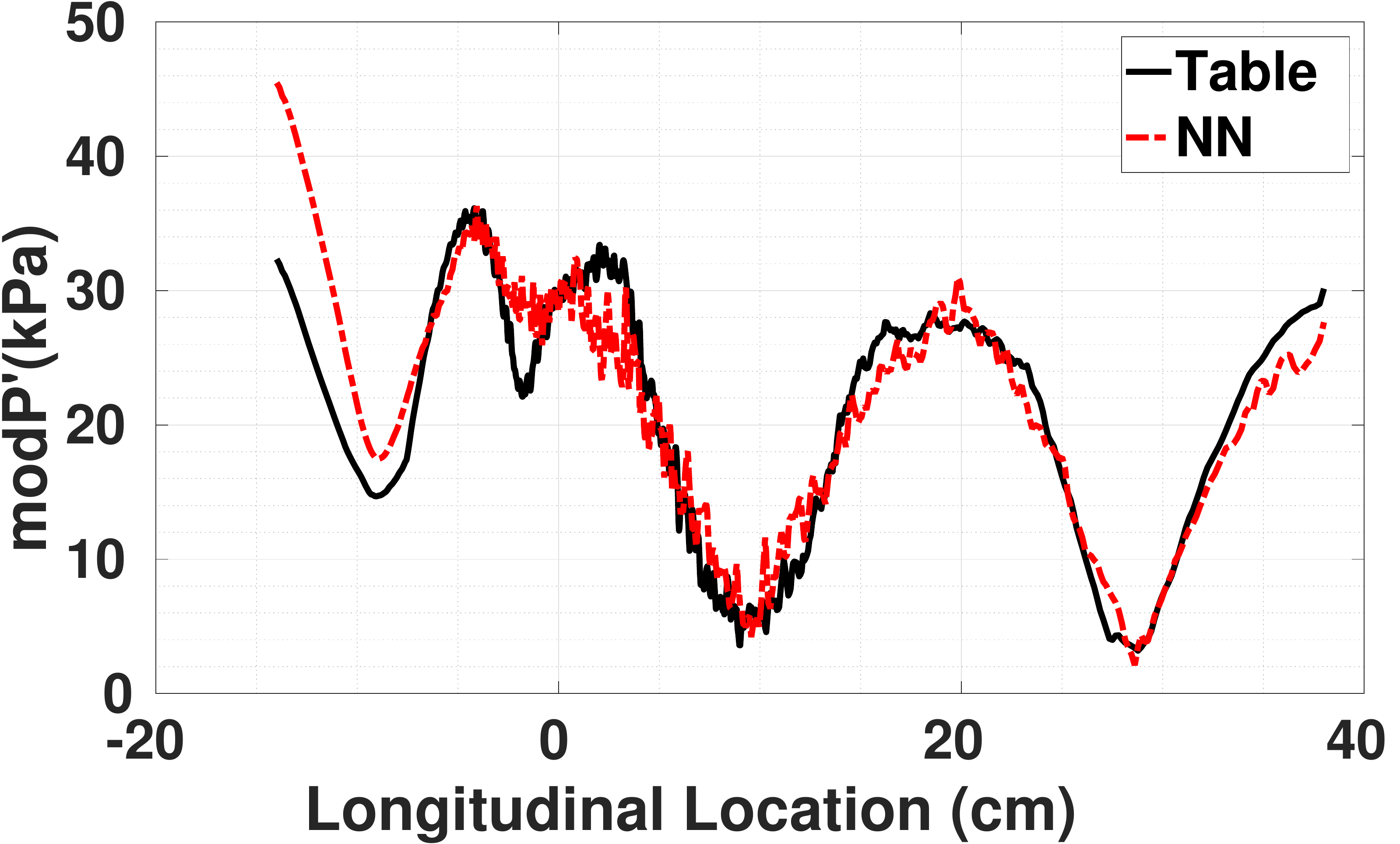}
		\caption{Second mode shape (\si{\kilo\pascal})}\label{trig_PMSsec6}		
	\end{subfigure}		
	\begin{subfigure}{.49\textwidth}
		\centering
		\includegraphics[width=.95\textwidth]{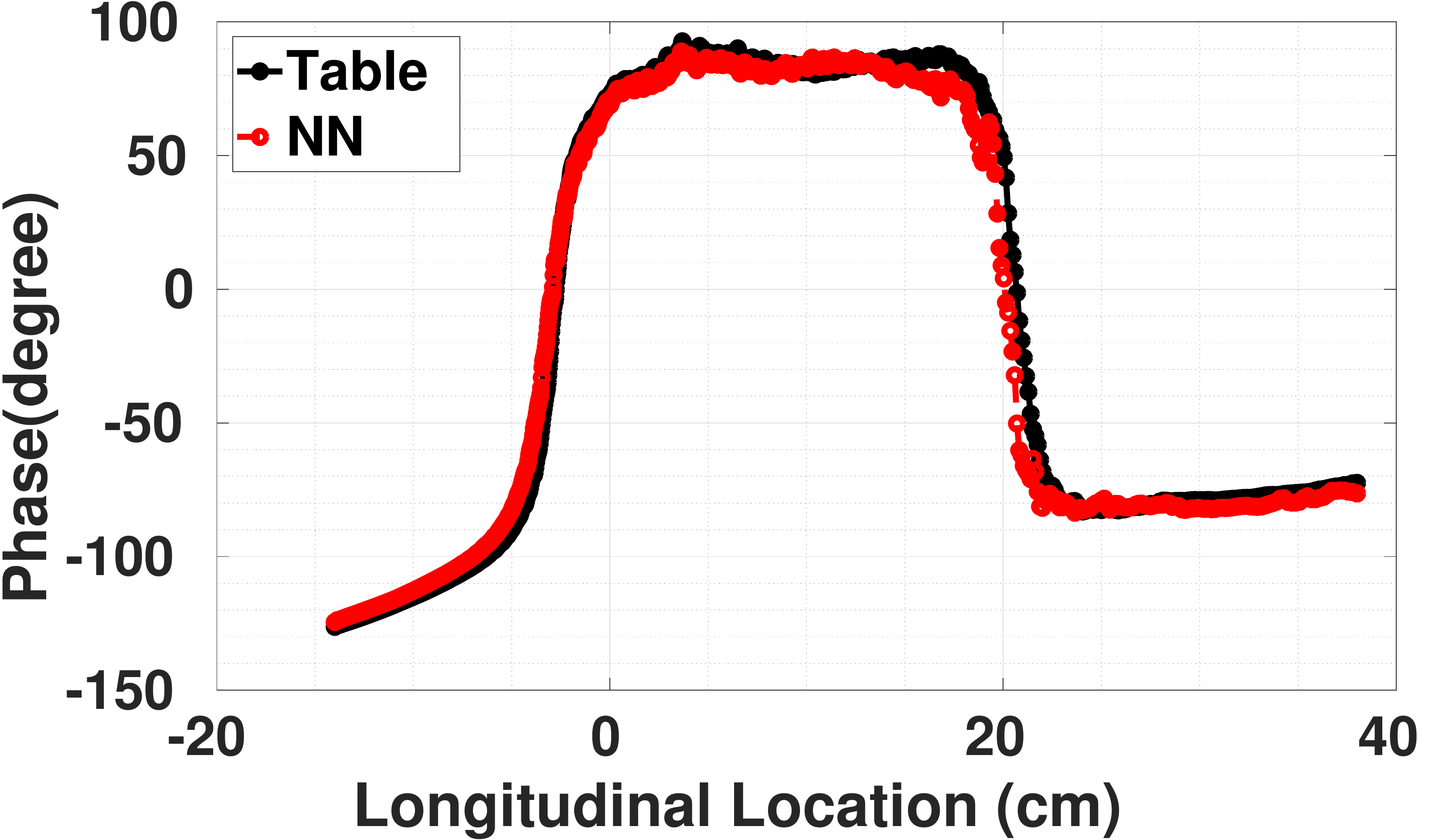}
		\caption{Phase of the first mode shape (\si{\kilo\pascal}) }\label{trig_PMSphasefirst6}			
	\end{subfigure}		
	\caption{TG: comparison of pressure mean, the first  longitudinal mode and its phase, and the second  longitudinal  mode shapes between NN-based and table-based simulations along  the  centerline}\label{trig_modshcompnew}
\end{figure}

The mean of pressure signal on the centerline, the amplitude and phase of the first harmonic of the Fourier spectrum, and the amplitude of the second  longitudinal mode are plotted in \figurename~\ref{trig_modshcompnew}. The  mean pressure calculated from the NN-based simulation is within 2.2\% of the table-based simulation. The amplitude of the first and second  longitudinal modes are highly consistent between the two simulations. Differences exist in the inlet of the oxidizer post and near the nozzle, mostly in the second  longitudinal mode. The phase graph of the NN-based simulation matches the one for the table-based simulation. The frequencies of the first and the second longitudinal modes across the combustor are around 1500 \si{\hertz} and 3000 \si{\hertz}, respectively.

\begin{figure}[H]
	\begin{subfigure}{.49\textwidth}
		\centering
		\includegraphics[width=0.95\textwidth]{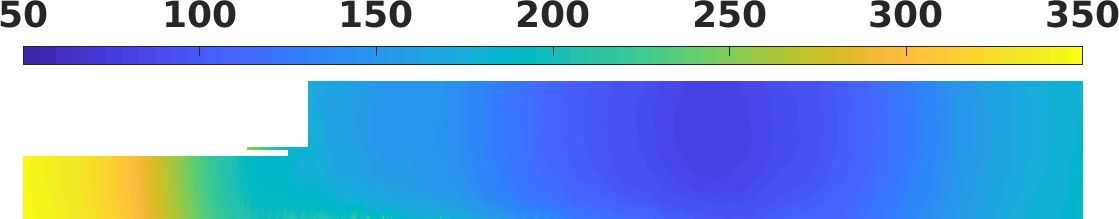}
		\caption{Table: rms}\label{trig_dynrmspowTAB}		
		\centering
		\includegraphics[width=0.95\textwidth]{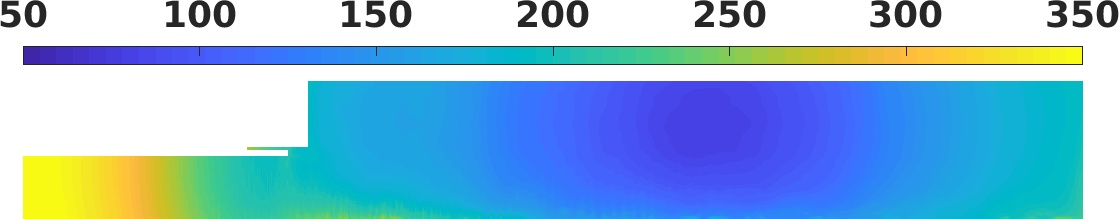}
		\caption{NN: rms}\label{trig_dynrmspowNN6}
	\end{subfigure}
	\begin{subfigure}{.49\textwidth}
		\centering
		\includegraphics[width=0.95\textwidth]{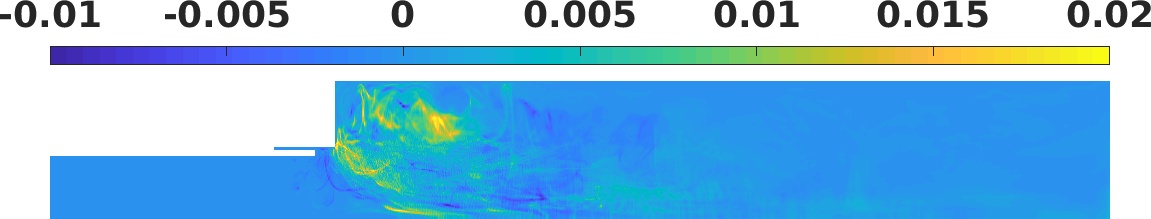}
		\caption{ Table: \textit{mRI}}\label{trig_WCRIrr}		
		\centering
		\includegraphics[width=0.95\textwidth]{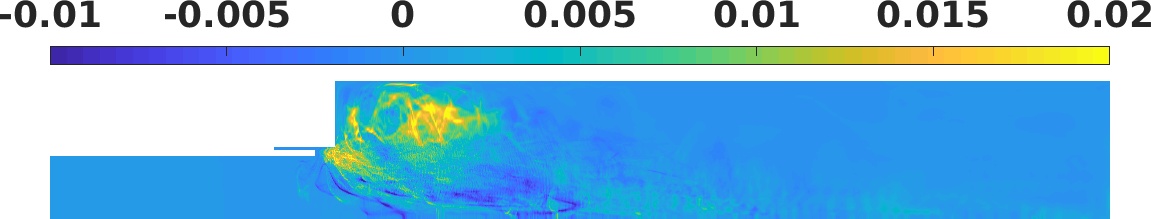}
		\caption{ NN: \textit{mRI} }\label{trig_WCRInna}										
	\end{subfigure}		
	\caption{ TG:  The rms of pressure signal fluctuations are compared between the simulations based on  NN  (\ref{trig_dynrmspowNN6}) and table(\ref{trig_dynrmspowTAB}). Also the \textit{mRI} from the NN-based (\ref{trig_WCRInna}) and the table-based (\ref{trig_WCRIrr}) simulations are compared}\label{trig_rmsratcomp3}
\end{figure}

Similar to DE scenario, the rms of pressure fluctuations across the grid points of the combustor are compared between the NN-based and table-based simulations in \figurename~\ref{trig_dynrmspowNN6} and \figurename~\ref{trig_dynrmspowTAB}, respectively. 

The two graphs from the two simulations are highly consistent. Slight differences exist in the  blue boundary in the combustion zone, indicating a slight overestimation of the pressure  fluctuations in the NN-based simulation. This overestimation can also be observed through the mRI quantity that is compared between the NN-based and table-based simulations in \figurename~\ref{trig_WCRInna} and \figurename~\ref{trig_WCRIrr}, respectively. The mRI shown in \figurename~\ref{trig_WCRInna} indicates higher values leading to a higher instability level.

\begin{figure}[H]
	\begin{subfigure}{.32\textwidth}
	\centering
	\includegraphics[width=0.98\textwidth]{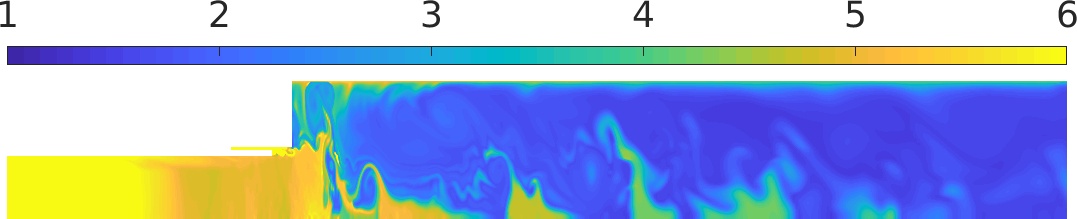}
	\caption{$\overline{\rho}$ (\si{\kilo\gram\per\meter\cubed}), $t_1$, table}\label{TGW8Wrhosigco_201}	
			\centering
	\includegraphics[width=0.98\textwidth]{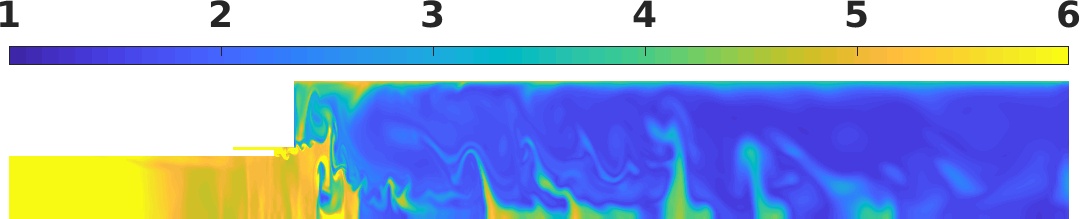}
	\caption{$\overline{\rho}$ (\si{\kilo\gram\per\meter\cubed}), $t_1$, NN}\label{TGNN8NNrhosigco_201}	
\end{subfigure}
	\begin{subfigure}{.32\textwidth}
	\centering
	\includegraphics[width=0.98\textwidth]{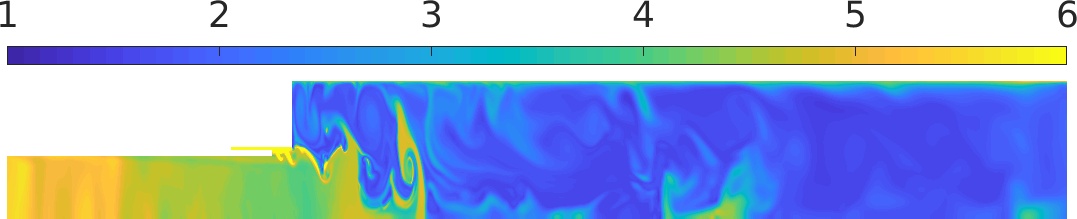}
	\caption{$\overline{\rho}$ (\si{\kilo\gram\per\meter\cubed}), $t_2$, table}\label{TGW8Wrhosigco_401}	
	\centering
	\includegraphics[width=0.98\textwidth]{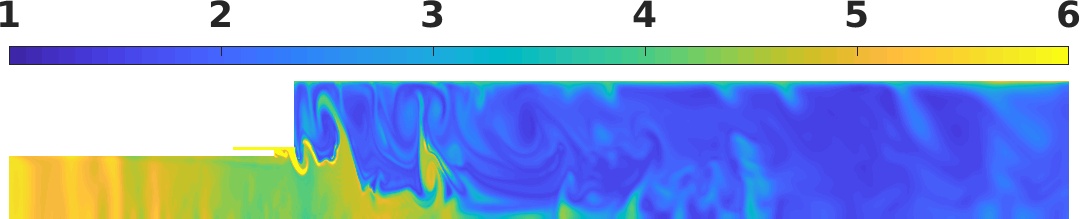}
	\caption{$\overline{\rho}$ (\si{\kilo\gram\per\meter\cubed}), $t_2$, NN}\label{TGNN8NNrhosigco_401}	
\end{subfigure}
\begin{subfigure}{.32\textwidth}
	\centering
	\includegraphics[width=0.98\textwidth]{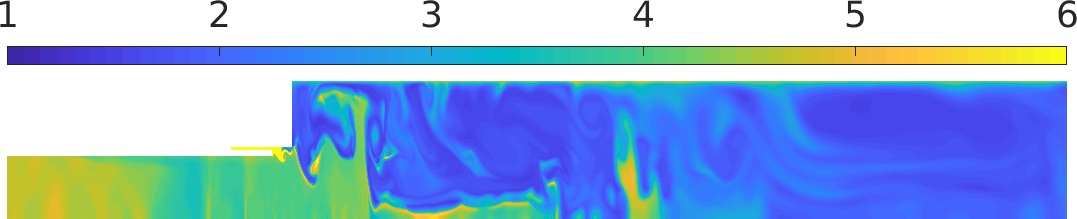}
	\caption{$\overline{\rho}$ (\si{\kilo\gram\per\meter\cubed}), $t_3$, table}\label{TGW8Wrhosigco_601}	
	\centering
	\includegraphics[width=0.98\textwidth]{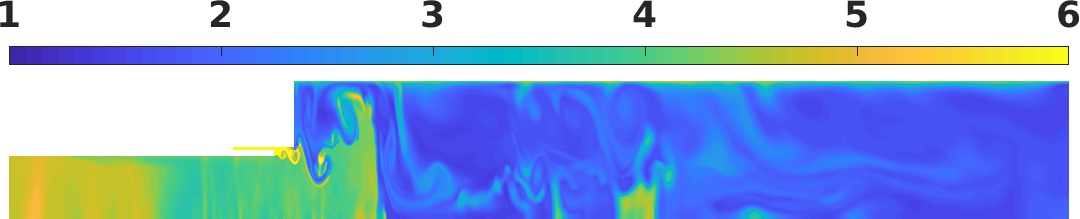}
	\caption{$\overline{\rho}$ (\si{\kilo\gram\per\meter\cubed}), $t_3$, NN}\label{TGNN8NNrhosigco_601}	
\end{subfigure}
\caption{TG: Density}\label{trig_rho}
\end{figure}

\begin{figure}[H]
	\begin{subfigure}{.32\textwidth}
		\centering
		\includegraphics[width=0.98\textwidth]{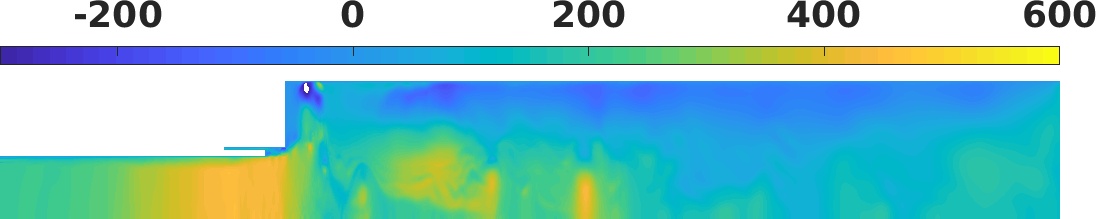}
		\caption{$\widetilde{U}$  (\si{\meter\per\second}), $t_1$, table}\label{TGW8WUsigco_201}	
		\centering
		\includegraphics[width=0.98\textwidth]{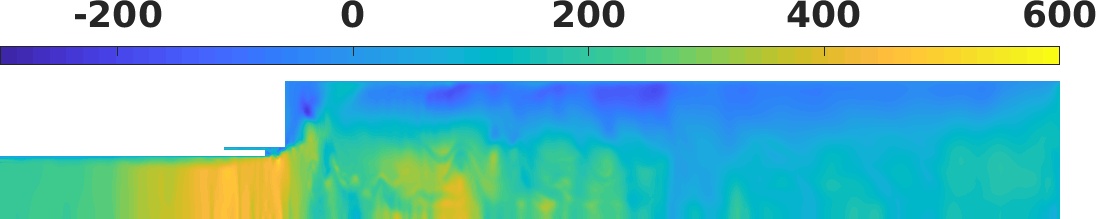}
		\caption{$\widetilde{U}$  (\si{\meter\per\second}), $t_1$, NN}\label{TGNN8NNUsigco_201}	
	\end{subfigure}
	\begin{subfigure}{.32\textwidth}
		\centering
		\includegraphics[width=0.98\textwidth]{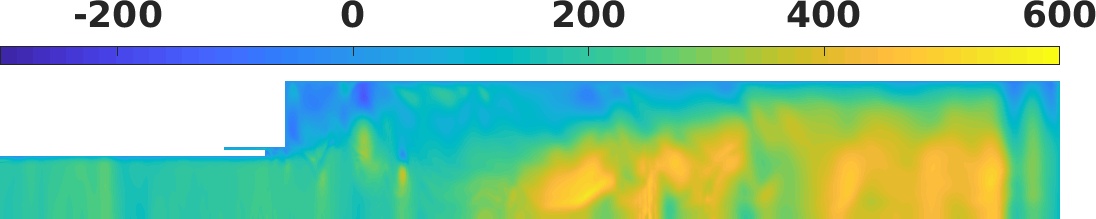}
		\caption{$\widetilde{U}$  (\si{\meter\per\second}), $t_2$, table}\label{TGW8WUsigco_401}	
		\centering
		\includegraphics[width=0.98\textwidth]{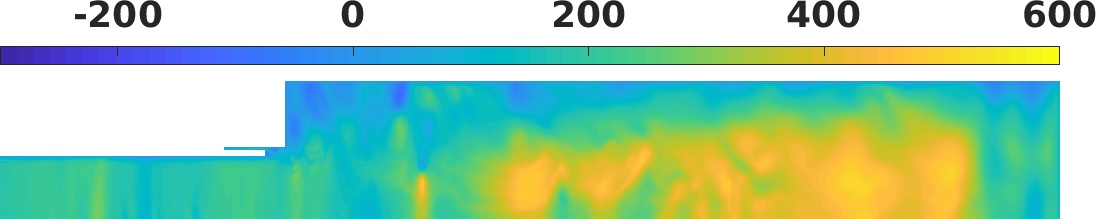}
		\caption{$\widetilde{U}$  (\si{\meter\per\second}), $t_2$, NN}\label{TGNN8NNUsigco_401}	
	\end{subfigure}
	\begin{subfigure}{.32\textwidth}
		\centering
		\includegraphics[width=0.98\textwidth]{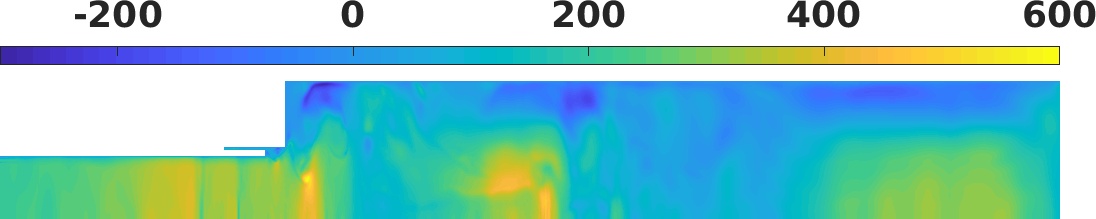}
		\caption{$\widetilde{U}$  (\si{\meter\per\second}), $t_3$, table}\label{TGW8WUsigco_601}	
		\centering
		\includegraphics[width=0.98\textwidth]{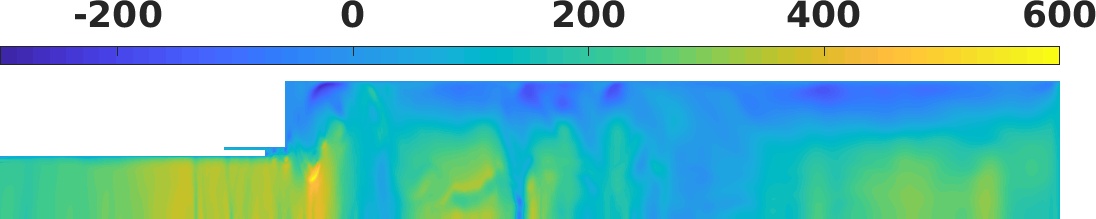}
		\caption{$\widetilde{U}$  (\si{\meter\per\second}), $t_3$, NN}\label{TGNN8NNUsigco_601}	
	\end{subfigure}
	\caption{TG: Axial velocity}\label{trig_U}
\end{figure}

\begin{figure}[H]
	\begin{subfigure}{.32\textwidth}
		\centering
		\includegraphics[width=0.98\textwidth]{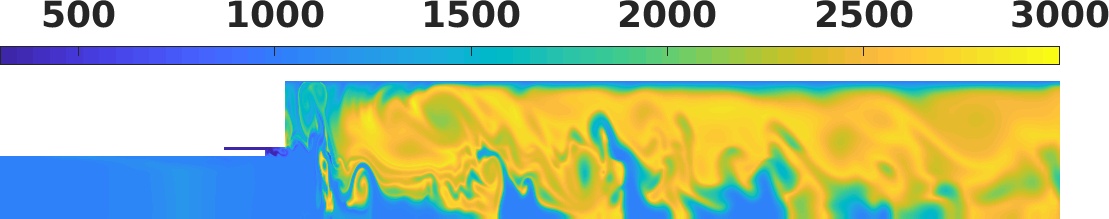}
		\caption{$\widetilde{T}$  (\si{\kelvin}), $t_1$, table}\label{TGW8WTsigco_201}	
		\centering
		\includegraphics[width=0.98\textwidth]{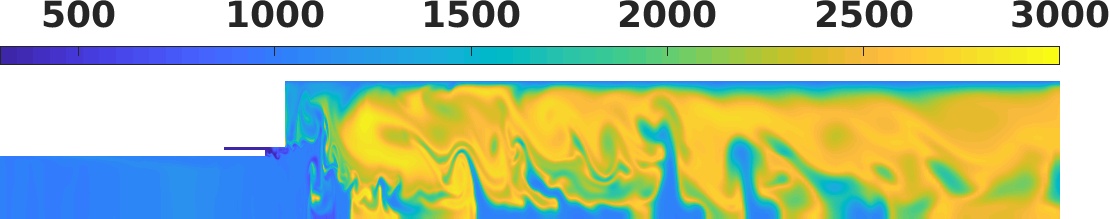}
		\caption{$\widetilde{T}$  (\si{\kelvin}), $t_1$, NN}\label{TGNN8NNTsigco_201}	
		\end{subfigure}
	\begin{subfigure}{.32\textwidth}
		\centering
		\includegraphics[width=0.98\textwidth]{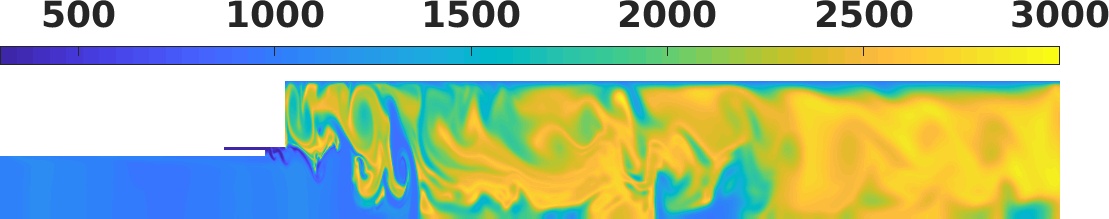}
		\caption{$\widetilde{T}$  (\si{\kelvin}), $t_2$, table}\label{TGW8WTsigco_401}	
		\centering
		\includegraphics[width=0.98\textwidth]{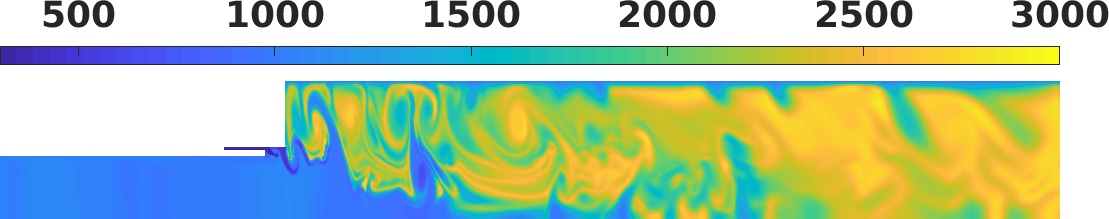}
		\caption{$\widetilde{T}$  (\si{\kelvin}), $t_2$, NN}\label{TGNN8NNTsigco_401}	
	\end{subfigure}
	\begin{subfigure}{.32\textwidth}
		\centering
		\includegraphics[width=0.98\textwidth]{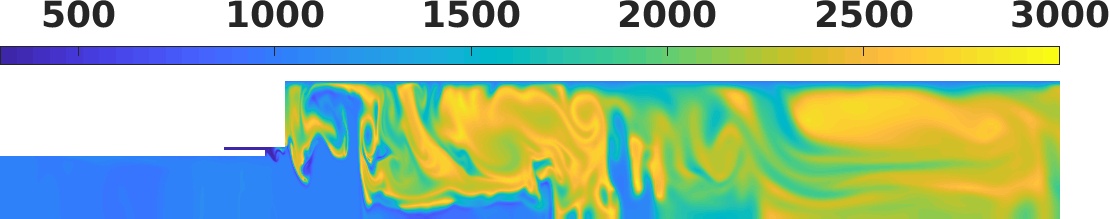}
		\caption{$\widetilde{T}$  (\si{\kelvin}), $t_3$, table}\label{TGW8WTsigco_601}	
		\centering
		\includegraphics[width=0.98\textwidth]{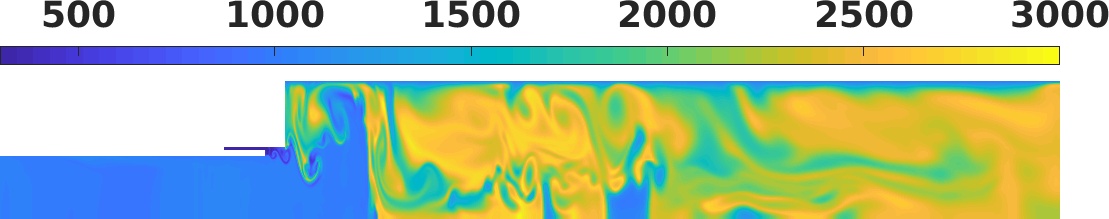}
		\caption{$\widetilde{T}$  (\si{\kelvin}), $t_3$, NN}\label{TGNN8NNTsigco_601}	
	\end{subfigure}
	\caption{TG: Temperature}\label{trig_T}
\end{figure}

\begin{figure}[H]
	\begin{subfigure}{.32\textwidth}
		\centering
		\includegraphics[width=0.98\textwidth]{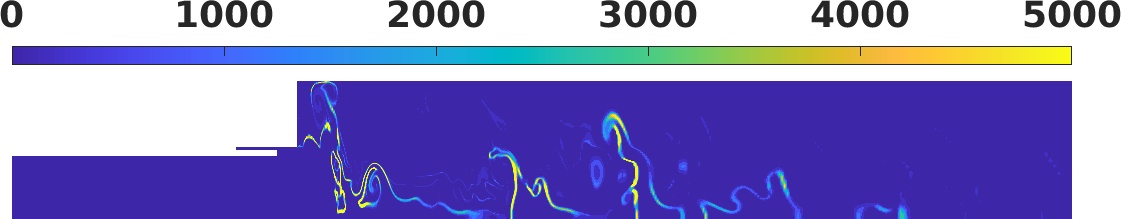}
		\caption{$\widetilde{\dot{\omega}}_C$ (\si{\kilogram\per\meter\cubed\per\second}), $t_1$, table}\label{ProdCGW8WProdCsigco_201}	
		\centering
		\includegraphics[width=0.98\textwidth]{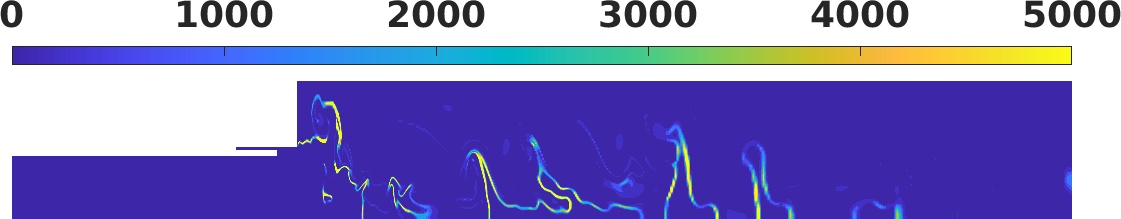}
		\caption{$\widetilde{\dot{\omega}}_C$ (\si{\kilogram\per\meter\cubed\per\second}), $t_1$, NN}\label{ProdCGNN8NNProdCsigco_201}	
		\centering		
	\end{subfigure}
	\begin{subfigure}{.32\textwidth}
		\centering
		\includegraphics[width=0.98\textwidth]{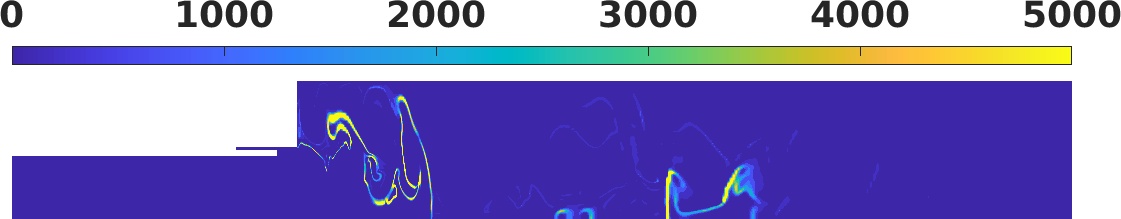}
		\caption{$\widetilde{\dot{\omega}}_C$ (\si{\kilogram\per\meter\cubed\per\second}), $t_2$, table}\label{ProdCGW8WProdCsigco_401}	
		\centering
		\includegraphics[width=0.98\textwidth]{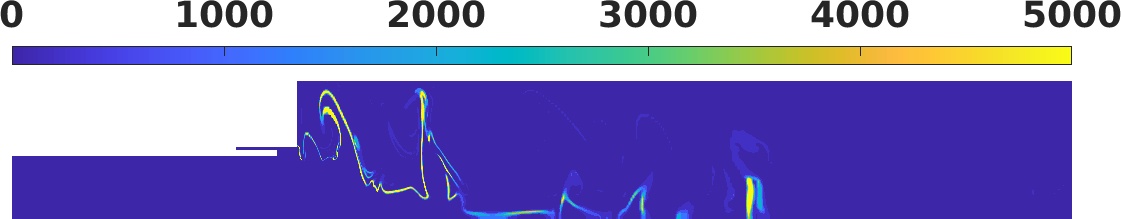}
		\caption{$\widetilde{\dot{\omega}}_C$ (\si{\kilogram\per\meter\cubed\per\second}), $t_2$, NN}\label{ProdCGNN8NNProdCsigco_401}	
		\centering
	\end{subfigure}
	\begin{subfigure}{.32\textwidth}
		\centering
		\includegraphics[width=0.98\textwidth]{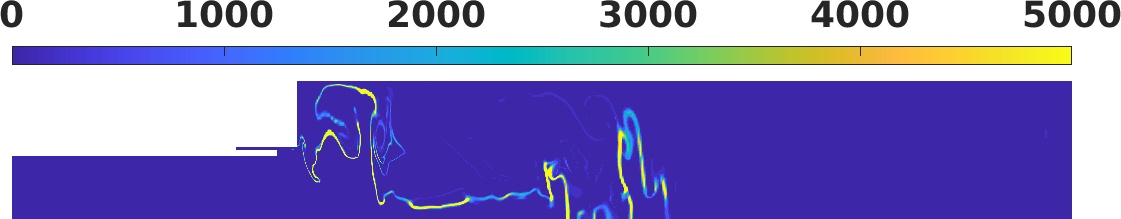}
		\caption{$\widetilde{\dot{\omega}}_C$ (\si{\kilogram\per\meter\cubed\per\second}), $t_3$, table}\label{ProdCGW8WProdCsigco_601}	
		\centering
		\includegraphics[width=0.98\textwidth]{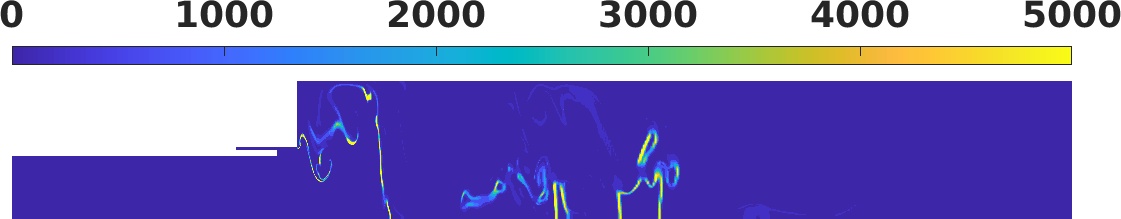}
		\caption{$\widetilde{\dot{\omega}}_C$ (\si{\kilogram\per\meter\cubed\per\second}), $t_3$, NN}\label{ProdCGNN8NNProdCsigco_601}	
		\centering
	\end{subfigure}
	\caption{TG: PVRR}\label{trig_ProdC}
\end{figure}

In \figurename~\ref{trig_rho}, the snapshots of density are compared at the three time points of 1, 2, and 3 \si{\milli\second}. The graphs are highly consistent between the two simulations. The first row in the figure shows the table-based simulation, and the second row shows the NN-based simulation. This consistency is also observed in the comparison of snapshots of axial velocity in \figurename~\ref{trig_U},  flow temperature in \figurename~\ref{trig_T}, and PVRR in \figurename~\ref{trig_ProdC}, at the same time points. Although the detailed behavior is different at each snapshot, the overall characteristics are consistent between the two simulations. For example, the location of the flame front, as shown in the graphs for PVRR, is similar between the two simulations.


\subsubsection{ Computational cost} 
 In this work, we have not attempted script optimization and parallelization of the NN implementation in the CFD simulation. However, we can still compare the computational cost of the NN-based and table-based simulations lasting for one millisecond (200 snapshots) of simulation time in the dynamic equilibrium test. The average required CPU times per time step for the NN-based simulation is 39.2 times for that the table-based simulation. The computational cost of the table-based simulation is discussed in \cite{Tuan1}, where a 0.28 core-hour per millisecond was reported for an axisymmetric simulation with $6.26e4$ grid points (the current work uses $1.375e5$ grid points) based on a flamelet model for combustion of 27 species. 
The flamelet model requires around 1.2 GB of hard disk space, almost 118 times the required memory by the NN model, which is around 102 kB. 
As discussed before, given the very low required memory by NN-based models, GPU implementation is more feasible for the parallelization of the code and might be able to  compensate for the higher required flops of the NN-based models. In 2010, one study showed that 16 to 137 times speed up in fluid mechanics related problems can be achieved when they are implemented on GPU \cite{GPUERA}.
\section{Conclusions} \label{conclu}
NN-based models are proposed to represent the flamelet model outputs. The novel aspect is that the NN models are trained from data that
are sampled from a CFD simulation of a transient scenario of a CVRC engine, in which pressure oscillations spontaneously develop from a low value to reach a high-amplitude limit cycle.
Knowledge of the physics of the problem was leveraged to enhance the training data set, giving rise to a physics-aware NN-based model. We achieve this by adding more points implied by the physical constraints, such as boundary conditions of the flame variables in the flamelet space, to the training set. We also use an output of the flamelet model as an input to the other variables; which helps to improve each model's accuracy and consistency between the model outputs. Our method contributed to developing significantly better models that obey the physical constraints of the problem.

 Relatively low cost, NN models are proposed with less than 0.6\% training error. For PVRR, a very deep and more computationally expensive NN model is required for accurate modeling. The training error is around 1.5\%, and the model fits great, especially for high values of PVRR, which are crucial for modeling combustion instability. 

These NN-based models are validated first by testing them on the whole flamelet table and then by implementing them in CFD simulations of two different stability configurations  of a highly unstable configuration with14-\si{\centi\meter} oxidizer post, in addition to the transient case, and comparing them with the table-based simulations, which are the dynamic equilibrium and triggered instability cases of CVRC experiment. 
In both cases, great consistency is observed in different measures such as pressure, modified Rayleigh index, and temporal behavior of other quantities. The model showed some weaknesses in modeling regions in which there is no combustion, i.e. the oxidizer post, due to the lack of sufficient training data collected through the CFD simulation.

Offline comparison of the NN models with the flamelet libraries confirms that the performance of a NN model is better in regions where more data is included in the training set. As the source simulation is  a fuel lean setup, higher values of $\widetilde{Z}$ were lacking in the training set. Similarly, regions with higher turbulence levels of the table (high $B$ values) have weaker performance because of the lack of data in the training set.

The goal of this work is to explore the possibilities of developing data-driven combustion models only from data generated from highly accurate numerical simulations or experiments. The significance is that each source of data observes only a subset of the actual model (which is assumed to be the flamelet library in this work). The information fusion power of NNs has been essentially tested in this work. Having a richer training set from multiple experiments allows having a more comprehensive NN model by providing better coverage of the actual dynamic behavior.
 Further investigation on selecting the right set of input and data can help in achieving better models.
 
While the training and validation activities in this work are still closely associated with the flamelet based method, the NN model developed with careful consideration of the flame characteristics simultaneously improve its fidelity and avoid some of the combustion model limitations. The success of the proposed approach paves the way for developing more descriptive by utilizing experiment and multiple sources of data.
In future research, the framework presented in this work can be applied to high-quality data from sources other than the flamelet table, such as high fidelity LES and DNS, where the cost saving of using machine learning models can be highly advantageous.
 \section*{Acknowledgments}
This research was supported by  the Air Force Office of Scientific 
Research under Grant FA9550-18-1-0392 with Dr. Mitat Birkan as the 
scientific officer.

\bibliography{myref2}
\end{document}